%% file: paper.tex
\documentclass[11pt,english]{article}
\usepackage{mathptmx}
\usepackage[T1]{fontenc}
\usepackage[utf8]{inputenc}
\usepackage{babel}
\usepackage{float}
\usepackage{dsfont}
\usepackage{amsmath}
\usepackage{amssymb}
\usepackage{graphicx}
\usepackage[letterpaper]{geometry}
\usepackage{rotfloat}
\usepackage{setspace}
\usepackage[authoryear]{natbib}
\usepackage{microtype}
\usepackage[bookmarks=true,bookmarksnumbered=false,bookmarksopen=false,
 breaklinks=true,pdfborder={0 0 0},pdfborderstyle={},backref=false,colorlinks=false]
 {hyperref}
\hypersetup{pdftitle={The Gendered Impacts of Perceived Skin Tone: Evidence from African-American Siblings in 1870–1940},
 pdfauthor={Ran Abramitzky,  Jacob Conway,  Roy Mill,  and Luke C.D. Stein}}

\makeatletter
\usepackage{amsthm}
\usepackage{bm}
\usepackage{setspace}
\usepackage{sectsty}
\usepackage{url}

\usepackage{datetime}
\newdateformat{ourdate}{\monthname[\THEMONTH] \THEDAY , \THEYEAR}
\usepackage{pdflscape}
\usepackage{babel}
\usepackage{times}
\usepackage{enumitem}
\usepackage{booktabs}
\usepackage{dsfont}
\usepackage[font=small,skip=0pt]{caption}
\usepackage[bottom,hang,flushmargin]{footmisc}
\usepackage[position=top,parskip=0pt,aboveskip=0pt,belowskip=0pt]{subfig}

\usepackage{multirow}

\sectionfont{\noindent\normalfont\large\bf}
\subsectionfont{\noindent\normalfont\normalsize\bf}
\subsubsectionfont{\noindent\normalfont\it}

\input{input/Scalars/scalars}
\input{input/Scalars/usct_complexion}

\setlist[itemize]{noitemsep, topsep=0pt}

\usepackage{mdwlist}%
\usepackage{booktabs}			%
\usepackage{xtab}			%
\usepackage{tabularx}			%
\usepackage{dcolumn}			%
\usepackage{threeparttable}
\usepackage[referable]{threeparttablex}
\usepackage{rotating}
\usepackage{multirow}
\usepackage{epstopdf} %
\usepackage{pdflscape}
\def\sym#1{\ifmmode^{#1}\else\(^{#1}\)\fi}

\ifdefined\showcaptionsetup
 \PassOptionsToPackage{caption=false}{subfig}
\fi
\usepackage{subfig}
\makeatother

\usepackage{latexml}

\begin{document}
\title{\textbf{The Gendered Impacts of Perceived Skin Tone:}\\
\textbf{Evidence from African American Siblings in 1870--1940}}
\iflatexml
\author{Ran Abramitzky\thanks{Email: ranabr@stanford.edu, jacob.conway@chicagobooth.edu, milmilon@gmail.com,
lcdstein@babson.edu. We are grateful to Ido Abramitzky, Lukas Althoff,
Nicholas Bloom, Leah Boustan, Levi Boxell, Louis Cain, Gregory Clark,
William Darity Jr., Lauren Davenport, Giacomo De Giorgi, Ellora Derenoncourt,
Dave Donaldson, Pascaline Dupas, Liran Einav, Katherine Eriksson,
Joseph Ferrie, Osea Giuntella, Matthew Gentzkow, Regina Grafe, Avner
Greif, Richard Hornbeck, Caroline Hoxby, Alex Imas, Lawrence Katz,
David Kennedy, Mark Koyama, Trevon Logan, Neale Mahoney, Robert Margo,
Joel Mokyr, Petra Moser, Evan Rose, Aliya Saperstein, Debra Satz,
Isaac Sorkin, Richard Steckel, Michael Tomz, Heidi Williams, Guillermo
Woo-Mora, Gui Woolston, Gavin Wright, Noam Yuchtman, and numerous
seminar participants for helpful comments and suggestions. We thank
Juan David Torres for excellent research assistance. We thank Joe
Price and the BYU Record Linking lab for their help linking individuals
across census years. We also thank Lukas Althoff for sharing alternative
cross-census links that were constructed in \citet{althoff2022intergenerational}.
We thank several anonymous referees for their helpful feedback on
an earlier working paper by two of the authors \citep{mill2016race},
which is superseded by this paper. Jacob Conway gratefully acknowledges
funding from the National Science Foundation (grant DGE-1656518).
The Union Army data collection was supported by NIH/NIA grant P01
AG10120. The views expressed are the authors' and do not necessarily
reflect those of NIH or NIA.}\\
\emph{Stanford University and NBER}\and
Jacob Conway\\
\emph{University of Chicago}\and
Roy Mill\and
Luke C.D. Stein\\
\emph{Babson College}}
\else
\author{Ran Abramitzky, \emph{Stanford University and NBER}\thanks{Email: ranabr@stanford.edu, jacob.conway@chicagobooth.edu, milmilon@gmail.com,
lcdstein@babson.edu. We are grateful to Ido Abramitzky, Lukas Althoff,
Nicholas Bloom, Leah Boustan, Levi Boxell, Louis Cain, Gregory Clark,
William Darity Jr., Lauren Davenport, Giacomo De Giorgi, Ellora Derenoncourt,
Dave Donaldson, Pascaline Dupas, Liran Einav, Katherine Eriksson,
Joseph Ferrie, Osea Giuntella, Matthew Gentzkow, Regina Grafe, Avner
Greif, Richard Hornbeck, Caroline Hoxby, Alex Imas, Lawrence Katz,
David Kennedy, Mark Koyama, Trevon Logan, Neale Mahoney, Robert Margo,
Joel Mokyr, Petra Moser, Evan Rose, Aliya Saperstein, Debra Satz,
Isaac Sorkin, Richard Steckel, Michael Tomz, Heidi Williams, Guillermo
Woo-Mora, Gui Woolston, Gavin Wright, Noam Yuchtman, and numerous
seminar participants for helpful comments and suggestions. We thank
Juan David Torres for excellent research assistance. We thank Joe
Price and the BYU Record Linking lab for their help linking individuals
across census years. We also thank Lukas Althoff for sharing alternative
cross-census links that were constructed in \citet{althoff2022intergenerational}.
We thank several anonymous referees for their helpful feedback on
an earlier working paper by two of the authors \citep{mill2016race},
which is superseded by this paper. Jacob Conway gratefully acknowledges
funding from the National Science Foundation (grant DGE-1656518).
The Union Army data collection was supported by NIH/NIA grant P01
AG10120. The views expressed are the authors' and do not necessarily
reflect those of NIH or NIA.}\\
Jacob Conway, \emph{University of Chicago}\\
Roy Mill\\
Luke C.D. Stein, \emph{Babson College}}
\fi
\date{\textbf{October 7, 2026}}
\maketitle
\begin{abstract}
\begin{singlespace}
We study differences in economic outcomes by perceived skin tone among
African Americans using full-count U.S.\ decennial census data from
the late-19\textsuperscript{th} and early-20\textsuperscript{th}
centuries. Comparing children coded as “Black” or “Mulatto” by census
enumerators and linking these children across population censuses,
we first document large gaps in educational attainment and income
among African Americans with darker and lighter perceived skin tones.
To disentangle the drivers of these gaps, we identify all $\nBMSBMDfamilies$
families in which enumerators assigned same-gender siblings different
Black\slash Mulatto classifications. Relative to sisters coded as
Mulatto, sisters coded as Black had lower educational attainment,
were less likely to marry, and had less-educated, likely lower-earning
husbands. These patterns are consistent with more severe contemporaneous
discrimination against African American women with darker perceived
skin tones. In contrast, we find similar educational attainment, marital
outcomes, and incomes among differently-classified brothers. Men perceived
as African Americans of any skin tone faced similar contemporaneous
discrimination, consistent with the ``one-drop'' racial classification
rule that grouped together individuals with \emph{any} known Black
ancestry. Lower incomes for African American men perceived as having
darker skin tone in the general population were driven by differences
in opportunities and resources that varied across families, likely
reflecting the impacts of historical or family-level discrimination.%
\end{singlespace}

\end{abstract}
\newpage{}

\section{Introduction}

It is well known that African Americans have faced severe discrimination
relative to White Americans throughout U.S.\ history. This discrimination
has led to large racial gaps in economic outcomes.\footnote{Differences in economic outcomes by race have generated a great deal
of attention from economic historians, economists, and other social
scientists. For studies by economic historians see, for example, \citet{Margo1990,Whatley1994,vedder1992,Maloney2002,boustan2012racial,wright2013sharing,althoff2022jim,collins2022african,derenoncourt2022wealth}.
For studies by other economists see, for example, \citet{Myrdal1944,card1993trends,Heckman2000,Becker1971,Altonji2001employer,Carneiro2005,Neal1996,Bertrand2004,List2004,chetty2020race}.
For studies by other social scientists, see, for example \citet{Allen1986,Smith1997Race,Hunt01062007,McConnell2001435,vera2007handbook}.} However, despite the importance of colorism and skin tone, less is
known about discrimination and gaps in economic outcomes between African
Americans perceived as having darker or lighter skin tones, and about
the drivers of these gaps.\footnote{We discuss this literature extensively in Section~\ref{subsec:literature-differences}.}
As \citet{monk2021unceasing} concludes: ``future research on colorism
will be essential to understand comprehensively the significance of
race/ethnicity in a demographically shifting United States.''

This paper starts by using full-count U.S.\ decennial census data
from the late-19th and early-20th centuries to systematically document
large gaps in educational attainment and income between African Americans
classified by census enumerators as having lighter and darker skin
tones, as well as between African American and White individuals.
Next, we focus on the degree to which African American men perceived
as having darker skin faced heightened labor market discrimination,
and African American women perceived as having darker skin suffered
worse marriage prospects.

Inferring the role of contemporaneous discrimination varying by perceived
skin tone in outcomes among African Americans is challenging because
perceived skin tone is typically correlated with other factors that
affect outcomes, including access to opportunities as well as characteristics
like education and occupation which can themselves influence perceived
race \citep[e.g., ][]{saperstein2013mulatto}. Differences in access
to opportunities and resources that vary across families themselves
likely reflect systematic, longstanding differences in racism and
discrimination. Indeed we find that African American children were
more likely to be perceived as lighter skinned in more urbanized and
``whiter'' counties, as well as in non-farming households, reflecting
differences in available resources that may trace back across generations.
For example, \citet{bodenhorn2015color} shows that enslaved people
with lighter complexions were more likely to have been assigned less
onerous tasks that led to the accumulation of skills, were likely
to have been freed earlier, and had better access to nutrition and
suffered less stunted growth.

To study the causal effect of contemporaneous discrimination, a ``controlled
experiment'' might randomize only the perceived skin tone of children
at birth and then compare siblings' economic outcomes later in life.
While such controlled randomization obviously does not occur, we use
a rare historical setting and newly constructed data from 1870--1940---a
period during which race was explicitly coded in U.S.\ population
censuses as “Black,” “Mulatto,” or ``White''---to conduct an observational
study related to this ``controlled experiment.''\footnote{This paper uses a number of terms (for example the word ``Mulatto'')
that were widespread in the time period we study, but can carry negative
connotations. To the extent that any readers find our inclusion of
such terminology offensive, we apologize. Fully motivated by good
will, we feel our subject is sufficiently important as to justify
careful research and discussion. We will often refer to individuals
being classified as White, Black, or Mulatto, in line with the classification
codes used by census enumerators at the time. We use the term ``African
American'' to designate Americans with known African ancestry, that
is inclusive of people categorized as either Black or Mulatto by census
enumerators during the period of our study.} In particular, we compare siblings who vary in their perceived skin
tone, reflecting differences in a bundle of physical features, social
cues, and in others' interpretation of these features and cues.\footnote{Our estimates capture differences by \emph{perceived} skin tone. A
large literature spanning sociology, economics, and other related
fields (discussed in more detail in Section~\ref{sec:literature})
has robustly demonstrated that perceptions of race respond not only
to physical attributes (e.g., skin tone and hair texture) but also
to a wide variety of contextual and social cues such as the way a
person dresses or the language they speak. In both the general population
and among differently-classified siblings, we find that Mulatto-coded
children were more likely than Black-coded children to be classified
as White adults in future census years (in which the Mulatto category
no longer existed) and when self-reporting race on Social Security
Number applications. These findings suggests that the bundle of physical
and\slash or social characteristics that differed between Mulatto-coded
and Black-coded children persisted even 20 years later and predicted
the perceptions of other census enumerators, as well as self-reported
race. Classification as Mulatto rather than Black by census enumerators
is positively associated at the individual level with lighter complexion
recorded in Freedman's Savings Bank records\ \citep{adukia2025social}
and Union Army enlistment records (see Appendix~\ref{sec:usct-complexion}).}

To construct our data, we started with all $\nchildren$ million children
classified as Black, Mulatto, or White by U.S.\ decennial census
enumerators in 1870--1920.\footnote{Only the population censuses of 1870, 1880, 1910, and 1920 both contain
the Black vs.\ Mulatto distinction and provide data for all people
(see Footnote~\ref{fn:mulatto-years} for detail). In most specifications,
our sample restricts to children age 3--18 in these four censuses.} Enumerators went door-to-door to all households, coding the perceived
race of respondents. Enumerators were instructed to consider individuals
as Mulatto if they observed “any perceptible trace of African blood”
and as Black if the respondent was ``evidently full-blooded.'' We
then group these children into families with the same set of parents,
and link children to their adult selves in future census years. Constructing
these links is a non-trivial task given the lack of Social Security
Numbers or other unique identifiers, relying instead on the similarity
of names, birth years, and other identifying information \citep[see, for example,][]{Abramitzky2012,abramitzky2021automated}.
Women are typically more difficult to link (and are often excluded
from linked historical studies) due to their more frequent name changes
at marriage. The primary linking method we use \citep{price2021combining}
successfully links a large number of women, even across name changes
at marriage, by incorporating information from genealogy links and
from additional records (including birth, marriage, and death records)
that often include maiden or parents' names for women. Because linking
across historical censuses is inevitably imperfect, we present results
based on a variety of automated and hand-linking methods.

When comparing childhood outcomes of the approximately $\nBMDdaughtersrounded$
differently-classified sisters, we find that sisters coded as Black
were about $\neggapmixedsibsdaughtersinschoolinteger$ percentage
points less likely to attend school and had lower levels of literacy.
Linking across censuses, we find that sisters with perceived darker
childhood skin tones were $\neggapmixedsibsmarriedplustwentywomen$
percentage points less likely to be married as adults and, when they
did marry, had less educated and likely lower-earning spouses.\footnote{The spouses of Black-coded sisters had $\neggapmixedsibsSyearseducfortywomen$
fewer years of schooling and likely earned $\neggapmixedsibslogincfortywomen$\%
less than the spouses of Mulatto-coded sisters on average. Relative
to sisters, in the general population we find larger perceived skin
tone gaps in educational attainment and similar magnitude gaps in
marital outcomes: By adulthood, Black-coded sisters had received $\neggapmixedsibsyearseducfortywomen$
fewer years of schooling than their Mulatto-coded sisters on average
(smaller than the analogous general population gap of $\neggappopyearseducfortywomen$
years), and their $\neggapmixedsibsmarriedplustwentywomen$ percentage-point
lower marriage rate is not significantly different from the $\neggappopmarriedplustwentywomen$
percentage-point analogous general population gap.} This suggests that African American women with perceived darker skin
tones faced significantly greater levels of discrimination on the
marriage market, consistent with a literature showing marriage-market
penalties for darker-skinned women due to discriminatory feminine
beauty standards \citep{hill_skin_2002,hamilton2009shedding}.

In contrast, when analyzing around $\nBMSsonsrounded$ differently-classified
brothers, we find small and generally statistically insignificant
differences in educational, marital, and labor market outcomes between
brothers coded as Black and Mulatto. There are, however, large differences
for men in the general population: African American sons with lighter
perceived skin tones were more literate and completed $\neggappopyearseducforty$
more years of schooling, their incomes were $\neggappoplogincforty$\%
higher, and they married spouses with $\neggappopSyearseducforty$
more years of schooling on average.\footnote{Our analysis also highlights that while \emph{Black--Mulatto} gaps
in the general population represent economically and statistically
meaningful differences, they are much smaller than analogous \emph{White--Mulatto}
differences ($\neggappopwmyearseducforty$\ more years of education
and $\neggappopwmlogincforty$\% higher earnings in 1940). In other
words, we show that throughout the late 19th and early 20th centuries,
the White vs.\ Mulatto\slash Black gaps in all outcomes studied were
much larger than the Mulatto vs.\ Black gaps, consistent with the
abundant qualitative and quantitative evidence of discrimination against
African Americans.} These differences are generally reduced by about a third when comparing
African Americans who grew up as children in the same area and by
another third when controlling for observable family characteristics
(such as parents' literacy and income scores, whether the household
head is listed as a parent, and whether the household owns their home),
themselves likely affected by historical discrimination. When comparing
brothers, these differences largely disappear. These results may reflect
the dominance of a ``one-drop'' racial classification rule, under
which individuals with \emph{any} known Black ancestry were grouped
together. In this historical setting, differences in discrimination
due to an individual’s perceived skin tone explain less than about
a third ($\neggapmixedsibslogincfortypoppct$\%) of the \emph{Black--Mulatto}
income gap observed in the general population, even at the upper bound
of a 95\% confidence interval.

The majority of income gaps by perceived skin tone among African American
men were likely driven by differences in families' access to opportunities
and resources, such as differences in geography, educational access,
and family wealth, which have been shown to meaningfully affect outcomes
\citep{doi:10.1086/690944,althoff2022jim,derenoncourt2022wealth}.
Since discrimination in previous generations likely created these
differences in access by perceived skin tone in the first place, family
differences included the effect of perceived skin tone discrimination
suffered by previous generations; this analysis of siblings should
thus be understood as measuring the effect of perceived individual
skin tone. Given our finding that unequal opportunity and resources
were the main drivers of observed outcome gaps for men, reducing outcome
gaps may require policies that seek to remedy the underlying inequality
rather than simply moderating contemporaneous discrimination. Such
policies might include affirmative action or reparations that seek
to address the differences in access to opportunities and resources,
which themselves have resulted from systematic racism and past discrimination.

While our data do not allow us to fully explain different patterns
between brothers and sisters, we discuss possible reasons such as
differential parental investment in girls \citep{landor2013exploring}
and higher skin tone discrimination against women with darker perceived
skin tones. Our analysis corroborates the literature on the gendered
nature of skin tone dynamics, which suggests that African American
women face harsher penalties for darker skin tone than African American
men (see \citealp{monk2014skin} for a summary). We show that these
penalties existed even when controlling for family background differences
by making comparisons among sisters.

There are a number of potential concerns with our identification strategy
and interpretation of our findings. First, we cannot know with certainty
that siblings shared the same biological parents. Our analysis partially
addresses this concern by defining two children as siblings only if
neither is listed as a stepchild or with an absent parent. Second,
perceived skin tone differences between Black-coded and Mulatto-coded
siblings could have been less stark than such perceived differences
in the population. We are reassured by our finding in Section~\ref{subsec:white-adult}
that gaps in the likelihood of future classification as White were
similar for differently-classified siblings and for the general population,
suggesting that this concern is likely not driving our main results.
Third, attenuation bias in estimates of the effect of perceived skin
tone may be exacerbated by the use of fixed effects in some specifications
\citep{Griliches1979}. We use an empirical specification comparing
the weighted populations of same-gender Black vs.\ Mulatto siblings
from differently-classified families, which helps limit this concern
and produces estimates similar to specifications including family
fixed effects when the latter have sufficient statistical power.

Fourth, we cannot know the precise bundle of unobservable characteristics
that differed between children coded as Black and Mulatto, including
differences in aspects of a child's physical and social presentation
as perceived by census enumerators. For this reason, our paper focuses
on the overall effects of \emph{perceived} skin tone rather than of
physical skin tone. At some level, the effects of perceived skin tone
are more interesting than the effects of physical skin tone, as perceptions
are what determine discrimination. We do, however, discuss and analyze
issues of presentation to census enumerators and endogenous classification.
For example, we show that Black- and Mulatto-coded siblings were similar
along other observable characteristics including racially distinctive
names, and motivate our focus on comparisons by the perceived skin
tone of children, who had not settled into occupations. These results
are consistent with physical features (e.g., skin tone) having been
a key driver of within-family variation in enumerators' classification
as Black or Mulatto. Finally, parents whose children varied substantially
in perceived skin tone may have reallocated resources within the family
to reinforce or counteract different social attitudes towards their
children. To the extent that this occurred, our estimates would still
correctly identify the causal effect of perceived individual skin
tone within differently-classified families (including the impact
of this family allocation effect) but might not extrapolate to the
effect in families with uniform classifications. We discuss these
concerns in more detail in Section~\ref{sec:interpretation}. Although
our ability to extrapolate the effects of skin tone from within families
to the population necessarily relies on assumptions typically required
in studies of siblings or twins \citep[e.g.,][]{Ashenfelter1994},
we argue that these assumptions are reasonable in our setting.

The remainder of the paper proceeds as follows. Sections~\ref{sec:background}
and \ref{sec:literature} put our paper in the broader historical
context and in the context of several related branches of literature
on race and economic outcomes, respectively. Section~\ref{sec:Data-and-matching}
describes our data sources, samples, and the process of assembling
the data set. Section~\ref{sec:specifications} lays out our conceptual
framework and empirical strategies. Section~\ref{sec:skincolor-analysis}
presents our main results, while Section~\ref{sec:interpretation}
discusses the interpretation of these results. Section~\ref{sec:conclusion}
concludes.

\section[Historical background]{\label{sec:background}Historical background on skin tones in a racially
segregated society}

To help motivate and interpret our estimated gaps in economic outcomes
by perceived skin tone classification, we now briefly describe some
historical context faced by individuals during our analysis period.
For more detailed discussion of segregation and African American economic
history, see \citet{boustan2012racial}, \citet{wright2013sharing},
\citet{bodenhorn2015color}, and \citet{boustan2016competition} among
others.

\subsection*{Segregation}

African Americans around the turn of the twentieth century inhabited
largely separate public realms due to the White majority's anti-Black
racism and reluctance to interact with African Americans in public
places. In the South, where 89\% of African Americans lived in 1910
\citep[Ch. 2, Table 14]{BureauoftheCensus1913}, segregation was set
by ``Jim Crow'' laws pertaining to the separate treatment of African
Americans. In the North, segregation was mostly \emph{de facto}, but
was nevertheless practiced extensively by private individuals and
companies. Surveying legal cases of segregation throughout the U.S.\ between
the abolition of slavery in 1865 and the reintroduction of Jim Crow
laws in the South in 1881, \citet{Stephenson1910} summarizes: “In
the absence of legislative authority, many of the public conveyance
companies had regulations of their own separating the races. The `Jim
Crow' laws {[}\ldots{]} did scarcely more than to legalize an existing
and widespread custom.”

White American society actively excluded African Americans from public
institutions such as schools, courts, and churches, as well as venues
such as restaurants, hotels, and theaters \citep{Stephenson1910,Margo1990,doi:10.1086/690944,cook2023evolution};
means of public transportation were regulated so that African Americans
had to ride separate cars or occupy separate sections \citep{Stephenson1910}.
At the same time, disenfranchisement excluded African Americans from
the political arena \citep{Kousser1974,Naidu2012}. Racial segregation
was endemic in markets as well, whether the result of individual or
collective action.

In particular, labor markets were largely segregated. \citet{Maloney1995}
and \citet{foote2003} describe how the Ford Motor Company exploited
discrimination against African Americans by other companies, but channeled
its own African American workers to more demanding and dangerous manufacturing
jobs. \citet{Sundstrom1994UrbMarketsDiscr} and \citet{Fishback1984}
highlight that even though junior roles were not segregated in some
occupations, African Americans were not allowed to supervise White
workers. Labor unions were another source of discrimination, as most
did not accept African Americans into their ranks.

African Americans faced discrimination in housing markets as well,
even if they migrated from the South to the industrialized North.
\citet{Cutler1999Ghetto} find evidence that “…variation in the level
of segregation in 1940 is due to collective action racism on the part
of Whites rather than a desire among Blacks to live in Black areas.”
White individuals' desire not to share their neighborhoods with African
Americans was important in determining the patterns of suburbanization
and urban development that occurred in the second half of the twentieth
century, even after the Civil Rights Acts of the 1960s \citep{Boustan01022010}.

\subsection*{Intermediate skin tones}

Whenever racial segregation emerged, the question of defining distinct
racial categories had to arise \citep{Stephenson1910}. While approximately
one-fifth of African Americans were classified as Mulatto in the 1910
census, the true proportion with some European ancestry may have been
as high as three-quarters \citep{Cummings1918}. Marriages between
African American and White individuals were very rare in the early
twentieth century \citep{Fryer2007}, with much of the variation in
the degree of European ancestry originating during the era of slavery
\citep{Williamson1980}.\footnote{\citet{bodenhorn2006colorism} shows that even among African Americans,
individuals were likely to marry other African Americans with relatively
similar skin tones.}

In the late nineteenth and early twentieth centuries, some states
considered those with only small proportions of Black ancestry to
be White. However, the standard approach to racial classification
shifted during the early twentieth century to the ``one-drop rule,''
which grouped together individuals with \emph{any} known Black ancestry
\citep{bodenhorn2015color}. This shift was embodied both in \emph{de
facto }norms and through \emph{de jure }political and legal changes;
between 1910 and 1930, seventeen states adopted or amended their laws
to define African American ancestry using one-drop rule equivalents
\citep{murray1997states}. Thus, even people of African American descent
who looked “fully European” were considered African American and were
excluded from the White public sphere if their ancestry was known.

Nevertheless, economic differences between lighter- and darker-skinned
African Americans date back at least to antebellum times, when free
African Americans were more likely to be light skinned, and among
free African Americans, those with light skin were on average more
educated and richer \citep{bodenhorn2007colourism,bodenhorn2015color}.
While mixed-race individuals sometimes tried to set themselves apart
culturally, this distinction faded somewhat during the first decades
of the twentieth century \citep{davenport2020fluidity}. Some of the
most prominent African American leaders such as Booker~T.\ Washington,
W.~E.~B.\ Du~Bois, and Walter Francis White were of mixed ancestry.
Horace Mann Bond, another prominent figure of mixed race, noted the
shift towards a binary treatment of skin tone and racial identity:
\begin{quote}
``Time was when there were blue-vein societies {[}social clubs admitting
only light-skinned African Americans{]} {[}…{]} among Negroes in this
country, but they seem largely to have disintegrated, owing to two
happy chances of fortune: The first has been that those who were so
much like the dominant group {[}…{]} have in great part folded their
tents and crept quietly into the ranks of the whites. The other {[}…{]}
has been the unyielding refusal of the dominant group to accept any
of its hybrid progeny, if known as such {[}…{]}. {[}The One Drop Rule{]}
has done countless good for the Negro, as it has served to focus his
energy and that of all his potential leaders upon the immediate task
of racial survival.'' \citep{Bond1931}
\end{quote}
The attempt to distinguish between light- and dark-skinned African
Americans in the U.S.\ census dates back to 1850; \citet{Hochschild2008}
study why the Mulatto category was first introduced. All censuses
from 1850 to 1920 (except 1900) asked enumerators to distinguish between
people of full and mixed African ancestry, despite contemporaneous
observers' recognition that such ancestry-based classification was
questionable and subjective \citep[p.~209]{Cummings1918}.

Census enumerators were instructed to classify African Americans according
to their appearance, distinguishing “persons who are evidently fullblooded
negroes” from those merely “having some proportion or perceptible
trace of negro blood” \citep[p.~48]{gauthier2002measuring}.\footnote{Complete instructions to census enumerators on racial classification
can be found in Appendix~\ref{sec:race-instructions}.} It appears that enumerators indeed applied an appearance-based (rather
than a “blood-based”) standard: Table~\ref{tab:bm-couple-stats}
shows that when one parent was classified as Mulatto and the other
as Black (so that a ``one-drop'' blood-based standard would categorize
\emph{all} children as Mulatto), between $\shBlackParBMmin$\% and
$\shBlackParBMmax$\% of a family's children were classified as Black.

\subsection*{Passing as White}

In a segregated public sphere, light-skinned African Americans could
sometimes “pass for White.” As long as nobody identified passers as
such, it was the choice of potential passers whether to present themselves
as White or Black. Passing could have been very casual if it was partial:
every bus ride or play at a theater offered an opportunity to pass
temporarily, only later to return to the African American world.

The types of passing more interesting for this paper are professional
and complete passing \citep[ch. 31, sec. 4]{Myrdal1944}. Professional
passing was the representation of oneself as White to colleagues,
supervisors, and clients, for professional purposes. Under segregated
arrangements where many White people were unwilling or unable to hire
or to trade with African Americans, professional passing opened many
doors.\footnote{The presumed disutility to White individuals of doing business with
African Americans was evident in the legal treatment of racial misidentification
as defamation \emph{per se} under the category of ``words disparaging
to a person \emph{in his trade, business, office, or profession}''
{[}emphasis added{]} \citep{Stephenson1910}. (Defamation \emph{per
se} meant that a White person accused of being African American could
sue the accuser for the recovery of damages even without proving actual
harm, but simply by proving that the false accusation was made.)} However, professional passing was not complete if the person returned
at the end of the day to a family or community that did not identify
him or her as White. Complete passing required both professional and
social passing. In many cases, professional passing was a step on
the way to complete passing.

African Americans considering passing needed not only sufficiently
light skin and the ability to convincingly ``act White,'' but also
that the potential benefits outweigh a variety of personal costs.
For example, \citet{Myrdal1944} quotes an African American who passed
for White in college but then reclaimed his African American identity
to become a teacher. He mentions four reasons not to pass: the psychological
burden of fearing that someone would reveal he was Black, the higher
relative social status he could enjoy in the African American community
versus the White community, the economic stability he gained from
the relative scarcity of skilled African American workers, and his
richer social life among the higher ranks of African Americans. \citet{hobbs2014chosen}
offers a rich history of the evolution of these factors, and the decisions
faced by African Americans on whether, when, and how to pass.

While census enumerators may have accidentally classified some African
Americans as White even if they did not intentionally pass, those
who did pass completely were very likely to have been accepted by
the enumerator---as by the community---as White. While the distinction
between Mulatto and Black had been eliminated from the censuses by
1930, the line between White and African American, however, did not
change: individuals who would have been classified as either Mulatto
or Black in earlier censuses were presumably later classified as Black
unless they passed. For more analysis of the economics of passing,
see \citet{bodenhorn2003economics}, \citet{mill2016race}, and \citet{dahis2020choosing}.

\section{\label{sec:literature}Related literature}

\subsection{\label{subsec:literature-perceived-race}Determinants of perceived
race and skin tone}

To help interpret our comparisons of differences in outcomes by perceived
race and skin tone, we draw insights from a rich literature theorizing
about racial categorizations and analyzing the determinants of race
and skin tone perceptions.

Debates over the concept of race in the social sciences revolve around
two primary theories of race: essentialism and constructivism. Essentialists
argue that humans are naturally divided into races that share inherent
or immutable defining qualities. This classification assigns different
biological and genetic traits to individuals from different ancestries
and phenotypes, and typically views race as an intrinsically defined
and immutable characteristic (for a discussion, see \citealp{morning2011nature}
and \citealp{sen2016race}).

On the contrary, constructivism argues that racial categories are
created products of social life. Constructivists point to the lack
of scientific basis for racial categories: rather than being based
on inherent differences, racial classifications and the relevance
assigned to certain traits respond to malleable social structures.
These structures reinforce hierarchies that provide socioeconomic
advantages to dominant groups \citep{morning2011nature,sen2016race,roseconstructivist}.

Theories of race are also a source of debate in economics. This is
the case for individualist and structuralist approaches to the relationship
between race and economic outcomes. Under an individualist approach,
differences between races are ascribed to individual behaviors and
cultural differences \citep{francis_black_2022}. Structuralist approaches,
such as stratification economics, suggest that intergroup differences
in outcomes between dominant and subordinate groups are due to the
uneven intergenerational transmission of resources and advantage over
time. In this case, which group dominates and which is subordinated
arises from differences in power, rather than culture \citep{darity2022position,francis_black_2022}.
Our design informs this debate by contrasting sibling and general-population
gaps by perceived racial classification in childhood, indicating the
possible role of family resources and opportunity without identifying
the underlying mechanisms.

Constructivism has challenged how race is often measured in the social
sciences. If race is taken as an exogenous characteristic, comparisons
of outcomes between racial groups reflect historically accumulated
disadvantages that may be misinterpreted as cultural differences.
This often ends up presuming inferiority for the underprivileged group
\citep{francis_black_2022}. \citet{sen2016race} argue that controlling
for characteristics already affected by race, such as education or
wages, can introduce post-treatment bias when estimating the effect
of race. Instead, they propose the understanding of race as a ``bundle
of sticks,'' with components ranging from societal values and institutional
power relationships to physical attributes and skin tone. Under this
setting, race is a fluid concept that can only be partially explained
or proxied by each of its components. Among these components, some
exhibit a higher degree of stability than others (e.g., skin tone)
\citep{sen2016race}.

The understanding of race as a mutable feature suggests that racial
categorizations are not determined only by stable components, such
as skin tone.%
{} As a whole, race is flexible and changes across time and context
\citep{davenport2020fluidity}. However, any understanding of race
must account for the beliefs, accurate or not, that shape how race
is perceived \citep{morning2011nature}.

These conceptions of race have practical implications for how race
is perceived. In the words of \citet{morning2011nature}, “racial
concepts---that is, the way we think about what race is and how racial
difference is demarcated---have the potential to shape our evaluations
of others and the courses of action we take with respect to them.”
Research in sociology documents the relationship between stereotypic
traits and racial perception. For instance, \citet{Penner2008} suggest
that racial perceptions are associated with perceptions of social
status. Using longitudinal data, they find that unemployed, incarcerated,
and poor individuals are more likely to be classified as Black, conditional
on being previously classified as White. Even when taking skin tone
into account, social status cues, such as a low-status attire, can
lead people to consider a light-skinned person as Black (\citealp{Freeman2011LookingThePart}).
Most closely related to our own time period and sample, \citet{saperstein2013mulatto}
analyze U.S.\ census data from 1870--1920 and show that Black vs.\ Mulatto
racial classification by census enumerators is associated with an
individual's occupational status.

Using linked Freedman's Savings Bank and 1870 Census records,\ \citet{adukia2025social}
show that, among individuals with the same light-brown skin tone,
wealthier or literate individuals were more likely to be classified
as White or Mulatto. They further show that endogenous racial assignment
can widen measured cross-race wealth gaps and induce selection when
conditioning on race. We complement their analysis by examining how
later-life outcomes vary with childhood racial classification, both
in the population and among siblings.

Any analysis of perceived skin tone can suffer from a resulting statistical
bias. This concern is partially mitigated in our data because we measure
racial classification in childhood, when occupation, individual income,
and related status cues have not yet been fully determined. However,
we are aware that social and physical cues of other family members
likely affected these perceptions. Perceptions of race can also be
driven by observer characteristics. Rather than being objective, physical
traits can be perceived differently if observers belong to a different
group. Using data from the early 1990s, \citet{hill_race_2002} finds
that White interviewers reported African American respondents' skin
tones as being darker than did African American interviewers. African
American interviewers show the opposite pattern when interviewing
White respondents. This analysis offers empirical support for viewing
perceived race and skin tone as flexible characteristics that depend
on individual and social contexts.

\subsection{\label{subsec:literature-differences}Empirical analyses of differences
by race and skin tone}

\subsubsection*{Across racial categories}

Numerous studies document differences in economic outcomes between
African American and White workers in American labor markets.\footnote{See \citet{lang2020race} for a review of the economic literature
on race discrimination. See also \citet{logan2022american} for a
discussion of the importance of qualitative and contextual information
in understanding racial discrimination and disparities.} For instance, 2010 data from the Bureau of Labor Statistics show
that White men earned 33\% higher wages than Black men on average
\citeyearpar[Table~14]{U.S.BureauofLaborStatistics2011}.

To identify the role of contemporaneous discrimination in wage gaps,
a common approach in observational studies is to control for observable
measures of productivity and ascribe remaining differences by race
to contemporaneous discrimination \citep[e.g.,][]{altonji,Altonji2001employer,hughes1990significance,keith1991skin,Kreisman2015}.\footnote{See also \citet{woo2022unveiling} for similar analysis of racial
disparities by skin tone in Latin America. This analysis shows that
income disparities by skin tone persist (but decrease by roughly two
thirds) even when controlling for both observed demographic differences
and unobserved geographic heterogeneity (via a comparison to a respondent's
immediate neighbor). These penalties are estimated to be nonlinear
in skin tone, with sharp differences among light-to-medium skin tones
coupled with strong but less varied penalties among medium-dark to
very dark skin tones. These nonlinear penalties are consistent with
our own results, in which we find much larger gaps when comparing
African American to White outcomes than when comparing Black-coded
to Mulatto-coded individuals.} It is nevertheless unclear whether remaining differences represent
discrimination, unobservable differences in productivity, or whether
some of the controls themselves capture aspects of discrimination
and thus underestimate discrimination; productivity differences themselves
may reflect historical or systemic discrimination.

One way to compare individuals who differ only by race is to conduct
experiments---specifically, correspondence and audit studies---where
researchers create fake agents who interact indirectly with real agents
in the market. Such studies allow researchers to manipulate the perceived
identity of the imaginary agent. For example, \citet{Bertrand2004}
submit similar (fictitious) resumes under names that are either typically
African American or typically White, and find that candidates with
“White-sounding” names get more calls back than similar candidates
with “Black-sounding” names. Other examples of this type of study
include \citet{Zussmanforthcoming}, who finds “Arab-sounding” names
disadvantage sellers in an Israeli online used car market, and \citet{Doleacforthcoming},
who find a visual cue that the seller is Black decreases responses
to online classified advertisements. Observational studies complement
such experiments: they are perhaps less able to cleanly identify discrimination,
but can follow people over time, analyzing more comprehensive consequences
of discrimination and other factors that may affect racial inequality.

A key identification challenge in such observational studies is that
families differ in their ability to access opportunities and resources,
and much of this variation is driven by historical or family-level
discrimination. \citet{althoff2022jim} and \citet{derenoncourt2022wealth}
show that historical racial wealth gaps have been and continue to
be driven in large part by the vastly different starting positions
of African American and White individuals under slavery. We are able
to control for differences among African American families' access
to opportunities and resources by comparing siblings of varying perceived
skin tones. Our sample is nearly unique in that it allows perceived
skin tone to vary within a family, whereas race or ethnicity is often
thought of as fully shared by all members of the family. \citet{Rangel2007}
uses a similar methodology of comparing siblings in a cross-section,
looking at self-reported skin tone classifications in the 1991 Brazilian
Census of Population and the 1987 Brazilian Survey of Nutrition. He
finds that lighter-skinned children received slightly more education.
\citet{francis2016light} similarly compares siblings using the 2010
Brazilian census and finds that skin tone correlates contemporaneously
with education and labor market outcomes.

Our analysis is complemented by recent work by \citet{dahis2020choosing},
who investigate the determinants of and possible returns to racial
passing in a large sample that links across several nineteenth and
twentieth century censuses. They document that many individuals formerly
classified as African American were later classified as White, and
that the rates of passing were correlated with factors that plausibly
relate to the returns to passing. In related work,\ \citet{noghanibehambari2025passing}
use family fixed effects to compare male siblings who later did and
did not pass as White, finding greater old-age longevity among passers;
their analysis focuses on longevity and defines the comparison by
passing in adulthood, whereas ours examines educational attainment,
marriage, and income using perceived racial classification recorded
in childhood. Additional evidence on the economic importance of changing
racial affiliation comes from \citet{cornwell2017wage}, who show
that in Brazil, the same worker receives higher pay from an employer
who classifies him as White than from another employer who classifies
him as non-White.

\subsubsection*{\label{sec:literature-colorism}Colorism and skin tone penalties
among African Americans}

Related to the analyses of inequalities across racial groups, a literature
in sociology and economics analyzes differences in outcomes by skin
tone, often within a given coarser racial group.

The study of skin tone penalties, a key component of colorism,\footnote{\citet{monk2021unceasing} defines colorism as ``as a discriminatory
practice by which lighter skin tones, straight hair, and relatively
more Eurocentric facial features are preferred over darker skin tones,
kinky hair, and more stereotypically Afrocentric facial features.''} offers an additional layer of analysis within racial gaps in the
US (see \citealp{monk2021unceasing} for an illuminating review).\footnote{See also \citet{dixon2017skin} for a review of colorism with a focus
outside of the U.S.} During recent decades, literature has shown that gaps by skin tone
partially explain inter- and intra-racial inequalities \citep{goldsmith2006shades},
and that skin tone racism persists as other explicit forms of racism
are condemned by society \citep{monk2021unceasing}.

The literature provides robust evidence of differential labor market
and related outcomes by skin tone dating back to at least antebellum
times. Skin tone gaps are documented within a perceived racial category,
usually among African Americans \citep{keith1991skin}. Most of the
research has focused on socioeconomic skin tone gaps, although these
also extend to the physiological, health, and legal realms \citep{adams_skin_2016,monk2021unceasing}.
Howard Bodenhorn and co-authors find skin tone gaps in several outcomes
(including occupational status, wealth, freedom, nutrition, and height)
among African Americans in the 19th century \citep{bodenhorn2002mulatto,bodenhorn2003complexion,bodenhorn2006colorism,bodenhorn2015color,bodenhorn2007colourism}.
\citet{gullickson_significance_2005} documents large skin tone gaps
in occupational status and education for cohorts of African Americans
born between 1900 and 1940. \citet{hughes1990significance} find that,
during the 1950--1980 period, the differences in socioeconomic status
between light- and dark-skinned African Americans were as high as
between African Americans and White Americans overall. Using survey
data for 1980, \citet{keith1991skin} suggest that among Black respondents,
skin tone can be more predictive of occupation or income than parental
socioeconomic status.\footnote{These findings were questioned by \citet{gullickson_significance_2005},
who highlighted that educational and labor market skin tone gaps actually
declined for cohorts born after 1950. The results of this study were
later contested because of its small sample of light-skinned Black
respondents in the later years \citep{goldsmith2006shades}.} \citet{goldsmith2007dark} show that, by the early 1990s, light-skinned
African Americans earned 23\% higher hourly wages than dark-skinned
African Americans.

These skin tone gaps extend to different measures of labor market
success, such as household income and occupational status, as observed
by \citet{monk2014skin} for data from the early 2000s. Furthermore,
\citet{monk2014skin} finds that skin tone does not correlate with
employment itself, suggesting skin tone discrimination may restrict
access to better jobs rather than to jobs overall. This robust literature
leads us to reasonably expect gaps in economic outcomes by perceived
skin tone among African Americans, which we confirm in the general
population with our census data.

The literature also highlights the intersectional (and in particular
often gendered) nature of colorism. \citet{monk2014skin} finds that
skin tone penalties among men in the labor market are driven by differences
among African Americans: dark-skinned African American men are almost
twice as likely to exhibit a lower occupational status than their
light-skinned counterparts. The paper points out that this could be
the product of stereotypes of criminality and troublemaking that dark-skinned
African American men face in a racially integrated labor market in
which White gatekeepers control their access to employment.

In contrast, African American women face important skin tone penalties
in the marriage market due to light-skinned preferences, usually associated
with femininity and discriminatory beauty standards \citep{hill_skin_2002,hamilton2009shedding}.\footnote{See also \citet{banerjee2013marry}, which estimates a marriage market
penalty for women with darker skin tones in the context of India based
on responses to matrimonial advertisements. \citet{fisman2008racial}
also estimate that African American women are on average less preferred
as potential partners than White women in a U.S.~experimental speed
dating market, and that this racial penalty appears to be driven by
a gap in perceived attractiveness. In contrast, African American men
and White men are perceived as similarly attractive on average in
this speed dating market.} Although there is not an overall strong relationship between skin
tone and marital status among African Americans in more recent years,\footnote{There are large gaps in marriage rates between White and African American
adults in recent years \citep{census2022marriage}. Among U.S.~individuals
15 years and older in 2022, White men and women were each more likely
to be currently married ($\marratenowwhitemen$\% and $\marratenowwhitewomen$\%,
respectively) than were Black men and women ($\marratenowblackmen$\%
and $\marratenowblackwomen$\%).} young light-skinned African American women exhibit a marriage premium
in that they marry African American men with higher average educational
attainment \citep{monk2014skin,hamilton2009shedding,hunter2013race}.
Also, \citet{bodenhorn2006colorism} and \citet{monk2014skin} find
assortative marital matching among African Americans of similar skin
tone. Our analysis of marital differences confirms and builds on this
important literature by showing that a significant part of these skin
tone effects holds even when controlling for unobserved family background
differences by comparing among differently-classified sisters.\footnote{Skin tone penalties for African American women extend to their physical
and psychological health. Using longitudinal health data, \citet{hargrove_light_2019}
finds that dark-skinned African American women exhibit higher cumulative
biological risk than light-skinned African American women. The results
are partially explained by socioeconomic characteristics and stressors.
In a similar spirit, \citet{diette_skin_2015} find that among unemployed
African American women, those with darker skin tones are more likely
to experience depression.}

Skin tone gaps are also traced to childhood and youth experiences,
and in particular to educational attainment. \citet{branigan_skin_2013}
find that, among African Americans between 1985 and 2000, a one standard
deviation increase in skin tone lightness was associated with a three-month
increase in educational attainment. \citet{monk2014skin} suggests
that educational inequality among African Americans with different
skin tones is as large as that between White and African Americans.
\citet{adams_skin_2016} argue that skin tone penalties affect boys
to a larger extent. Stereotypic perceptions of the intellectual abilities
and aggressiveness of dark-skinned boys can truncate their academic
achievement \citep{maddox_cognitive_2002}. For instance, \citet{hannon_relationship_2013}
find that dark-skinned African American children face a higher probability
of school suspension. Finally, skin tone biases translate into legal
consequences for African Americans. \citet{levinson_different_2010}
find that dark-skinned African Americans are more likely to be considered
guilty than lighter-skinned African Americans and that skin tone cues
become more important in the absence of clear evidence. Analyzing
court records, \citet{blair_influence_2004} find that feature-based
stereotypes about inmates, such as darker skin and other Afrocentric
physical features, are correlated with longer sentences. \citet{white_salience_2015}
and \citet{monk_color_2019} find similar results among other dimensions
of the criminal justice system. \citet{branigan_complicating_2017}
find that skin tone penalties also exist among lighter- and darker-skinned
White Americans in quick, low information decisions, such as arrests.

There is a large body of literature that studies \emph{how} contemporary
skin tone discrimination operates. \citet{goldsmith2006shades} highlight
that preferences and biases for light skin tones (colorism) have persisted
in the U.S.\ over centuries. They suggest that light-skinned African
American workers are favored because of their physical proximity to
White workers. This skin-tone-based proximity activates racialized
stereotypes and beliefs about worker competence \citep{adams_skin_2016,goldsmith2007dark,maddox_cognitive_2002}.
The conflation between light-skin tones and desirable characteristics,
when internalized, can bias how people infer the socioeconomic status
of others. \citet{paul_about_2022} find that, although people can
accurately gauge wealth and income from facial cues for light-skinned
individuals, they perform poorly when doing so with high-income African
American and Latino populations. These patterns extend to perceptions
of beauty, even among African Americans \citep{hill_skin_2002,landor2013exploring}.

Skin tone affects social outcomes in direct and indirect ways. For
the former, the effect of colorism on outcomes is mediated by how
the additional skin-tone-related discrimination changes their experiences
in school, marriage, the labor market, etc. \citep{branigan_skin_2013}.
For instance, stereotypes about skin tone can end up as self-fulfilling
prophecies: positive or negative beliefs about a given group's ability
can affect their performance, regardless of the belief's accuracy
\citep{darity2022position}.

\citet{adams_skin_2016} point out that discrimination is interwoven
with individual and contextual factors. For instance, a cohesive racial
identity at home can prevent dark-skinned African Americans from being
largely affected by colorism. Also, if there is skin tone variation
among children in the household, families can tailor their parenting
practices to the level and type of discrimination they expect the
children to face \citep{adams_skin_2016}. By comparing variation
in skin tones within families, we can examine how families might respond
to expected discrimination with differential investments in their
children. However, as \citet{branigan_skin_2013} note, linking discrimination
to skin tone biases, leaving aside other components of racism, proves
challenging.

On the other hand, skin tone can affect social outcomes in indirect
ways when it becomes a proxy for characteristics that are historically
associated with certain gradients of skin tone \citep{branigan_skin_2013}.
In this case, inequalities from skin tone accumulate over time and
are intergenerationally transmitted, even if the discrimination that
caused those gaps in the past has waned \citep{branigan_skin_2013}.
We point to this indirect mechanism by comparing skin tone gaps among
differently-classified brothers. We find \emph{Black--Mulatto} perceived
skin tone gaps in earnings and educational attainment among siblings
that are significantly and economically smaller than those observed
in the general population. This points to the explanatory power of
differences in families’ access to opportunity that accumulates over
time, driving intra-group skin tone inequality.

Finally, the literature on skin tone penalties has shed light on the
experience of other racial and ethnic groups where skin tone discrimination
is salient. Among Latin American immigrants, dark-skinned immigrants
face higher penalties in the labor market \citep{hersch_profiling_2008,rosenblum2016looking,bonilla2006racism}.
These patterns even extend, in some cases, to Asian Americans \citep{monk2021unceasing}.

To summarize the key takeaways from this literature as it pertains
to our analysis, we first note that there is ample evidence of skin
tone socioeconomic differences among African Americans. Higher penalties
for darker-skinned African Americans start at school and persist over
the life-cycle through limited labor market opportunities and, in
the case of women, reduced marriage prospects and health.

We also highlight the two key mechanisms proposed by the literature
through which skin tone penalties operate among African Americans:
(a)~through contemporaneous discrimination in many contexts (e.g.,
in labor markets, marriage markets, and other bias-ridden interactions)
in which individuals are perceived, treated, and stereotyped differently
based on the tone of their skin (which can in turn affect individuals'
own identity and investment decisions); and (b)~through the intergenerational
transmission of stratification and disadvantage, due to differential
access to resources and opportunities correlated with skin tone due
to discrimination at the family-level or in the past \citep{monk2021unceasing}.
The existing literature has had difficulty differentiating between
these two mechanisms as potential drivers of observed differences
by skin tone. Our paper seeks to decompose the relative effects of
these two mechanisms through our analysis of differently-classified
siblings, who share the same family-level access to opportunities
but who differ in their perceived skin tone.\footnote{Our analysis is related to the decomposition of ``direct'' and ``systemic''
discrimination analyzed theoretically and experimentally by \citet{bohren2022systemic}.} We also study the potentially gendered nature of perceived skin tone
effects (as suggested by the literature) by assessing these effects
separately for men and women.

\section{\label{sec:Data-and-matching}Data and linking}

\subsection{\label{subsec:child-cohorts}Child cohorts}

We analyze complete-count, decennial U.S.\ census data from 1870--1940
\citep{ruggles2021ipums}. Our primary analysis focuses on cohorts
of children age 3--18 and coded as White, Black, or Mulatto in 1870,
1880, 1910, or 1920.\footnote{\label{fn:mulatto-years}This is the full set of available census
years including all African Americans and a Black and Mulatto classification
distinction. Data from the 1890 census is no longer extant. The 1900,
1930, and 1940 census years do not distinguish between Black-coded
and Mulatto-coded individuals, and so we use this data only when defining
linked adult outcomes or in analysis of White vs.\ Non-White children.
We exclude the 1850 and 1860 census years from our analysis, as only
free individuals (an unrepresentative and small portion of African
Americans, $\sharefreeeightfifty$\% in 1850) are enumerated in this
pre-Civil War data.} We exclude children below the age of 3 and above the age of 18 since
very young children have typically not reached their adult skin tone,
and older children who remain in their parents' household are highly
selected. This gives a full sample of $\nsons$ sons and $\ndaughters$
daughters. We enrich data on the children themselves with characteristics
of their household head and parents (when present within the household).\footnote{We locate a child's father and mother when present within the household
using the \emph{\href{https://usa.ipums.org/usa-action/variables/POPLOC\#description_section}{poploc}}
and\emph{ \href{https://usa.ipums.org/usa-action/variables/MOMLOC\#description_section}{momloc}}
variables constructed by IPUMS.} We then group children within a household into families. Given our
interest in comparing siblings with identical family opportunity access,
we define two sons to be members of the same family only if they shared
the same general race\footnote{We initially considered also analyzing the small number of households
containing both White and Mulatto-coded children. However, examination
of the original census records for such households in 1910 suggested
that virtually all of them contained either enumeration errors (e.g.,
household code corrections were visible on the original census record)
or data entry errors. The latter were frequently driven by the similarity
of handwritten codes used to distinguish White from Mulatto: ``W''
and ``M.'' (Many enumerators ignored the instruction to identify
Mulatto-coded individuals with the more distinctive ``Mu.'') Because
of the small sample and high error rate, we decided to limit analysis
of differently-classified siblings to households with both Black-coded
and Mulatto-coded children.} and if we are reasonably certain that they shared the same pair of
biological parents. Specifically, we define children to have been
members of the same family only if they had: the same household; the
same general race (coded as White vs.\ Black\slash Mulatto); the
same parents listed; both parents present within the household; no
parent listed as a step-parent; and a nonzero number of siblings.
The possible consequences of illegitimacy, remarriage, adoption, and
relationship misclassification are discussed in Section~\ref{subsec:Full-siblings-or-not}.

Given this family definition, two key subsets we use in our analysis
are the set of sons from families that had both a Black-coded and
a Mulatto-coded son age 3--18, and an analogous sample for daughters
from differently-classified families. We refer to these samples throughout
as BMS (``Black-and-Mulatto-sons'') and BMD (``Black-and-Mulatto-daughters'').\footnote{Another sample that may be of interest ignores the gender of children
when looking for families with both Mulatto or Black siblings. This
sample would include the union of the BMS and BMD samples, along with
additional families where race varies only across the gender of siblings.
We proceed only with the BMS and BMD samples as the considerations
in investments in sons and daughters during childhood in this era
were very different.} We find $\nBMSsons$ sons in $\nBMSfamilies$ families that had some
sons classified as Black and the others as Mulatto (BMS), and $\nBMDdaughters$
daughters in $\nBMDfamilies$ families that met the equivalent condition
for girls (BMD).\footnote{We note that the same family could be counted multiple times if it
contains at least one Black-coded and one Mulatto-coded same-gender
sibling both age 3--18 in multiple distinct decennial census years.}

Figure~\ref{fig:census-examples} illustrates our empirical exercise
with two families. In the 1910 census, Tom Best was coded as Black,
while his brothers John and Pete were coded as Mulatto. The 1940 census
records their 1939 wage and salary income as \$900, \$600, and \$1,050,
respectively.\footnote{In 1940, Tom and Pete are working as farm laborers and remain in their
childhood township (Stony Creek, North Carolina), while John works
as a carpenter and has moved to Goldsboro, North Carolina.} In the 1920 census, Dorothy and Rebecca Ridley were coded as Mulatto,
while their sister Texanna was coded as Black. By 1940, Dorothy and
Rebecca were married, while Texanna was single.\footnote{The Best and Ridley households also each contain an additional same-gender
sibling that is excluded from their respective child cohort due to
their age (Dock Best, coded as Mulatto and age 19 in 1910; Lulu Ridley,
coded as Mulatto and age 25 in 1920). The Best household also includes
an additional eligible-age child, Chape (Black, age 15 in 1910). We
do not have a linked census record for Chape in 1940, as Chape appears
to have passed away in 1912. As such, Chape would be included in our
analysis of childhood outcomes, but would be excluded from our analysis
of linked adult outcomes.}

Table~\ref{tab:bm-couple-stats} shows the distribution of skin tone
classifications of African American couples overall and in the BMS
and BMD samples. While $\couplesBlackBlack$\% of all African American
couples had both members classified as Black, these couples represented
only $\bmsParentsBlackBlack$\% of BMS households, while $\bmsParentsMulatto$\%
of BMS households had at least one parent classified as Mulatto. (The
remaining $\bmsParentsRemainder$\% primarily reflect households classified
as Black--White couples.)\footnote{One might be concerned that some census enumerators could have classified
multiracial children as Mulatto if their parents came from at least
two different perceived races, even when neither parent was perceived
as African American. Relatedly, one might be concerned that Native
American or other individuals not necessarily of African ancestry
might have been misclassified as Mulatto. Parent race frequencies
for our BMS and BMD samples (Table~\ref{tab:bm-couple-stats}) and
among all families with at least one Mulatto-coded child (available
upon request) both show a vanishingly small fraction of families in
which neither parent is coded as Black\slash Mulatto, suggesting that
consistent with instructions, enumerators generally did not code non-African
American but multiracial individuals as Mulatto. Individuals classified
as Native American also represent a much smaller share of the general
population than individuals coded as Mulatto or Black, limiting the
scope for misclassification of entire families as African American
to impact our estimates or interpretation.} Differently-classified Black- and Mulatto-coded parents were especially
overrepresented, accounting for $\bmsParentsBlackMulattoUnordered$\%
and $\bmdParentsBlackMulattoUnordered$\% of parents in our BMS and
BMD families, respectively, but only $\couplesBlackMulattoUnordered$\%
of African American couples overall. This higher frequency of differently-classified
couples likely reflects either greater variation in perceived skin
tone among the children of differently-classified parents, and\slash or
enumerators who were particularly responsive to this variation.

Although it was rare for brothers to receive different racial classifications,
Appendix Figure~\ref{fig:Geographic-frequency-of-sons} shows that
these families were not highly geographically concentrated. We know,
therefore, that many enumerators gave different racial classifications
to siblings in the same household, which also increases our confidence
that the existence of such households represented genuine differences
in perceived skin tone. This is also consistent with the fact that
Mulatto-coded siblings were more likely than Black-coded siblings
to be classified as White adults in future census years (in which
the Mulatto category no longer existed), as discussed in Section~\ref{subsec:white-adult}.

Census records do not indicate which household members were physically
present when enumerators visited, and we cannot know how exactly enumerators
assessed ``color or race,'' especially for individuals they did
not observe (see Appendix~\ref{sec:race-instructions} for complete
racial classification instructions to enumerators).\footnote{The 1940 census---in which a Black\slash Mulatto distinction no longer
exists---does include information about ``the respondent who provided
most of the information for that household.''} When spouses or siblings \emph{were} coded differently, this was
presumably either because both were present and the enumerator perceived
them differently, or because the enumerator was told by a respondent
about differences in household members' skin tone. In both cases,
coding reflected skin tone as perceived by a third party, and our
estimates capture the effects of perceptions rather than physical
skin tone. Regardless of how enumerators assessed skin tone, we note
that differences in perceived skin tone persisted across decades and
third-party observers (as shown by differences in future classification
as White) and therefore allow us to consistently estimate the effects
of contemporaneous discrimination based on perceived skin tone.

Table~\ref{tab:attribs} Panel~A provides a variety of summary statistics
for the BMS sample and for households with multiple all-Black- or
all-Mulatto-coded sons. The characteristic that varied the most across
the samples was the number of sons, which varied in part because of
the way we constructed the samples: The greater the number of sons
in a household, the more chances there were for one of them to have
a different skin tone classification, and indeed households in the
siblings sample had the highest average number of sons. Even though
we restrict the same-race sample to families with at least two sons,
the number of sons was still lower than in the siblings sample. The
average age of children followed the same pattern, perhaps since having
more sons in the household also meant that younger or older households
(with fewer children, or only youngest son present) were less likely
to appear, and the younger households were more numerous because of
population growth. Besides these two mechanical differences in characteristics,
the sibling sample was not selected evenly across geographic locations.
Children in the siblings sample came from more rural counties than
the general population. This geographic selection contributed to differences
in other characteristics such as literacy and occupational structure.
Panel~B shows similar results for the samples of daughters.

\subsection{\label{subsec:Matching-across-censuses}Linking across censuses}

To assess outcomes later in life, we link individuals from the child
cohorts described above to their records as adults in future census
years.\footnote{Focusing our analysis on perceived racial and skin tone gaps by childhood
(rather than adult) race lessens the extent to which census enumerator
racial classifications responded endogenously to an individual's occupation,
income, or other outcomes of interest, as these characteristics had
not yet been realized for children.} We link all cohorts to their adult selves 20 and\slash or 30 years
later, depending on data availability. All cohorts are linked at least
to their adult selves 20 years later, with the exception of the 1870
cohort (which is instead linked to the 1900 census); we refer to these
as \emph{``t+20 }links.''\footnote{We link children from the 1870 cohort to the 1900 census because the
1890 census is no longer extant. Our main analyses include age fixed
effects which are allowed to vary flexibly by cohort and comparisons
are made only within a cohort, thus accounting for potential life
cycle differences after 20 vs.\ 30 years.} We also link the 1910 cohort to their adult selves in 1940 (as well
as 1930), since 1940 is the only currently available census with full-count,
individual wage-income data.\footnote{We do not analyze 1940 outcomes among the 1870 and 1880 child cohorts,
as most of these individuals would no longer have been of typical
working age in 1940. The 1950 full-count census is available in preliminary
form, but its income questions were asked only of sample-line respondents
in the contiguous United States.} We refer to the union of 1910--1940 and 1920--1940 links as ``1940
links.''

Given the lack of Social Security Numbers or other unique identifiers,
linking children to their adult selves is a non-trivial task; we use
links identified by the Census Tree data set \citep{price2021combining}
as our baseline specification, and show robustness to using links
identified by users of a genealogy platform (FamilySearch), by the
Census Linking Project \citep{abramitzky2020census}, by information
from Social Security Number applications \citep{althoff2022intergenerational},
by alternative publicly released Census Tree links \citep{buckles2025censustree},
and by IPUMS Multigenerational Longitudinal Panel (MLP) V2.0 links
\citep{ruggles2025mlp}.

The Census Tree data set combines links from a variety of methods
and sources. These include links from a genealogy platform (FamilySearch),\footnote{Our genealogy series uses the subset marked by the public Census Tree
release's family-tree flag.} from machine learning predictions trained on these genealogy links,
from the method used in the Census Linking Project \citep{Abramitzky2012},
from the IPUMS Multigenerational Longitudinal Panel \citep{helgertz2021mlp},
and from hand-linking by a large team of research assistants and volunteers.%
{} Importantly for our analysis of daughters' adult outcomes, the Census
Tree data set is able to form links for women. The Census Tree data
set successfully links a large number of women across name changes
at marriage by incorporating information from genealogy links and
additional records (including birth, marriage, and death records)
that often include maiden or parents' names for women.\footnote{Our paper contains two distinct versions of the Census Tree links.
Our baseline specification uses a version of the Census Tree links
received directly from the Census Tree authors, which rely less heavily
on racial concordance than a public version of the links (V1.0). The
alternative public release more heavily incorporates MLP links that
favor racial concordance and XGBoost links that exclude Black\slash Mulatto-classified
child to White-classified adult transitions. This passing as White
phenomenon was both important historically and closely tied to our
perceived skin tone mechanism of interest, which leads us to prefer
our baseline Census Tree links in our setting.}

The Census Linking Project is a research effort which uses a variety
of automated algorithms to identify records in different census years
which represent the same individual. Two records are linked if they
have the same state/country of birth, if they have similar birth years,
first names, and last names, and if these records better align on
these criteria than all other possible links for these records \citep{Abramitzky2012}.\footnote{See \citet{abramitzky2020census} for more detail on the specifics
of the linking algorithms implemented in the Census Linking Project
and see \citet{abramitzky2021automated} for discussion of linking
considerations, these algorithms, and their relative performance.
We use the ``ABE Exact Conservative'' links provided by the Census
Linking Project. See \citet{price2021combining} for more detail on
the Census Tree data set.} Since most women changed their last names upon marriage, the Census
Linking Project does not attempt to link daughters to their adult
selves.

\citet{althoff2022intergenerational} similarly links individuals
across census years by using a version of the linking algorithm from
\citet{Abramitzky2012} that has been modified to incorporate additional
information from individuals' applications for a Social Security Number
(SSN). This additional information includes applicants' maiden names,
as well as the names and maiden names of applicants' parents. This
additional information helps to improve the match rate for applicants
and their parents, and importantly links women across potential name
changes at marriage. Relative to the other linking methods we use,
it also reduces linking rates and statistical power by restricting
to individuals (and their parents) that applied for SSNs and, for
data privacy reasons, that were known to have died before 2007.

Table~\ref{tab:match-characteristics} reports the results of linking
sons from our child cohorts to their adult selves. Two linking rules
are considered. The first set of links, reported in columns (2)--(5),
comes from the Census Tree and links sons at a relatively high $\MatchRatebyuauto$\%
linking rate ($\MatchRatewomenbyuauto$\% for daughters\footnote{See Appendix Table~\ref{tab:match-characteristics-daughters} for
linking statistics for daughters.}). A more conservative method, reported in columns (6)--(9), uses
the Census Linking Project's ``ABE Exact Conservative'' links and
results in a linking rate of $\MatchRateABEexcon$\% for sons (and
does not link daughters), relying only on information determined at
birth.\footnote{These linkage rates of $\MatchRateABEexcon$ to $\MatchRatebyuauto$\%
exceed or are comparable with other papers using historical individual-level
data \citep{abramitzky2012europe,Abramitzky2012,abramitzky2021automated,Ferrieforthcoming,price2021combining}.} We use the more complete (Census Tree) links in our baseline specification,
and show robustness of our key analyses to alternative linking algorithms
in Appendix Figures~\ref{fig:algorithm-robustness} and~\ref{fig:algorithm-robustness-women}.\footnote{Most of these alternative linking algorithms, including the alternative
public Census Tree, IPUMS MLP, and SSN-application-based links (and
possibly also the genealogy links), rely more heavily on racial concordance
when determining links than do our baseline Census Tree or ABE links.
This presents a particular identification challenge in our setting.
These alternatives relying on racial concordance differ from our preferred
links in that they sometimes show no economically or statistically
significant difference between Black- vs. Mulatto-classified children
or siblings in their future classification as White, suggesting that
selection into these links can eliminate the difference in average
perceived skin tone between our Black- vs. Mulatto-classified samples.
Given this lack of a first stage, it is perhaps not surprising that
we then also sometimes do not find significant differences for other
adult outcomes between Black vs. Mulatto-coded children, in contrast
to our preferred Census Tree-based results for women.}

Individuals are not randomly selected into the linked sample. Table~\ref{tab:match-characteristics}
columns (2)--(3) and (6)--(7) report attributes of linked sons,
and differences in those attributes between linked sons and the full
initial sample. Linking rates are higher among sons who were lighter-skinned,
literate, and those in households with literate heads or more children.\footnote{Appendix Table~\ref{tab:match-characteristics-daughters} shows similar
results when linking daughters.} Literate and richer households may have used more unique names which
make an individual more likely to be successfully linked; most of
the observations that we fail to link have more than one potential
link that we cannot sufficiently distinguish from the best potential
match, rather than lacking even a single sufficiently close match.

To make the linked sample more representative of the baseline sample,
we reweight observations according to how likely they are to be linked.
As suggested in the linking literature (e.g., \citealp{perez2017south,zimran2019sample,bailey2020simple}),
we predict the probability of an observation being linked given a
set of background variables.\footnote{\label{fn:pweight-estimation}We predict a link using a probit regression.
We include the following controls as indicator or factor variables:
the numbers of same-gender children in the family; household head
is child's parent; head home ownership; head, father, and mother literacy;
head, father, and mother presence in the household; head, father,
and mother 1-digit occupation; child age; child birth order; child
is Black- or Mulatto-coded. We also include continuous controls for
the head, father, and mother's income score (see Section~\ref{subsec:income-prediction}).
We include missing values as a distinct outcome level for each of
these variables.} We then assign weights equal to the reciprocal of this probability,
representing the average number of observations each linked observation
represents. The resulting weighted characteristics of the linked samples,
and the differences between weighted linked observations and the full
initial sample, are presented in columns (4)--(5) and (8)--(9).
The weighted linked sample is now much closer to the full initial
sample. In our baseline specification, we weight all observations
of linked outcomes by these inverse link probability weights.\footnote{We also winsorize weights at the 99th percentile to limit the influence
of outliers. When analyzing linked specifications that include family
fixed effects (Equation~\ref{eq:within-hh-estimation}), only cases
in which both an individual and their same-gender sibling are linked
will inform our coefficient estimates. As such, for these specifications
we weight by the inverse of the predicted probability of successfully
linking an individual \emph{and} at least one of their same-gender
siblings.}

\subsection{\label{subsec:income-prediction}Measuring income scores in adulthood}

Data on individual incomes was collected as part of the decennial
census starting in 1940, by which point our 1870 and 1880 cohorts
would no longer have been of working age. To enable an analysis of
individual earnings which covers these child cohorts, we follow \citet{abramitzky2021intergenerational}
in constructing proxies of individual income (which we refer to as
``income scores''). To do so, we first predict log wage income among
men age 20--50 in 1940 census data based on an individual's 3-digit
occupation, state, and age.\footnote{The precise controls included when predicting an individual's log
wage income in 1940 are: occupation; age; squared age; state; and
(1-digit occupation)$\times$(region). We predict and analyze only
income earned by men due to lower female labor force participation
rates during this time period.} As the 1940 income variable excludes income from self-employment
(a category which includes most farmers), we adjust this income prediction
for farmers using a method developed by \citet{collins2022african}.
This adjustment makes use of the facts that the 1940 census records
the income of farm laborers, while the 1960 census records the incomes
of both farmers and farm laborers. We thus impute the (non-wage, unrecorded)
income of farmers in 1940 by multiplying the income of farm laborers
in 1940 with the ratio of earnings for farmers relative to farm laborers
in the 1960 census, by region and home-ownership (when available,
starting in 1900). We also follow \citet{abramitzky2021intergenerational}
and \citet{collins2022african} in scaling up the incomes of farm
managers and farm laborers to account for in-kind transfers.

We also show the robustness of our income score-based results to alternative
constructions of this income score in Appendix Figure~\ref{fig:incscore-robustness}.
In these alternatives, we either include race as an income predictor,\footnote{The precise controls added when predicting an individual's log wage
income in 1940 are: race indicators (Black/Mulatto vs.\ White); (race)$\times$(1-digit
occupation); (race)$\times$(region).} predict log total income (as defined in Section~\ref{subsec:adult_income}
to include imputed non-wage income, rather than predicting log wage
income), use the 1900 Census of Agriculture to adjust incomes for
farm managers and farmer laborers (rather than the 1960-decennial-census-based
adjustment suggested by \citealp{collins2022african}), or use the
1901 Cost of Living survey (rather than decennial census data) as
our income data source.\footnote{We follow \citet{abramitzky2021intergenerational} in the construction
of these latter two alternative income scores, where this construction
is described in greater detail. Relative to the 1940--1960 decennial
censuses, the 1900 Census of Agriculture and the 1901 Cost of Living
surveys occurred closer in time to when adult outcomes are measured
for our earlier cohorts, but cover fewer individuals or contain less
granular information on occupations and wages.}

\section{\label{sec:specifications}Empirical strategy}

A naïve assessment of the effects of discrimination might simply compare
the outcomes of individuals in different groups: for example, did
individuals perceived as being darker-skinned earn more than those
perceived as lighter? In Table~\ref{tab:descriptive-means}, we report
average characteristics by perceived child race and gender. We see
that children perceived as darker-skinned had lower levels of literacy
and school attendance as children on average coupled with lower educational
attainment. We also see gaps in earnings and predicted income as adults
among sons, along with gaps among daughters in marital rates and in
the spouse's education and earnings. However, we also see that a child's
perceived skin tone was correlated with myriad confounding factors,
including home geography, parental literacy, and other family characteristics.

A more informative assessment of contemporaneous discrimination will
attempt to control econometrically for (some of) these factors. We
therefore first estimate an equation of the form:
\begin{equation}
y_{i}=\alpha+\beta_{1}Black_{i}+\beta_{2}(Black\cup Mulatto)_{i}+\mathbf{x}_{i}\mathbf{\gamma}+u_{i}\label{eq:naive-discrimination-estimation}
\end{equation}
In this specification, we regress an outcome $y_{i}$, on indicators
of individual $i$'s perceived skin tone and on $\mathbf{x_{i}}$,
a vector of observable attributes that are expected to affect the
outcome in question and which may be correlated with race (e.g., age,
geography, or family characteristics). Our estimated $\beta$ coefficients
will then measure mean differences in $y_{i}$ between racial groups
conditional on these other controls, with $\beta_{1}$ measuring the
``\emph{Black--Mulatto}'' gap and $\beta_{2}$ measuring the ``\emph{Mulatto--White}''
gap. We use this approach to measure the initial gap in outcomes by
child race in the general population, along with the extent that this
gap is explained by observable characteristics. We pool observations
across years in our baseline specification, including cohort fixed
effects and allowing all non-race controls to vary flexibly by year
(via interactions with cohort fixed effects). In alternative specifications,
we also separately estimate race and perceived skin tone gaps by child
cohort.

Even an extensive set of controls cannot control for the unobservable
factors that might vary systematically with race. In particular, there
are many attributes of the environment in which an individual grows
up that affect their adult outcomes. When comparing Black vs.\ Mulatto
children, we can control for even unobservable differences in children's
access to resources and opportunities varying across families by comparing
Black-coded and Mulatto-coded children from differently-classified
(BMS\slash BMD) families. These siblings grew up in the same family
but differ in their perceived skin tone to the extent that the enumerator
classified them differently.

One natural way to measure the gap across African American skin tones
within the BMS sample would be to reformulate Equation~\ref{eq:naive-discrimination-estimation}
by decomposing the error term for individual $i$ (raised in family
$f$) into a family and an individual-specific component:\footnote{See \citet{abramitzky2020discrimination} and \citet{kreisman2022distinctively}
for papers that use a similar identification strategy to disentangle
family circumstances from within-household differences. Both use household
fixed effects to compare the outcomes of siblings with more distinctively
foreign or Black names, respectively. Similar to most of our results
for men, both papers find much smaller gaps in outcomes among siblings
than in the general population.}
\begin{equation}
y_{i}=\alpha+\beta_{1}Black_{i}+\mathbf{x}_{i}\mathbf{\gamma}+\underset{u_{i}}{\underbrace{\phi_{f}+\varepsilon_{if}}},\label{eq:within-hh-estimation}
\end{equation}
In this family fixed effect specification, sons coded as Black are
compared only to their Mulatto-coded brothers.\footnote{Our family definition (see Section~\ref{subsec:child-cohorts}) precludes
White and Non-White siblings, so our ``\emph{Mulatto--White}''
comparison is subsumed by our family fixed effects.} This requirement compounds the considerable difficulty of linking
individuals across census years, as two differently-classified brothers
will contribute to our ``\emph{Black--Mulatto}'' gap estimate only
if \emph{both} brothers are linked across census years. This results
in a much smaller sample size and lower power when analyzing linked
outcomes.

Our preferred specification for estimating the \emph{Black--Mulatto}
outcome gap among BMS families compares the weighted mean of $y_{i}$
among Black-coded BMS sons to that of Mulatto-coded BMS sons, conditional
on other controls (with an analogous specification for BMD daughters).
We implement this approach by adding race indicators interacted with
BMS family indicators to Equation~\ref{eq:naive-discrimination-estimation}:

\begin{multline}
y_{i}=\alpha+\beta_{1}Black_{i}+\beta_{2}(Black)\times(Not\:BMS)_{i}+\beta_{3}(Black\cup Mulatto)_{i}\times(BMS)_{i}\\
+\beta_{4}(Black\cup Mulatto)_{i}\times(Not\:BMS)_{i}+\mathbf{x}_{i}\mathbf{\gamma}+u_{i}\label{eq:within-BMS-estimation}
\end{multline}

In this specification, $\beta_{1}$ measures the \emph{Black--Mulatto}
gap among sons in differently-classified families. In the BMS sample,
a given family may have contained an unequal number of Black-coded
and Mulatto-coded sons. We therefore weight observations within a
BMS family by the inverse proportion of their racial classification
within a family.\footnote{BMS individuals of racial classification $j\in{\{Black,Mulatto\}}$
are weighted proportionally to $\frac{{min\{N_{Black},N_{Mulatto}\}}}{N_{j}}$,
where $N_{j}$ denotes the number of sons of classification $j$ within
an individual's family. Weights in our BMS sample are then scaled
to have mean 1, so that BMS and non-BMS individuals receive the same
average weight.} For example, if a given BMS family contained one Black-coded and
two Mulatto-coded sons, the Black-coded son is assigned twice the
weight of each of the Mulatto-coded sons, so that this family contributes
equally to the weighted Black and Mulatto averages within the BMS
sample. This ensures that our initial sample of Black-coded and Mulatto-coded
children in BMS families have identical (weighted) distributions of
family characteristics, including access to opportunities and resources
that vary across families. This approach is closely related to controlling
for differences in the perceived skin tone distribution across families
by using inverse propensity score weights for racial group membership
when predicting race using family fixed effects. It is also conceptually
similar to randomly sampling matched pairs of Black- and Mulatto-coded
siblings from within differently-classified families. See Appendix~\ref{sec:appendix-weighted-estimator}
for a finite-sample unbiasedness result for a simplified, complete-outcome
version of our weighted mixed-sibling specification that treats the
covariate adjustment as known, and for conditions under which its
reported coefficient consistently estimates\ $\beta_{1}$ under the
data-generating process in Equation~\ref{eq:within-hh-estimation}.

By comparing these two groups of individuals with identical family
opportunity access distributions but different perceived individual
skin tones, we are able to plausibly estimate the causal effect of
skin tone perceptions themselves on economic outcomes (see Section~\ref{sec:interpretation}
for further discussion). In Appendix Table~\ref{tab:bms-match-rate-diff},
we show that Black- and Mulatto-coded siblings in our BMS and BMD
samples are linked to their adult selves at similar rates by our preferred
Census Tree links, which suggests that differential attrition is unlikely
to significantly influence our estimates.\footnote{When using our alternative links, we do sometimes find small but statistically
significant differences in linking rates; MLP links match Black-coded
siblings more frequently than Mulatto-coded siblings ($\gaplinkmlp$
and\ $\gaplinkmlpwomen$ percentage points for brothers and sisters,
respectively), while genealogical links match Black-coded daughters
less frequently than their Mulatto-coded siblings ($\neggaplinkgenealwomen$
percentage points).%
} To account for any small differences in these populations which occur
due to minor differences in selection into our regression sample,
we also include our full set of granular geographic and family characteristic
controls in this weighted difference specification. In Sections~\ref{subsec:child-education}--\ref{subsec:adult_educ_married}
we also confirm that our weighted difference and family fixed effect
specifications produce quantitatively similar results when analyzing
child outcomes and both estimate no statistically significant differences
in our main adult economic outcomes among Black vs.~Mulatto siblings
in our BMS and BMD samples.

Relative to the family fixed effect specification, our preferred BMS\slash BMD
weighted specification dramatically increases (generally by a factor
ranging from 2--12) the number of differently-classified siblings
used in the analysis by relaxing the requirement that at least two
siblings are successfully linked. This significantly increases the
precision of these estimates, and Appendix Figure~\ref{fig:algorithm-robustness}
shows that our weighted BMS\slash BMD specification is also more robust
than is the family fixed effect specification to our choice of linking
algorithm.

\section{\label{sec:skincolor-analysis}Perceived skin tone and economic outcomes}

\subsection{\label{subsec:child-education}Gaps in child education}

Table~\ref{tab:inschool} shows race and perceived skin tone gaps
in school attendance among children age 5--18, defining \emph{$y_{i}$
}as an indicator for reported school attendance within a recent specified
period. Note that analysis of this subsection does not rely on linking,
because childhood race classification, education, and siblings are
observed in the same census. Comparing sons in Panel~A, we see large
initial gaps in the general population correlated with perceived skin
tone. $\meaninschoolsonswhite$\% of White-coded children attended
school, which is larger than the corresponding share among Mulatto-coded
($\meaninschoolsonsmulatto$\%) and Black-coded ($\meaninschoolsonsblack$\%)
children. Controlling only for age and birth order fixed effects (interacted
with child cohort) in column~(1), we observe a \emph{Black--Mulatto}
school attendance gap of $\gappopinschool$ percentage points and
a \emph{Mulatto--White} school attendance gap of $\gappopwminschool$
percentage points. These gaps shrink to $\gapstateinschool$ percentage
points and $\gapstatewminschool$ percentage points, respectively,
when controlling for state fixed effects, to $\gapenuminschool$ percentage
points and $\gapenumwminschool$ percentage points after adding granular
geographic controls, and shrink further to $\gapfamcharinschool$
percentage points and $\gapfamcharwminschool$ percentage points when
adding controls for a rich set of family characteristics.\footnote{\label{fn:family-characteristics}Controls for family characteristics
include the following predictors: fixed effects for the number of
same-gender children within the family (including those younger than
3); an indicator for the household head being the child's parent;
and indicator for home ownership by the household head (available
starting in 1900). The remaining family characteristic controls include
the following predictors associated with a child's household head,
father, and mother: predicted log wage income score; indicators for
literacy; indicators for presence within the home; fixed effects for
general (1-digit) occupation. We include missing values as a distinct
outcome level for each of these controls. All non-race controls, including
family characteristics, are allowed to vary flexibly by child cohort.} While smaller in magnitude than the White vs.\ non-White gap, the
gap between perceived lighter- and darker-skinned African Americans
was economically and statistically significant in the general population
and across observably similar families.

In contrast, column~(5) shows that among BMS families the weighted
\emph{Black--Mulatto} gap was only $\gapmixedsibsinschool$ percentage
points. This estimate is very close to that of our specification with
family fixed effects (column~6), which estimates a gap of $\gapfamfeinschool$
percentage points. Both of these estimates are not different from
zero at a standard 95\% significance level and are an order of magnitude
smaller than the \emph{Black--Mulatto} school attendance gap in the
general population. The 95\% confidence intervals on these differently-classified
brother estimates have lower bounds of $\gapmixedsibsinschoollb$
percentage points and $\gapfamfeinschoollb$ percentage points, respectively.
This suggests that the vast majority of the \emph{Black--Mulatto}
school attendance gap in the population was driven by differences
across families, rather than by skin tone discrimination itself. We
visualize our estimated racial gaps in child education in Figure~\ref{fig:child-education}.

Table~\ref{tab:inschool} Panel~B similarly estimates gaps in school
attendance among \emph{daughters} by child race. \emph{Black--Mulatto}
gaps were similar to those among sons in columns~(1)--(4), ranging
from $\gappopdaughtersinschool$ percentage points when controlling
only for age and birth order to $\gapfamchardaughtersinschool$ percentage
points when controlling for granular geographic and family characteristics.
Our \emph{Mulatto--White} estimates in these same regressions range
from $\gappopwmdaughtersinschool$ percentage points to $\gapfamcharwmdaughtersinschool$
percentage points. In contrast to our analysis of sons, a smaller
but significant \emph{Black--Mulatto} gap in school attendance remains
among differently-classified families in columns (5)--(6); we estimate
gaps of $\gapmixedsibsdaughtersinschool$ percentage points and $\gapfamfedaughtersinschool$
percentage points using our weighted and family fixed effect specifications
respectively, which are both statistically significant at standard
significance levels.\footnote{These estimates for daughters are statistically significantly different
than our estimates for sons, with p-values for these differences of
$\pgenderinschoolweighted$ and $\pgenderinschoolfamfe$, respectively.} Even within the same African American family, daughters coded as
lighter-skinned were about $\neggapfamfedaughtersinschoolinteger$
percentage points more likely to attend school than their darker-skinned
sisters.

Table~\ref{tab:literacy} similarly estimates racial gaps in literacy
(defined as the ability to read and write) among children age 10 to
18. Panels~A and B report results for sons and daughters, respectively,
which largely parallel our findings regarding school attendance. We
again find similar initial \emph{Black--Mulatto} gaps in literacy
among sons and daughters ($\gappopliterate$ percentage points and
$\gappopdaughtersliterate$ percentage points, respectively), which
largely then disappear among differently-classified brothers ($\gapmixedsibsliterate$
percentage points among BMS families) and which shrink but significantly
persist among differently-classified daughters ($\gapmixedsibsdaughtersliterate$
percentage points among BMD families). Our 95\% confidence interval
on the \emph{Black--Mulatto} gap among BMS sons has a lower bound
of $\gapmixedsibsliteratelb$ percentage points, again suggesting
that most of the \emph{Black--Mulatto} literacy gap in the general
population was driven by factors which varied across families.

It is impossible to tell with certainty why racial gaps in child education
persisted to a significant degree among differently-classified daughters
but not among differently-classified sons. Much of the research on
colorism and our own analysis suggests that African American women
faced harsher penalties for darker skin tone than African American
men, especially on the marriage market (see Section~\ref{subsec:adult_educ_married}
and \citealp{monk2014skin}).\footnote{See \citet{hamilton2009shedding} and \citet{arceo2014race} for empirical
evidence of this phenomenon in more recent times. \citet{Kotlikoff1979}
also shows light skin was associated with the price for female but
not male enslaved people. For contemporaneous cultural references
see, for example, \citet{hurston2010their}.} Higher investment, then, may have yielded higher returns in the marriage
or labor markets for lighter-skinned girls and led to this gap in
educational investment (see \citealp{goldin1992meaning} for discussion
of the historical returns to educational investment for women).\footnote{An alternative explanation is that reverse causality may have driven
some of this educational gap if census enumerators were more likely
to classify African American girls as Mulatto (rather than Black)
if they were more educated. For this effect to drive our observed
differences among differently-classified daughters but not among differently-classified
sons, enumerator classification would need to have responded to the
education of daughters but not of sons. This explanation seems somewhat
unlikely.}

We show educational gaps by child cohort in Figure~\ref{fig:child-education-bycohort}
(\emph{Black--Mulatto} gaps), Appendix Figure~\ref{fig:child-education-bycohort-mulattolesswhite}
(\emph{Mulatto--White} gaps), and Appendix Figure~\ref{fig:child-education-bycohort-bmlesswhite}
(\emph{Black\slash Mulatto--White} gaps). We see that gaps among
perceived darker-skinned children in the general population partially
closed over time, with a particularly dramatic decrease in gap magnitudes
relative to White children (as in \citealp{collins2006historical}).
In contrast, we do not observe clear time trends in the (much smaller)
\emph{Black--Mulatto} gap among differently-classified siblings.

\subsection{\label{subsec:white-adult}Gaps in classification as White adult}

We next analyze adult outcomes, which requires linking children forward
to their adult census records as discussed in Section~\ref{subsec:Matching-across-censuses}.
In Table~\ref{tab:white-plus20} and Figure~\ref{fig:white-yearseduc}
Panel~A, we first analyze differences (by perceived child race and
gender) in an individual's likelihood of being classified as a White
adult by future census enumerators. To do so, we define $y_{i}$ to
be 1 if a child from a cohort in census year $t$ was classified as
White in \emph{t+20} and 0 if this child was identified as Black in
\emph{t+20}, noting that Mulatto was not an available racial category
in any of the linked census years from which we draw adult outcomes
(1900, 1930, and 1940).\footnote{We exclude the rare cases in which a child was later classified as
a race other than White or Black in adulthood ($\raceotherdropped$\%
and $\raceotherdroppedwomen$\% of linked sons and daughters, respectively).} Unsurprisingly, we find that Black- and Mulatto-coded children were
much less likely than White-coded children to be classified as White
adults.

Importantly, we also find that Mulatto-coded sons were significantly
more likely than Black-coded sons to be classified as a White adult
in the general population ($\neggappopwhiteplustwenty$ percentage
points more likely), conditional on geography and family characteristics
($\neggapfamcharwhiteplustwenty$ percentage points more likely),
among sons in our weighted BMS specification ($\neggapmixedsibswhiteplustwenty$
percentage points more likely), and conditional on family fixed effects
($\neggapfamfewhiteplustwenty$ percentage points more likely).\footnote{\citet[Table 4]{dahis2020choosing} similarly find that individuals
(including adults) coded by census enumerators as Mulatto in 1910
were more likely to be coded as White in the following 1920 census,
relative to individuals originally coded as Black. In related work,
\citet{saperstein2013mulatto} analyze transitions in racial classifications
among Black- and Mulatto-coded individuals in U.S.\ census data from
1870--1920. They find that Mulatto-coded individuals are more likely
to appear as Mulatto in future census years, relative to individuals
originally coded as Black.} We see similar differences for daughters, with Mulatto-coded daughters
having been more likely than Black-coded daughters to be classified
as a White adult in the general population ($\neggappopwhiteplustwentywomen$
percentage points more likely), conditional on geography and family
characteristics ($\neggapfamcharwhiteplustwentywomen$ percentage
points more likely), among daughters in our weighted BMD specification
($\neggapmixedsibswhiteplustwentywomen$ percentage points more likely),
and conditional on family fixed effects ($\neggapfamfewhiteplustwentywomen$
percentage points more likely). This difference in future classification
as White provides strong empirical support for the claim that Black
vs.\ Mulatto classification by a census enumerator reflected real
differences in skin tone or other features influencing one's perceived
race and skin tone, with likely similar difference magnitudes in the
general population and among differently-classified siblings.\footnote{The difference between the gaps in the general population vs.\ among
differently-classified siblings varies in direction and magnitude
across different linking methods, and is generally not significantly
different at standard 95\% significance levels. See~Appendix Figures~\ref{fig:algorithm-robustness}
and~\ref{fig:algorithm-robustness-women} for details.}

In Appendix Table~\ref{tab:white-ssn}, we similarly analyze gaps
(by childhood race) in the likelihood that an individual self-reports
their race as White when applying for a Social Security Number (SSN).
To do so, we start with our child cohorts linked to their adult selves
in 1940. We then use data and links provided by the CenSoc Project
\citep{goldstein2021censoc} to link adults in the 1940 Census to
their SSN application information. We again find that individuals
classified as Mulatto during childhood are more likely than Black-coded
children to be listed as White (rather than Black) on their SSN applications
in the general population ($\neggapnonenumwhitefforty$ percentage
points and $\neggapnonenumwhiteffortywomen$ percentage points more
likely among sons and daughters, respectively, statistically significant
only among sons).\footnote{We note that in general, Black\slash Mulatto-coded daughters are much
more likely to be classified as White on their linked SSN applications
than in their adult census records, and also more likely to be classified
as White than are Black\slash Mulatto-coded sons. This may reflect
a higher rate of false positive matches for women in the CenSoc Project
links between adult census records and SSN applications. Gaps among
differently-classified siblings are less precisely estimated and statistically
insignificant due to the difficulty of linking children to both their
adult census records and to their SSN applications.}

\subsection{\label{subsec:adult_educ_married}Gaps in adult educational attainment
and marital outcomes}

Table~\ref{tab:years-educ-1940} and Figure~\ref{fig:white-yearseduc}
Panel~B report differences in the total number of years of schooling
attained by adulthood. Sons identified as White in 1910 or 1920 had
on average attended $\neggappopwmyearseducforty$\ more years of
schooling by 1940 than Mulatto-coded children. As we add controls
for geography and family characteristics, this \emph{White--Mulatto
}gap among sons decreases to $\neggapfamcharwmyearseducforty$\ years.
Comparing differences by child skin tone within the African American
population, we see that perceived darker-skinned children attended
$\neggappopyearseducforty$\ fewer years of education than lighter-skinned
African American children. This gap gradually decreases in magnitude
to $\neggapstateyearseducforty$, $\neggapenumyearseducforty$, and
$\neggapfamcharyearseducforty$ years as we add controls for a child's
state, enumeration district, and family characteristics, respectively.\footnote{Enumeration districts are small, sub-county geographies that could
be completely covered by a single enumerator within two weeks in cities
and within four weeks in rural areas. Enumeration district data are
not available in the 1870 cohort, for which we therefore compare within
counties rather than enumeration districts.} Analyzing differences among differently-classified siblings in column~(5),
we see that Black-coded and Mulatto-coded brothers had similar educational
attainment as adults. This \emph{Black--Mulatto }gap of $\gapmixedsibsyearseducforty$\ years
among siblings is statistically insignificant, and is orders of magnitude
smaller and in the opposite direction of the general population gap.
Adding family fixed effects in column (6) shows a similarly small
and insignificant gap of $\gapfamfeyearseducforty$ between differently
classified brothers. This finding regarding adult educational attainment
is consistent with our analysis of child literacy and school attendance
in Section~\ref{subsec:child-education}, where we also found insignificant
and very small differences between Black- and Mulatto-coded sons within
our sample of differently-classified families.

Comparing daughters in this same figure and table, we see more persistent
educational differences as we add controls, consistent with our analysis
of childhood educational outcomes. We initially see a \emph{Black--Mulatto}
gap among daughters of $\neggappopyearseducfortywomen$ years in the
general population, which decreases in magnitude to $\neggapstateyearseducfortywomen$,
$\neggapenumyearseducfortywomen$, and $\neggapfamcharyearseducfortywomen$
years as we add controls for a child's state, enumeration district,
and family characteristics, respectively. This gap persists at $\neggapmixedsibsyearseducfortywomen$\ years
even among sisters in our preferred weighted BMD specification, but
is less precisely estimated and not statistically significant. Adding
family fixed effects decreases the magnitude of this gap to a much
smaller $\neggapfamfeyearseducfortywomen$ but also decreases the
precision noticeably (se=$\gapfamfeyearseducfortysewomen$), as our
estimate in this specification is not statistically different from
either that of our preferred weighted BMD specification or from zero.
As discussed in Section~\ref{sec:specifications}, specifications
including family fixed effects rely on many fewer observations because
of the need to link both an individual and at least one of their same-gender
siblings. When combined with our more precisely estimated gaps in
childhood educational outcomes that do not rely on imperfect linking,
we can reasonably conclude that there were significant gaps in education
among sisters from differently-classified (BMD) families with perceived
lighter and darker skin tones, but that the magnitude of these gaps
was smaller than in the general population.

We next turn to comparisons of marital outcomes in Tables~\ref{tab:married-plus20}
and~\ref{tab:Syears-educ-1940} and in Figure~\ref{fig:married-Syearseduc}.
First analyzing differences among African American sons, we find that
individuals with perceived darker skin tones (Black-coded) as children
were more likely than Mulatto-coded sons to be married as adults in
the general population ($\gappopmarriedplustwenty$ percentage points
more likely) or conditional on controls for geography and family characteristics
($\gapfamcharmarriedplustwenty$ percentage points). This gap becomes
near-zero and statistically insignificant in our preferred weighted
BMS specification ($\gapmixedsibsmarriedplustwenty$ percentage points),
and remains statistically insignificant conditional on family fixed
effects (with a point estimate of $\gapfamfemarriedplustwenty$ percentage
points). Next analyzing differences in the total number of years of
schooling attained by the sons' spouses (conditional on marriage),
we estimate initially significant \emph{Black--Mulatto }gaps in the
general population ($\neggappopSyearseducforty$ more years on average
for the spouses of Mulatto-coded sons) and conditional on geography
and family characteristics ($\neggapfamcharSyearseducforty$ more
years). This gap also shrinks dramatically and become insignificant
when comparing siblings in our weighted BMS ($\neggapmixedsibsSyearseducforty$
more years) and family fixed effect ($\neggapfamfeSyearseducforty$
more years) specifications. Together, these results suggest that African
American men with darker perceived skin tones did not face significantly
greater contemporaneous discrimination on the marriage market than
did African American men with lighter perceived skin tones.

We find much starker perceived skin tone penalties on the marriage
market when repeating this same analysis for women in Panel~B of
these same tables and in the same figure. We find that daughters with
darker perceived skin tones were less likely than Mulatto-coded daughters
to be married as adults in the general population ($\neggappopmarriedplustwentywomen$
percentage points less likely) or conditional on controls for geography
and family characteristics ($\neggapfamcharmarriedplustwentywomen$
percentage points). This gap actually grows slightly and remains highly
statistically significant among differently classified sisters in
our preferred weighted BMD specification ($\neggapmixedsibsmarriedplustwentywomen$
percentage points, se=$\gapmixedsibsmarriedplustwentysewomen$ percentage
points). This gap becomes less precise and shrinks conditional on
family fixed effects ($\neggapfamfemarriedplustwentywomen$ percentage
points, se=$\gapfamfemarriedplustwentysewomen$ percentage points),
and is not statistically significantly different from our weighted
BMD gap, population gap, or an estimate of zero given this decreased
precision. Next analyzing differences in the total number of years
of schooling attained by the daughters' spouses (conditional on marriage),
we see significant \emph{Black--Mulatto }gaps in the general population
($\neggappopSyearseducfortywomen$ more years on average for the spouses
of Mulatto-coded daughters), conditional on geography and family characteristics
($\neggapfamcharSyearseducfortywomen$ more years), and when comparing
siblings in our preferred weighted BMD specification ($\neggapmixedsibsSyearseducfortywomen$
more years). This gap has a similar magnitude ($\neggapfamfeSyearseducfortywomen$
more years) conditional on family fixed effects, but with less precision
(se=$\gapfamfeSyearseducfortysewomen$ years). In sharp contrast to
our results for men, these results suggest that African American women
with perceived darker skin tones faced substantially more discrimination
than lighter-skinned African American women on the marriage market.\footnote{The p-values for gender differences between our preferred weighted
BMS vs.\ BMD estimates are $\pgendermarriedweighted$ for marriage
and $\pgenderspouseeducweighted$ for spouse education outcomes, respectively.} These results resonate strongly with the existing literature, which
often finds that skin tone discrimination is gendered and more prevalent
against women, particularly on the marriage market due to discriminatory
beauty standards \citep{hill_skin_2002,hamilton2009shedding}. Our
findings of significant differences in marital outcomes among sisters,
and of differences in the likelihood of future classification as White
among both brothers and sisters, also suggest that our data and weighted
BMS\slash BMD method are able to detect significant differences when
such gaps existed, supporting our interpretation of estimates of otherwise
similar outcomes among brothers as ruling out large effects of contemporaneous
discrimination.

In Appendix Table~\ref{tab:Swhite-plus20} and Appendix Figure~\ref{fig:white-other}
Panel~B, we show that African American children with lighter perceived
skin tones (both sons and daughters) were more likely to be married
to spouses classified as White adults (conditional on marriage). This
result is not particularly surprising given the differences in the
future racial classification of these children themselves as adults
and the historical context, in which interracial marriages were often
particular targets of discrimination and were illegal in some states.

\subsection{\label{subsec:adult_income}Gaps in adult income}

We now turn to our analysis of income differences by perceived child
race and gender. To do so, we first analyze differences in actual
log wage income among sons in 1940 in Table~\ref{tab:loginc-1940}
Panel~A and Figure~\ref{fig:income} Panel~A. Darker-skinned African
American men earned $\neggappoplogincforty$\% less than lighter skinned
ones in the general population (column~1). This gap decreases to
$\neggapstatelogincforty$\%, $\neggapenumlogincforty$\%, and $\neggapfamcharlogincforty$\%
as we control for childhood state, enumeration district, and family
characteristics respectively, but remains economically and statistically
significant. In contrast, Black-coded brothers in our BMS sample earned
similar amounts relative to Mulatto-coded brothers in our preferred
weighted BMS specification ($\gapmixedsibslogincforty$\% higher,
se=$\neggapmixedsibslogincfortyse$\%) or conditional on family fixed
effects ($\gapfamfelogincforty$\% higher). In other words, controlling
for childhood access to opportunities at the family-level by comparing
brothers, we find similar incomes among African American brothers
with lighter vs.\ darker perceived skin tones. While there is significant
uncertainty in this estimate due to the difficulty of linking children
to their adult selves, this gap among differently-classified siblings
is different from the general population gap from both an economic
and statistical perspective. With a standard error of $\neggapmixedsibslogincfortyse$\%
on our preferred weighted BMS estimate, our 95\% confidence interval
rules out a gap larger than $\neggapmixedsibslogincfortylb$\%. This
suggests that most of the \emph{Black--Mulatto} observed-wage gap
in the general population was driven by opportunity access that varies
across families, rather than by contemporaneous discrimination that
differentiates among African American skin tones. The \emph{Mulatto--White}
gap in earnings is measured as $\gappopwmlogincforty$\% in the general
population and $\gapfamcharwmlogincforty$\% conditional on geography
and family characteristics, much larger than the income gaps between
lighter- and darker-skinned African Americans.\footnote{See \citet{chetty2020race}, \citet{collins2022african}, \citet{jacome2021mobility},
and \citet{ward2021intergenerational} for related analysis showing
lower rates of intergenerational mobility among African American relative
to White individuals. Racial gaps in income due to discrimination
also contributed to important differences in other important outcomes
such as health \citep{eli2019physician}.}

One concern with an analysis of labor-market income (recorded on the
census as ``amount of money wages or salary received (including commissions)'')
is that the reported values exclude income from one's farm, business,
or other sources. Appendix Table~\ref{tab:loginctot-1940} therefore
repeats the analysis in Table~\ref{tab:loginc-1940}, but adds in
non-wage income. This is imputed using data from the 1950 census,
in which a sample of respondents separately reported their income
from labor and other sources. We calculate the median non-wage income
for each occupation in the South and outside of the South from the
1950 census, and use the deflated values to approximate each individual's
non-wage income in 1940 given his occupation and location.\footnote{We note that significant wage compression occurred between the 1940
and 1950 censuses \citep{goldin1992great}.} Imputing income for farmers and others who did not receive income
from wages or salaries also adds observations to our analysis. Overall,
the results for total income are qualitatively similar to those for
labor income.

While this income analysis uses actual wage income in 1940, this restricts
our analysis to our 1910 and 1920 child cohorts as our 1870 and 1880
cohorts are unlikely to still be working in 1940. In Table~\ref{tab:logincscore-plus20}
Panel~A and Figure~\ref{fig:income} Panel~B, we now analyze differences
in log wage income scores for all four cohorts in census year \emph{t+20.}
As described in Section~\ref{subsec:income-prediction}, log wage
income scores are predicted by occupation, location, and other characteristics.
Our conclusions are qualitatively similar using income scores and
all four cohorts. We again find \emph{Mulatto--White} gaps among
sons which shrink as we add controls (from a $\gappopwmincscoreplustwenty$\%
difference in predicted income scores to\ $\gapfamcharwmincscoreplustwenty$\%)
and which are larger than the \emph{Black--Mulatto} gaps (which range
from $\gappopincscoreplustwenty$\% in the general population to $\gapfamcharincscoreplustwenty$\%
when controlling for enumeration districts and family characteristics).
Among our differently-classified brothers sample, we again find small
and statistically insignificant gaps in our preferred weighted BMS
specification ($\gapmixedsibsincscoreplustwenty$\%, se=$\neggapmixedsibsincscoreplustwentyse$\%)
and in our family fixed effects specification ($\gapfamfeincscoreplustwenty$\%
in the opposite direction, se=$\neggapfamfeincscoreplustwentyse$\%).
Taken together, these results suggest that any gaps in earned or predicted
income among differently-classified brothers were small relative to
the overall population gaps.

We next analyze these same income and income score outcomes for daughters'
spouses, conditional on marriage. We focus on analyzing the earnings
of spouses (men) rather than of the women themselves because female
labor force participation rates were low during this period. These
results are reported in Tables~\ref{tab:loginc-1940}--\ref{tab:logincscore-plus20}
Panel~B and in Figure~\ref{fig:income}. We find a \emph{Black--Mulatto}
gap in the 1940 income of daughters' spouses in the general population
(the spouses of Black-coded daughters earned $\neggappoplogincfortywomen$\%
less on average). This gap decreases to $\neggapstatelogincfortywomen$\%,
$\neggapenumlogincfortywomen$\%, and $\neggapfamcharlogincfortywomen$\%
as we control for childhood state, enumeration district, and family
characteristics, respectively. When comparing among differently-classified
sisters in our weighted BMD specification, this gap mostly persists
in magnitude but becomes less precisely estimated ($\neggapmixedsibslogincfortywomen$\%,
se=$\neggapmixedsibslogincfortysewomen$\%). Adding family fixed effects
dramatically decreases the precision and changes the sign of this
estimate, as it ultimately relies on a comparison of just $\mixedsibscountSlogincfamfewomen$
differently-classified sisters who are both linked and have spousal
income.\footnote{Our estimates are similar when incorporating imputed non-wage income
in Appendix Table~\ref{tab:loginctot-1940} Panel~B.}

When similarly analyzing the predicted log wage income of daughters'
spouses, we see similar gaps in the general population (the spouses
of Black-coded daughters earned $\neggappopincscoreplustwentywomen$\%
less on average). This gap decreases to $\neggapstateincscoreplustwentywomen$\%,
$\neggapenumincscoreplustwentywomen$\%, and $\neggapfamcharincscoreplustwentywomen$\%
as we control for childhood state, enumeration district, and family
characteristics respectively, and remains significant (and is somewhat
larger) in our preferred weighted BMD specification $(\neggapmixedsibsincscoreplustwentywomen$\%,
se=$\neggapmixedsibsincscoreplustwentysewomen$\%). Adding family
fixed effects leads to a somewhat smaller and less precisely estimated
gap $(\neggapfamfeincscoreplustwentywomen$\%, se=$\neggapfamfeincscoreplustwentysewomen$\%).\footnote{Appendix Figure~\ref{fig:incscore-robustness} shows that our conclusions
are fairly robust to alternative income score constructions.} Together, these results are somewhat underpowered but suggest that
differently-classified sisters with perceived darker skin tones may
have married lower-earning men on average (conditional on marriage)
than did their sisters with lighter perceived skin tones. These results
provide additional evidence consistent with a marriage market penalty
for African American women with darker perceived skin tones, though
less definitively than in the case of marriage rates or spouse's education.

\subsection{\label{subsec:migration}Gaps in migration}

African Americans with different perceived skin tones might also have
been differentially likely to move and/or to choose different destinations
given differences across geographies that interact with perceived
skin tone, potentially including variation in discrimination, the
feasibility of passing as White, or in employment or social opportunities.
Appendix Tables~\ref{tab:mvstate-plus20}--\ref{tab:south-plus20}
and Appendix Figure~\ref{fig:migration} report differences by perceived
skin tone in migration patterns, considering two outcome measures:
the propensity to have moved to a different state between a child's
cohort year and \emph{t+20}, and whether individuals lived in the
South in \emph{t+20}.\footnote{Southern states are defined as: Texas, Louisiana, Mississippi, Alabama,
Georgia, South Carolina, Florida, North Carolina, Tennessee, Virginia,
Maryland, Arkansas, or the District of Columbia.} In the general population and conditional on child geography and
family characteristics, Mulatto-coded children were more likely to
migrate to a different state than were Black-coded or White-coded
children. This \emph{Black--Mulatto} gap in migration in the general
population %
shrinks dramatically and becomes statistically insignificant when
we restrict this analysis to differently-classified siblings using
either our weighted BMS\slash BMD specification %
or family fixed effect specification. Black-coded sons were more
likely than Mulatto-coded sons to live in a southern state in \emph{t+20},
with small and statistically insignificant gaps among differently
classified siblings in both our weighted BMS and family fixed effect
specifications. Black-coded daughters were also more likely than Mulatto-coded
daughters to live in a southern state in \emph{t+20 }in the general
population, but among differently classified sisters this gap significantly
changes direction with Black-coded sisters being less likely than
Mulatto-coded sisters to be in a southern state in \emph{t+20}. We
see no such gap among differently classified-sisters conditional on
family fixed effects. Together these results suggest that perceived
skin tone differences didn't significantly affect whether siblings
migrated away from their initial state, and provide suggestive and
somewhat mixed evidence regarding effects on which region individuals
move to when migrating.%

\section{Discussion of identification and interpretation\label{sec:interpretation}}

\subsection{\label{subsec:Full-siblings-or-not}Did siblings share the same biological
parents?}

We use a stringent family definition for which we can be fairly confident
that two siblings within a family shared the same set of biological
parents. We do so because we want to compare siblings who shared the
same family-level access to resources and opportunities, and might
be concerned that this access potentially differed for half, step,
or otherwise non-biological siblings. The high frequency of differently-classified
parents in our BMS\slash BMD samples (discussed in Section~\ref{subsec:child-cohorts})
is also consistent with real differences in perceived skin tone among
biological siblings; we might reasonably expect that households with
differently-classified parents contained siblings with relatively
large variation in perceived skin tone and\slash or that these households
were coded by enumerators who were particularly responsive to variation
in their racial classifications. Still, the possibility remains that
at least some siblings in our sample might not have shared the same
biological parents despite this observed overrepresentation and despite
the sample-selection steps described in Section~\ref{subsec:child-cohorts}.

The possibility that some individuals whom we identify as siblings
in fact had different parents does not in itself constitute a threat
to our differently-classified-family identification strategy. This
would threaten the interpretation only if biological parentage were
systematically correlated with both classification and outcomes, either
through inherited traits or through differential treatment by non-biological
parents (e.g.,\ \citealp{case2000hungry}). Our strict family construction
and parent definition reduce but cannot fully eliminate this concern.%

\subsection{\label{subsec:external-validity}Inferring the role of families'
opportunity access}

By comparing outcome differences by perceived skin tone in the general
population to those among differently-classified siblings, we can
infer the role that systematic differences in families' access to
opportunities and resources played in explaining the population gaps.
Such inference, however, requires that the effects of being perceived
as Mulatto versus Black, conditional on perfect measures of family
opportunity access, were the same in the general population as they
were between differently-classified siblings. This may have been true
under certain assumptions that we discuss below. Under these assumptions,
if contemporaneous skin tone discrimination drove outcome differences
rather than systematic differences in family opportunity access, we
would expect perceived skin tone stratification among siblings to
have been roughly equal to stratification in the general population:
$\hat{\beta}^{f}\approx\hat{\beta}^{p}$. In contrast, if systematic
differences in families' opportunity and resource access drove outcome
differences in the population, we would expect $\hat{\beta}^{f}\approx0$.
Our results for sons are broadly consistent with the latter case.
We therefore consider three conditions that might have caused skin
tone to operate differently across families rather than within our
BMS\slash BMD samples.

\paragraph{Differences in perceived skin tone between Black- and Mulatto-coded
siblings not equal to those in the population}

The difference in perceived skin tone between the average Black-coded
and Mulatto-coded siblings may have differed from the perceived skin
tone difference between the average Black-coded and Mulatto-coded
children in the population. If Black- and Mulatto-coded siblings had
smaller perceived skin tone differences (presumably because siblings
tended to look alike), we should expect smaller outcome differences
than in the population, where perceived skin tone differences were
larger. Small skin tone differences between siblings would also exacerbate
a concern, as laid out in \citet{Griliches1979}, that comparisons
among siblings take away most of the true variation in the independent
variable of interest (schooling in \citealp{Griliches1979}, perceived
skin tone here), leaving the remaining total variation with a higher
share of measurement error. However, enumerators had a strong tendency
to classify all sons or daughters in a household with the same skin
tone. This suggests that large differences in physical appearance
or other social cues may have been required for siblings to be classified
differently, in which case Black-coded and Mulatto-coded siblings
in our sample may actually have had quite distinct skin tones.

Consistent with the idea of similar perceived skin tone differences
among Black- vs.$\:$Mulatto-coded children within differently-classified
families and in the general population, we show in Table~\ref{tab:white-plus20}
and discussed in Section~\ref{subsec:white-adult} that the \emph{Black--Mulatto}
gap in future classification as White was similar in magnitude in
our BMS\slash BMD samples and in the general population. Similar differences
in Black- vs.$\:$Mulatto-coded perceived skin tone variation within
differently-classified families offer support for our differently-classified
family estimates as reasonable estimates for the causal effect of
contemporaneous perceived skin tone discrimination on economic outcomes
in the general population among African Americans.

We also use an empirical specification comparing the weighted populations
of Black- vs.\ Mulatto-coded brothers from differently-classified
families. This helps limit the attenuation concern associated with
fixed effects described in \citet{Griliches1979}, and also generally
produces estimates similar to specifications including family fixed
effects when the latter have sufficient statistical power in our analysis.

\paragraph{Endogenous skin tone classification exacerbated within families}

\citet{Penner2008}, \citet{Freeman2011LookingThePart}, \citet{saperstein2013mulatto},
and \citet{woo2022unveiling} show that racial classification may
depend on signals such as incarceration, unemployment, the way a person
dresses, occupational status, and language. Skin tones themselves
could also have in part reflected whether an individual worked outdoors
or their clothing. If enumerators' racial classification was partially
determined by factors that correlate with our outcomes of interest,
estimates of the effect of perceived skin tone on outcomes can be
biased. Our focus on comparisons by \emph{childhood} perceived skin
tone partially remedies this concern as children had not settled into
occupations.

In line with this logic, in Appendix Figures~\ref{fig:age-robustness-sons}
(for sons) and \ref{fig:age-robustness-daughters} (for daughters)
we compare adult outcome gaps estimated in our baseline samples to
gaps in subsamples based on the age at which childhood race was enumerated
(separating children age 3--10 and 11--18). If an important part
of enumerators' classifications were driven endogenously by aspects
of a child's social presentation that developed over time and which
may be correlated with future outcomes (such as perceptions of intelligence
or markers of perceived future attractiveness), we might expect to
see systematically larger gaps among the subsample classified after
having had more time to develop these characteristics. We find similar
patterns across all three samples, without systematic evidence of
larger gaps among those classified at older ages (in either the general
population or among differently-classified siblings).

Of course, we are unable to determine exactly which characteristics
drove differences in racial classification among differently-classified
siblings, as actual skin tone pigmentation and many other characteristics
are unobservable to researchers. We can however sharpen our interpretation
by testing for significant differences among differently-classified
siblings in observable, \emph{non-physical} characteristics that might
drive persistent differences in perceived race, such as in the racial
distinctiveness of a child's first name (following \citealp{cook2014distinctively}).
Defining names' racial distinctiveness as the share of children coded
as Black or Mulatto among all same-gender children in a child's cohort,
Appendix Table~\ref{tab:distinctive-names} shows that the first
names of Black-coded and Mulatto-coded siblings are similarly distinctive
on average in both our BMS and BMD samples.

One key factor that does systematically affect physical skin tone,
even conditional on family, is age. Skin pigment typically continues
to develop during the early months (and sometimes years) of life.
Appendix Figure~\ref{fig:share-mulatto-sons-byage} Panel~A illustrates
that among families with Black- and\slash or Mulatto-coded sons, the
youngest children are disproportionately likely to be classified as
Mulatto. We therefore restrict our analysis throughout to children
whose racial classification was conducted between the ages of 3 and
18. In Appendix Figure~\ref{fig:share-mulatto-sons-byage} Panel~B,
we show the share of Mulatto-coded sons by age among families with
both a Black- and Mulatto-coded son age 3--18, where skin tone does
not appear to have been strongly correlated with age. Regardless,
we include age and birth order fixed effects in all baseline regressions.
We also note that endogenous skin tone classification would (if anything)
likely bias our estimates towards finding significant differences
by perceived skin tone among differently-classified siblings, and
we find no such differences (at the 5\% significance level) in main
outcomes among differently-classified brothers and only small childhood
differences among differently-classified sisters. We also show qualitatively
similar results across distinct age groups in Appendix Figures~\ref{fig:age-robustness-sons}
and \ref{fig:age-robustness-daughters}.

\paragraph{Systematic reallocation of resources or peer effects in the family}

Having a sibling of a different skin tone may have given rise to differential
treatment that would not have arisen if there were no lighter- or
darker-skinned sibling in the same household. If parents reallocated
familial resources from darker-skinned children to lighter-skinned
ones (perhaps perceiving higher returns on their investments), our
within-family estimates of the effect of perceived skin tone would
overstate the effects of discrimination as it operated in the general
population. The observed gaps among sisters in educational attainment
and marital outcomes could plausibly be explained by such a mechanism,
with parents potentially investing less in the education of their
daughters with perceived darker skin due to anticipation of discrimination
toward these daughters on the marriage market. On the other hand,
if parents wished to equalize the economic welfare of their children,
they might have tried to counteract discrimination outside of the
family by reallocating resources from lighter- to darker-skinned children,
and our within-family estimates would understate the effect of perceived
skin tone. To the extent that either response occurred, our estimates
would still correctly identify the causal effect of perceived individual
skin tone within differently-classified families (including the impact
of this family allocation effect) but might not extrapolate to the
effect in families with uniform classifications.

Peer effects between siblings may be another channel through which
skin tone had a different effect on outcomes inside the family vs.\ in
the population. The specific family trait of having children with
different skin tone cannot be extrapolated to the general population.
Children who grew up with a sibling of different perceived skin tone
could have changed their own behavior, and\slash or their siblings'.%

Both reallocation by parents and peer effects across siblings are
unobservable to us as researchers. They are common potential confounders,
however, in this type of sibling study.

\section{\label{sec:conclusion}Conclusion}

We use full-count U.S.\ decennial census data from 1870--1940 to
analyze differences in economic outcomes by race and perceived skin
tone. To do so, we construct a panel data set that follows children
from their childhood households to their adult outcomes and we compare
children coded as White, Black, or Mulatto by census enumerators.
We first document large educational and income gaps between African
American and White individuals that continued throughout the period,
as would be expected given the severe discrimination African Americans
have faced. The paper then documents smaller but economically significant
gaps in these outcomes among African Americans with darker skin tones
relative to lighter-skinned African Americans (as perceived by enumerators).

We next focus on estimating the extent to which these economic gaps
among African Americans can be explained by contemporaneous racism
and discrimination that differentiates across perceived skin tone
differences, or instead by differences in families' access to opportunities
and resources. These differences in access themselves likely reflect
racism and discrimination in earlier generations or at the family-level.
Our primary analysis answering this question compares differently-classified
siblings from families in which census enumerators classified at least
one child as Black while coding their same-gender sibling as Mulatto.
Comparing the outcomes of differently-classified siblings allows us
to estimate the causal impact of being perceived as Black relative
to Mulatto during this time period, holding family opportunity and
resource access constant so that we can separately account for the
relative role of these two mechanisms. As discussed throughout the
paper, these differences in perceived skin tone likely reflect a bundle
of not only physical characteristics but also social and contextual
cues. We find (in both the general population and among differently-classified
siblings) that children coded as Mulatto were more likely than Black-coded
children to be classified as White adults in future census years (in
which the Mulatto category no longer existed). This difference in
future classification as White suggests that whatever the bundle of
physical and\slash or social characteristics that differed on average
between Mulatto-coded and Black-coded children, this difference persisted
across different time periods and census enumerators.

We find significantly lower educational attainment, marriage rates,
and spouse's education and likely earnings among African American daughters perceived
as having darker skin tones in differently-classified families relative
to their Mulatto-coded sisters. The educational gaps are much smaller
in magnitude than \emph{Black--Mulatto} differences in the general
population, suggesting that differences in families' opportunity and
resource access explained the majority of the educational gap among
African American girls with darker perceived skin tones (relative
to African Americans perceived as lighter-skinned) but that ongoing
discrimination still played a significant role in determining this
gap. Importantly, the differences in marital outcomes show that Black-coded
sisters faced a large penalty on the marriage market relative to their
sisters with perceived lighter skin tones. This suggests different
anticipated returns to educational investment as a potential mechanism
for the observed gap in educational attainment among sisters. The
gap in marital outcomes among differently-classified sisters is similar
in magnitude to the general population gap, suggesting contemporaneous
perceived skin tone discrimination as the primary driver of these
marriage market gaps, due perhaps to discriminatory beauty standards
\citep{hill_skin_2002,hamilton2009shedding}.

In contrast, we find that African American brothers who differed in
perceived skin tone had similar outcomes on average in terms of education,
marriage prospects, and earnings. Our point estimates of this gap
are generally near zero, and their 95\% confidence intervals can rule
out that the \emph{Black--Mulatto }income gap among differently-classified
brothers was more than about a third as large as the analogous \emph{Black--Mulatto}
gap in the population. These results suggest that conditional on being
perceived as African American, the differences in 1870--1940 outcomes
among men with different third-party categorization of skin tones
were likely driven primarily by variation in families' access to resources
and opportunities, rather than resulting from perceptions of an individual's
skin tone. Differences in families' opportunity access likely reflect
variation in discrimination faced by earlier generations or at the
family-level, and alleviating these gaps in outcomes may require policies
specifically targeting these underlying disparities in access to opportunities
and resources.

We note that skin tone variation among African Americans could have
larger impacts during other time periods or in other societal contexts
less beholden to ``one-drop'' racial classification rules, or through
the long-term accumulation of even small effects over generations.
Our analysis also shows that throughout the late 19th and early 20th
centuries, White vs.\ Mulatto\slash Black gaps in all outcomes studied
were much larger than Mulatto vs.\ Black gaps, consistent with the
abundant qualitative and quantitative evidence of discrimination towards
African Americans.

Our findings regarding the gendered and intersectional nature of perceived
skin tone effects among African Americans resonate with a rich literature,
which we contribute to by quantifying differences in economic outcomes
across perceived skin tones while controlling for family background
differences by comparing differently-classified siblings. The gendered
nature of these perceived skin tone effects could reflect different
stereotypical gender roles in society during the time period we study.
More broadly, our paper contributes to our understanding of racial
equality, and of the important and understudied topics of colorism
and skin tone. As \citet{monk2021unceasing} writes: ``in a society
with increasing intermarriage and `multiracial' children---in which
everyone will look mixed---the importance of skin tone should only
increase over time as ethnoracial categories become even more heterogenous
with respect to phenotype.''

\newpage{}

{\footnotesize\begin{singlespace}\bibliographystyle{chicago}
\bibliography{references}
\end{singlespace}}{\footnotesize\par}

\begin{singlespace}
\pagebreak{}
\end{singlespace}

\section{Figures and tables}

\input{figures.tex}\pagebreak\input{tables.tex}

\appendix
\newgeometry{bottom=1cm}

\setcounter{figure}{0}
\renewcommand{\figurename}{Appendix Figure}
\renewcommand\thefigure{\thesection\arabic{figure}}
\setcounter{table}{0}
\renewcommand{\tablename}{Appendix Table}
\renewcommand\thetable{\thesection\arabic{table}}

\section{\label{sec:appendix-exhibits}Additional figures and tables}

\input{appendix_figures.tex}\pagebreak\input{appendix_tables.tex}

\pagebreak{}

\section{\label{sec:race-instructions}Race instructions to census enumerators,
by year}

\input{appendix_race_instructions.tex}

\restoregeometry

\pagebreak{}

\section{\label{sec:appendix-weighted-estimator}Balancing family effects
with sibling weights}

We first give intuition and a finite-sample unbiasedness result for
a simplified version of Equation~\ref{eq:within-BMS-estimation}
(our weighted mixed-sibling specification) under the data-generating
process in Equation~\ref{eq:within-hh-estimation} (our family fixed
effects specification). We then state conditions connecting this benchmark
to the reported estimator. Under the measurement assumptions in the
text, we interpret the common Black--Mulatto classification effect
as a perceived skin tone effect.

The family fixed effect data-generating process in Equation~\ref{eq:within-hh-estimation}
is

\[
y_{if}=\alpha+\beta_{1}Black_{i}+\mathbf{x}_{i}\mathbf{\gamma}+\phi_{f}+\varepsilon_{if}.
\]

Within a census-year-by-gender cell, let\ $B_{f}$ and\ $M_{f}$
be the sets of Black- and Mulatto-coded children in differently-classified
family\ $f$, and let\ $N_{Bf}$ and\ $N_{Mf}$ be their sizes.
Set\ $q_{f}=\min\{N_{Bf},N_{Mf}\}$ and assign child\ $i$ in classification\ $d\in\{B,M\}$
the sibling weight\ $\omega_{if}=q_{f}/N_{df}$. Because the total
weight assigned to Black-coded and Mulatto-coded siblings in family\ $f$
is\ $q_{f}$ for each group, the family effect cancels:

\[
\sum_{i\in B_{f}}\omega_{if}\phi_{f}-\sum_{i\in M_{f}}\omega_{if}\phi_{f}=q_{f}\phi_{f}-q_{f}\phi_{f}=0.
\]

This equality holds for every value of\ $\phi_{f}$. It requires
neither equal family effects nor independence between family effects
and family composition.

In the proposition below, we show this sibling-weighting argument
more formally using a complete-outcome, oracle-residualized benchmark.\footnote{To isolate this argument, we condition on the childhood roster, treat\ $\mathbf{x}_{i}\mathbf{\gamma}$
as known, and abstract from estimation and winsorization of the inverse-link
weights and from the non-mixed observations used to estimate shared
controls. These implementation features generally preclude exact finite-sample
unbiasedness of the feasible coefficient, which we instead interpret
using the conditions below.}

\textbf{Proposition C.1 (finite-sample unbiasedness).}\emph{ Suppose
Equation~\ref{eq:within-hh-estimation} is the data-generating process
with a common\ $\beta_{1}$. Condition on\ $\mathcal{G}$, consisting
of the complete family rosters, classifications, covariates, and arbitrary
realized family effects\ $\{\phi_{f}\}$. Suppose every child used
to construct the sibling weights has a valid observed outcome,\ $\mathbf{x}_{i}\mathbf{\gamma}$
is known or has already been partialled out, the two estimators below
are identified, and\ $\mathbb{E}[\varepsilon_{if}\mid\mathcal{G}]=0$.
Let\ $y^{*}_{if}=y_{if}-\mathbf{x}_{i}\mathbf{\gamma}$,\ $Q=\sum_{f}q_{f}$,
and define the simplified weighted mixed-sibling estimator}

\emph{
\[
\widehat{\beta}^{(3)}_{1,\circ}\equiv\frac{1}{Q}\sum_{f}\left[\sum_{i\in B_{f}}\omega_{if}y^{*}_{if}-\sum_{i\in M_{f}}\omega_{if}y^{*}_{if}\right].
\]
}

\emph{Let\ $\widehat{\beta}^{(2)}_{1}$ denote the corresponding
ordinary-least-squares, complete-outcome family fixed effect estimator
in Equation~\ref{eq:within-hh-estimation} on the same mixed-family
sample. Then}

\emph{
\[
\mathbb{E}\!\left[\left.\widehat{\beta}^{(2)}_{1}\right|\mathcal{G}\right]=\mathbb{E}\!\left[\left.\widehat{\beta}^{(3)}_{1,\circ}\right|\mathcal{G}\right]=\beta_{1}.
\]
}

\emph{Proof.} Substituting the data-generating process into the simplified
weighted estimator gives

\[
\begin{aligned}\widehat{\beta}^{(3)}_{1,\circ}-\beta_{1} & =\frac{1}{Q}\sum_{f}\Bigg[\underbrace{(q_{f}-q_{f})(\alpha+\phi_{f})}_{0}\\
 & \qquad\qquad+\sum_{i\in B_{f}}\omega_{if}\varepsilon_{if}-\sum_{i\in M_{f}}\omega_{if}\varepsilon_{if}\Bigg].
\end{aligned}
\]

The conditional expectation of the remaining terms is zero. For the
family fixed effect estimator, let\ $D_{if}=Black_{i}$ and let\ $\ddot{D}_{if}$
be its residual after projection on the family indicators and\ $\mathbf{x}_{i}$.
Frisch--Waugh--Lovell gives

\[
\widehat{\beta}^{(2)}_{1}-\beta_{1}=\frac{\sum_{f}\sum_{i\in f}\ddot{D}_{if}\varepsilon_{if}}{\sum_{f}\sum_{i\in f}\ddot{D}^{\,2}_{if}},\qquad\mathbb{E}\!\left[\left.\widehat{\beta}^{(2)}_{1}-\beta_{1}\right|\mathcal{G}\right]=0.\qquad\square
\]

\textbf{Relation to the implemented estimator.} For linked outcomes,
Equation~\ref{eq:within-BMS-estimation} also uses estimated and
winsorized inverse-link weights. Across outcomes, its shared controls
are estimated using both mixed- and non-mixed-family observations.
Let\ $R^{y}_{if}$ indicate that outcome\ $y$ is observed.\footnote{For an adult outcome,\ $R^{y}_{if}=1$ requires both a successful
match to a future census and a valid recorded value for that outcome.
For a childhood outcome, it requires a valid recorded value and satisfaction
of the outcome's age restriction.} Let\ $a_{if}$ be the final realized regression weight for a mixed-sibling
observation, incorporating\ $\omega_{if}R^{y}_{if}$, the inverse-link
weight, and any normalization or winsorization. Define\ $A_{df}=\sum_{i\in d_{f}}a_{if}$,\ $p_{df}=A_{df}/\sum_{h}A_{dh}$,
and the weighted difference in means

\[
\Delta_{a}z\equiv\frac{\sum_{f}\sum_{i\in B_{f}}a_{if}z_{if}}{\sum_{f}A_{Bf}}-\frac{\sum_{f}\sum_{i\in M_{f}}a_{if}z_{if}}{\sum_{f}A_{Mf}}.
\]

Let\ $\widehat{\mathbf{\gamma}}$ be the coefficient on the shared
controls. The group-indicator normal equations imply\ $\widehat{\beta}^{(3)}_{1}=\Delta_{a}(y_{if}-\mathbf{x}_{i}\widehat{\mathbf{\gamma}})$.
Equation~\ref{eq:within-hh-estimation} therefore yields the exact
decomposition

\[
\widehat{\beta}^{(3)}_{1}-\beta_{1}=\underbrace{\sum_{f}(p_{Bf}-p_{Mf})\phi_{f}}_{\text{post-link family-share imbalance}}+\Delta_{a}\varepsilon+\Delta_{a}\!\left\{ \mathbf{x}_{i}(\mathbf{\gamma}-\widehat{\mathbf{\gamma}})\right\} .
\]

Under complete outcomes and the raw sibling weights,\ $p_{Bf}=p_{Mf}$
family by family, so the first term is zero for arbitrary family effects.
With observed outcomes and feasible weights, a sufficient balance
condition is

\[
\mathbb{E}\!\left[A_{Bf}-A_{Mf}\mid\mathcal{F}_{f}\right]=0,
\]
 where\ $\mathcal{F}_{f}$ contains all pre-observation family information,
including\ $\phi_{f}$. Because\ $\mathcal{F}_{f}$ includes\ $\phi_{f}$,
this condition restricts observation and weighting, not family-effect
levels or their relation to family composition. With positivity, finite
moments, laws of large numbers for\ $A_{df}$ and\ $\phi_{f}A_{df}$,
and group weight totals sharing a positive limit, it makes the direct
family-share term vanish asymptotically. True, untruncated inverse
probabilities of\ $R^{y}_{if}=1$ conditional on\ $\mathcal{F}_{f}$
satisfy this condition. For the implemented inverse-link weights,
residual dependence of outcome observation on unobserved family factors
must be common to Black- and Mulatto-coded children within a family;
outcome validity, age eligibility, weight estimation, and winsorization
must also preserve the balance. Sibling observations may be dependent.
The full estimator also requires the displayed control-adjustment
term to vanish. Family-share balance covers family-constant controls;
child-level controls such as age and birth order require weighted
balance or consistent estimation of\ $\mathbf{\gamma}$.

\emph{Corollary. If the conditional family-balance condition above
and these regularity conditions hold, and the residual difference
and shared-control adjustment term converge in probability to zero,
then\ $\widehat{\beta}^{(3)}_{1}\xrightarrow{p}\beta_{1}$.}

\emph{Proof.} Under the stated conditions, all three terms in the
exact decomposition are\ $o_{p}(1)$.\ $\square$

For linked outcomes, empirically Equation~\ref{eq:within-BMS-estimation}
generally uses 2--12 times as many differently-classified siblings
and has smaller standard errors than Equation~\ref{eq:within-hh-estimation}.

The assumptions also clarify when estimates can differ. Differential
outcome observation within families, misspecified inverse-link weights,
or balance-breaking winsorization can give families different Black
and Mulatto shares, allowing their family effects to enter the estimate.
Misspecified shared controls can leave a nonzero control-adjustment
term. If\ $\beta_{1}$ is heterogeneous, the estimators may recover
different weighted averages; spillovers or other nonadditivity invalidate
the common-effect benchmark.

\pagebreak{}

\section{\label{sec:usct-complexion}Union Army complexion and Census classification}

In this appendix, we test whether Union Army records list darker complexions,
on average, for individuals classified as Black in the 1870 Census
than for those classified as Mulatto. We use the Expanded U.S. Colored
Troops (USCT) data\ \citep{costa2015usct}. Army personnel, including
recruiting officers, recorded complexion as part of recruits' physical
descriptions.

Appendix Table~\ref{tab:usct-complexion} reports the primary comparison
for\ \usctN{} veterans. Army records list ``brown''\slash ``light'' complexions
(as opposed to ``black''\slash ``dark'') for\ $\usctShareMulatto$\% of those classified
in the Census as Mulatto, but for only\ $\usctShareBlack$\% of those
classified in the Census as Black. This positive correlation between
Mulatto (vs.\ Black) Census classification and lighter complexions
in Army records persists across Census-link quality restrictions and
with Army company fixed effects.

\setcounter{table}{0}
\begin{table}[H]
\begin{singlespace}
\begin{centering}
\caption{\label{tab:usct-complexion}Gaps in classification as brown/light
in Union Army records, by 1870 Census classification}
\par\end{centering}
\centering{}\input{input/Tables/usct_complexion.tex}
\end{singlespace}
\end{table}

\end{document}

%% file: input/Scalars/scalars.tex
\newcommand{\nsons}{49,609,678}
\newcommand{\ndaughters}{49,034,333}
\newcommand{\nchildren}{99}
\newcommand{\sharefreeeightfifty}{12}
\newcommand{\nBMSsons}{50,333}
\newcommand{\nBMSfamilies}{17,320}
\newcommand{\nBMDdaughters}{53,762}
\newcommand{\nBMDfamilies}{18,672}
\newcommand{\nBMSBMDfamilies}{35,992}
\newcommand{\nBMDdaughtersrounded}{54,000}
\newcommand{\nBMSsonsrounded}{50,000}
\newcommand{\couplesBlackBlack}{79}
\newcommand{\couplesBlackMulattoUnordered}{10}
\newcommand{\shBlackParBMmin}{34}
\newcommand{\shBlackParBMmax}{53}
\newcommand{\bmsParentsBlackBlack}{29}
\newcommand{\bmsParentsMulatto}{70}
\newcommand{\bmsParentsBlackMulattoUnordered}{59}
\newcommand{\bmsParentsRemainder}{1}

\newcommand{\bmdParentsBlackMulattoUnordered}{55}

\newcommand{\MatchRatebyuauto}{52}
\newcommand{\MatchRategeneal}{24}
\newcommand{\MatchRatessnapp}{10}
\newcommand{\MatchRatectany}{61}
\newcommand{\MatchRatemlp}{43}
\newcommand{\MatchRateABEexcon}{17}
\newcommand{\MatchRatewomenbyuauto}{35}
\newcommand{\MatchRatewomengeneal}{24}
\newcommand{\MatchRatewomenssnapp}{8}
\newcommand{\MatchRatewomenctany}{41}
\newcommand{\MatchRatewomenmlp}{16}
\newcommand{\gappopinschool}{-7.8}
\newcommand{\gappopwminschool}{-18.6}
\newcommand{\gapstateinschool}{-5.6}
\newcommand{\gapstatewminschool}{-10.2}
\newcommand{\gapenuminschool}{-5.0}
\newcommand{\gapenumwminschool}{-10.7}
\newcommand{\gapfamcharinschool}{-3.1}
\newcommand{\gapfamcharwminschool}{-5.4}
\newcommand{\gapmixedsibsinschool}{-0.6}
\newcommand{\gapfamfeinschool}{-0.5}
\newcommand{\gapmixedsibsinschoollb}{-1.3}
\newcommand{\gapfamfeinschoollb}{-1.2}
\newcommand{\gappopdaughtersinschool}{-8.0}
\newcommand{\gapfamchardaughtersinschool}{-3.4}
\newcommand{\gappopwmdaughtersinschool}{-14.9}
\newcommand{\gapfamcharwmdaughtersinschool}{-3.0}
\newcommand{\gapmixedsibsdaughtersinschool}{-2.2}

\newcommand{\neggapmixedsibsdaughtersinschoolinteger}{2}
\newcommand{\gapfamfedaughtersinschool}{-2.0}

\newcommand{\neggapfamfedaughtersinschoolinteger}{2}
\newcommand{\gappopliterate}{-12.6}
\newcommand{\gappopdaughtersliterate}{-12.6}
\newcommand{\gapmixedsibsliterate}{-0.3}
\newcommand{\gapmixedsibsdaughtersliterate}{-2.4}
\newcommand{\gapmixedsibsliteratelb}{-1.2}
\newcommand{\neggappopyearseducforty}{0.9}
\newcommand{\neggappopSyearseducforty}{0.8}
\newcommand{\neggappoplogincforty}{13.6}

\newcommand{\gapmixedsibslogincforty}{0.6}
\newcommand{\neggapmixedsibslogincfortyse}{2.7}
\newcommand{\neggapmixedsibslogincfortylb}{4.8}
\newcommand{\neggapmixedsibslogincfortypoppct}{35}
\newcommand{\neggappopwmyearseducforty}{2.8}
\newcommand{\neggappopwmlogincforty}{68.6}
\newcommand{\neggappopwhiteplustwenty}{2.8}
\newcommand{\neggapfamcharwhiteplustwenty}{2.3}
\newcommand{\neggapmixedsibswhiteplustwenty}{1.6}
\newcommand{\neggapfamfewhiteplustwenty}{1.0}
\newcommand{\neggapfamcharwmyearseducforty}{1.2}
\newcommand{\neggapstateyearseducforty}{0.8}
\newcommand{\neggapenumyearseducforty}{0.7}
\newcommand{\neggapfamcharyearseducforty}{0.3}

\newcommand{\gapmixedsibsyearseducforty}{0.004}

\newcommand{\gapfamfeyearseducforty}{0.1}

\newcommand{\gappopmarriedplustwenty}{2.6}

\newcommand{\gapfamcharmarriedplustwenty}{0.9}

\newcommand{\gapmixedsibsmarriedplustwenty}{0.4}

\newcommand{\gapfamfemarriedplustwenty}{0.9}

\newcommand{\neggapfamcharSyearseducforty}{0.4}
\newcommand{\neggapmixedsibsSyearseducforty}{0.1}
\newcommand{\neggapfamfeSyearseducforty}{0.1}

\newcommand{\neggapstatelogincforty}{10.7}
\newcommand{\neggapenumlogincforty}{9.1}
\newcommand{\neggapfamcharlogincforty}{5.1}

\newcommand{\gapfamfelogincforty}{1.7}

\newcommand{\gappopwmlogincforty}{-68.6}
\newcommand{\gapfamcharwmlogincforty}{-44.3}
\newcommand{\gappopwmincscoreplustwenty}{-45.7}
\newcommand{\gapfamcharwmincscoreplustwenty}{-20.9}
\newcommand{\gappopincscoreplustwenty}{-9.1}

\newcommand{\gapfamcharincscoreplustwenty}{-2.6}

\newcommand{\gapmixedsibsincscoreplustwenty}{-1.3}

\newcommand{\gapfamfeincscoreplustwenty}{0.4}

\newcommand{\neggapmixedsibsincscoreplustwentyse}{0.8}
\newcommand{\neggapfamfeincscoreplustwentyse}{1.1}

\newcommand{\neggapnonenumwhitefforty}{1.6}

\newcommand{\neggappopyearseducfortywomen}{0.7}
\newcommand{\neggappopSyearseducfortywomen}{0.6}
\newcommand{\neggappoplogincfortywomen}{11.6}

\newcommand{\neggapmixedsibslogincfortywomen}{2.0}

\newcommand{\neggapmixedsibslogincfortysewomen}{6.8}

\newcommand{\neggappopwhiteplustwentywomen}{3.0}
\newcommand{\neggapfamcharwhiteplustwentywomen}{2.5}
\newcommand{\neggapmixedsibswhiteplustwentywomen}{2.3}
\newcommand{\neggapfamfewhiteplustwentywomen}{2.2}

\newcommand{\neggapstateyearseducfortywomen}{0.7}
\newcommand{\neggapenumyearseducfortywomen}{0.6}
\newcommand{\neggapfamcharyearseducfortywomen}{0.3}
\newcommand{\neggapmixedsibsyearseducfortywomen}{0.2}
\newcommand{\neggapfamfeyearseducfortywomen}{0.03}
\newcommand{\gapfamfeyearseducfortysewomen}{0.3}
\newcommand{\neggappopmarriedplustwentywomen}{3.0}

\newcommand{\neggapfamcharmarriedplustwentywomen}{4.4}

\newcommand{\neggapmixedsibsmarriedplustwentywomen}{4.3}

\newcommand{\gapmixedsibsmarriedplustwentysewomen}{1.4}
\newcommand{\neggapfamfemarriedplustwentywomen}{0.2}

\newcommand{\gapfamfemarriedplustwentysewomen}{2.4}

\newcommand{\neggapfamcharSyearseducfortywomen}{0.2}
\newcommand{\neggapmixedsibsSyearseducfortywomen}{0.5}
\newcommand{\neggapfamfeSyearseducfortywomen}{0.3}
\newcommand{\gapfamfeSyearseducfortysewomen}{0.5}

\newcommand{\neggapstatelogincfortywomen}{7.7}
\newcommand{\neggapenumlogincfortywomen}{7.0}
\newcommand{\neggapfamcharlogincfortywomen}{2.4}

\newcommand{\neggappopincscoreplustwentywomen}{9.4}

\newcommand{\neggapstateincscoreplustwentywomen}{5.3}

\newcommand{\neggapenumincscoreplustwentywomen}{4.8}

\newcommand{\neggapfamcharincscoreplustwentywomen}{2.6}

\newcommand{\neggapmixedsibsincscoreplustwentywomen}{6.2}

\newcommand{\neggapfamfeincscoreplustwentywomen}{1.8}
\newcommand{\neggapmixedsibsincscoreplustwentysewomen}{2.0}
\newcommand{\neggapfamfeincscoreplustwentysewomen}{4.3}

\newcommand{\neggapnonenumwhiteffortywomen}{1.1}

\newcommand{\gaplinkmlp}{0.9}
\newcommand{\gaplinkmlpwomen}{0.6}
\newcommand{\neggaplinkgenealwomen}{0.4}

\newcommand{\mixedsibscountSlogincfamfewomen}{69}
\newcommand{\meaninschoolsonsblack}{39.9}
\newcommand{\meaninschoolsonsmulatto}{49.8}
\newcommand{\meaninschoolsonswhite}{67.3}
\newcommand{\raceotherdropped}{0.04}
\newcommand{\raceotherdroppedwomen}{0.03}
\newcommand{\marratenowwhitemen}{54}
\newcommand{\marratenowwhitewomen}{53}
\newcommand{\marratenowblackmen}{36}
\newcommand{\marratenowblackwomen}{32}
\newcommand{\pgenderinschoolweighted}{0.0007}
\newcommand{\pgenderinschoolfamfe}{0.0018}
\newcommand{\pgendermarriedweighted}{0.0037}
\newcommand{\pgenderspouseeducweighted}{0.0569}
\newcommand{\algorithmrobustnessmincount}{100}

%% file: input/Scalars/usct_complexion.tex
\newcommand{\usctN}{2,330}
\newcommand{\usctShareBlack}{9.1}
\newcommand{\usctShareMulatto}{20.0}

%% file: figures.tex
\begin{figure}[H]
\begin{centering}
\caption{\label{fig:census-examples}The Best brothers and Ridley sisters in
childhood and adulthood}
\par\end{centering}
\begin{centering}
\includegraphics[width=1\textwidth]{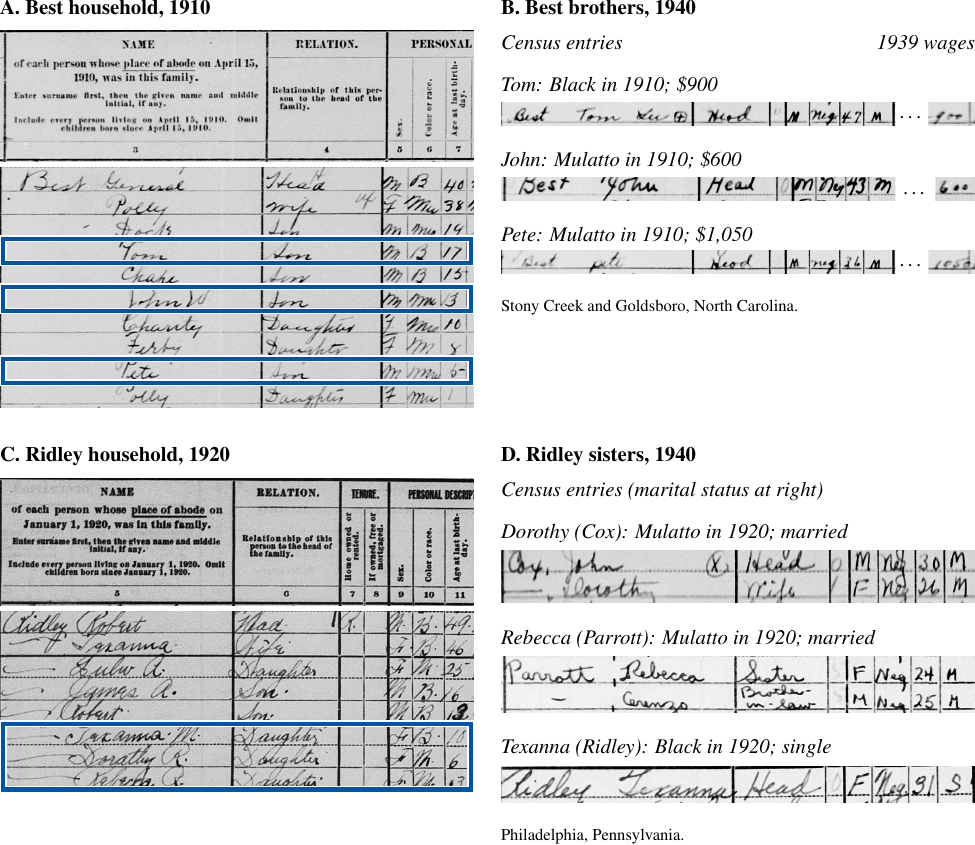}
\par\end{centering}
{\small Note: Panels A and C show childhood census records; boxes mark
the siblings discussed in the text. Panels B and D show the corresponding
adult records from the 1940 census. Wages are annual wage and salary
income in 1939 dollars. Ellipses separate nonadjacent columns from
the same census row. In Panel D, M denotes married and S denotes single.
Dorothy and Rebecca are shown with their husbands, and are linked
across last name changes with evidence from marriage records (not
shown) in 1932 and 1935, respectively.}{\small\par}
\end{figure}

\pagebreak{}
\begin{figure}[H]
\begin{centering}
\caption{\label{fig:child-education}Gaps in childhood education, by perceived
child race and gender}
\par\end{centering}
\begin{centering}
\subfloat[\label{subfig:inschool}School attendance]{\begin{centering}
\par\end{centering}
\centering{}\includegraphics[width=0.8\textwidth]{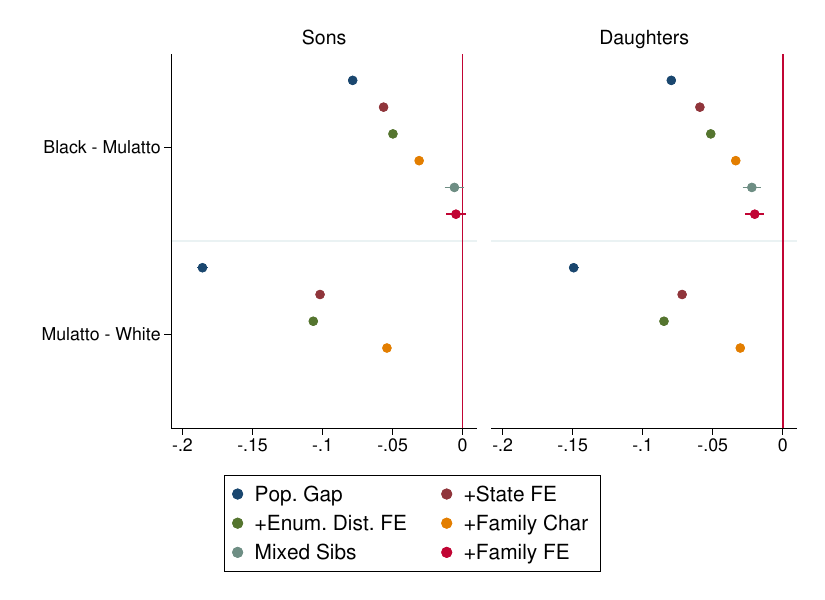}}
\par\end{centering}
\begin{centering}
\subfloat[\label{subfig:literate}Literacy]{
\centering{}\includegraphics[width=0.8\textwidth]{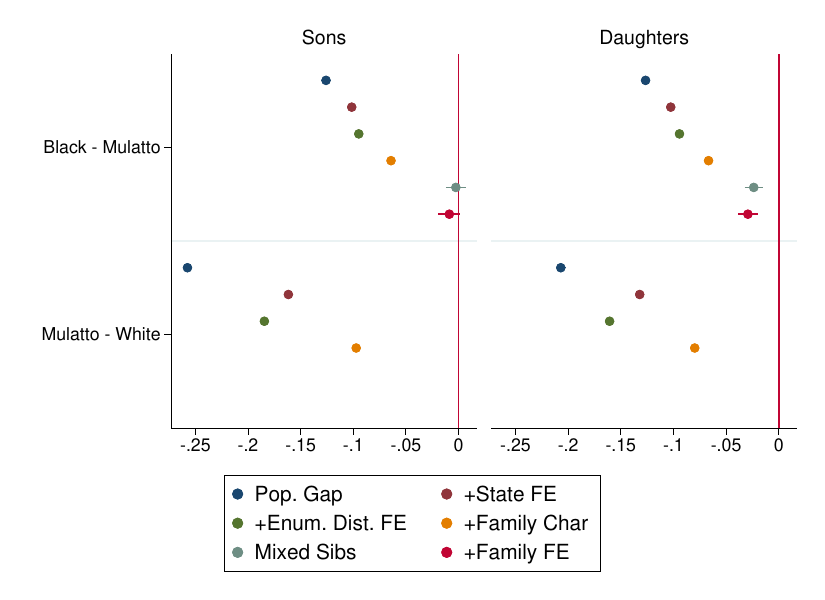}}
\par\end{centering}
{\small Note: Figure shows regression coefficients and 95\% confidence
intervals documenting racial gaps in childhood school attendance (Panel
A) and literacy (Panel B). See Section~\ref{subsec:child-education}
for detail and Tables~\ref{tab:inschool} and \ref{tab:literacy}
for corresponding regression tables.}{\small\par}
\end{figure}

\pagebreak{}
\begin{figure}[H]
\begin{centering}
\caption{\label{fig:child-education-bycohort}\emph{Black--Mulatto }gaps in
childhood education, by child cohort and gender}
\par\end{centering}
\begin{centering}
\subfloat[\label{subfig:inschool-bycohort}School attendance]{
\centering{}\includegraphics[width=0.8\textwidth]{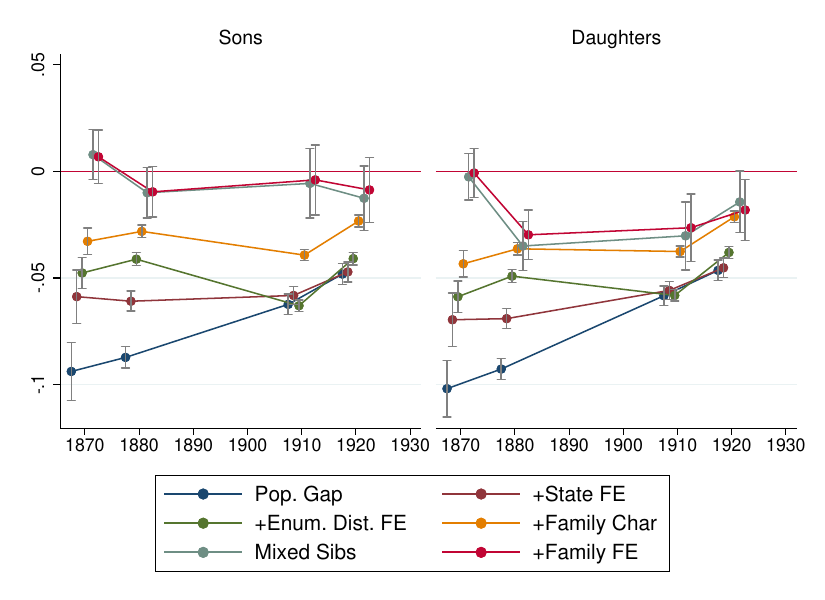}}
\par\end{centering}
\begin{centering}
\subfloat[\label{subfig:literate-bycohort}Literacy]{
\centering{}\includegraphics[width=0.8\textwidth]{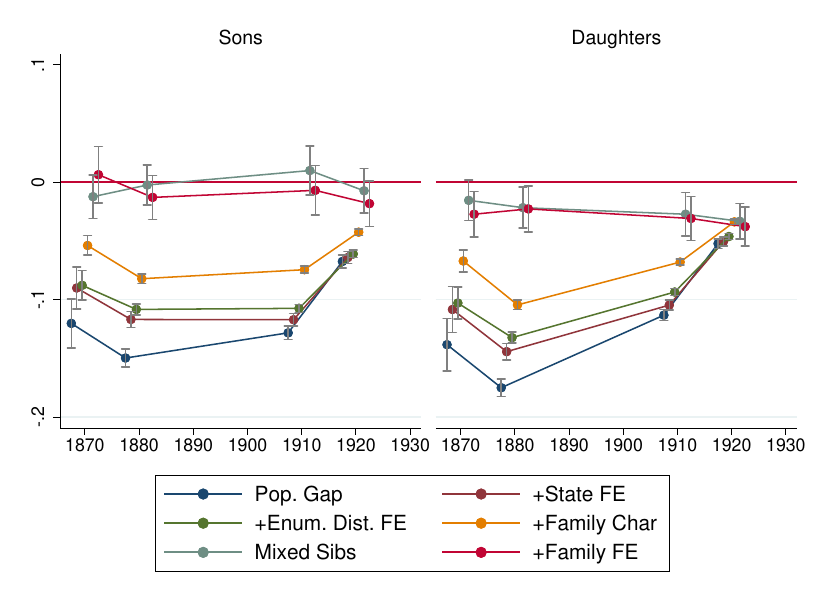}}
\par\end{centering}
{\small Note: Figure shows regression coefficients and 95\% confidence
intervals documenting Black--Mulatto gaps in childhood school attendance
(Panel A) and literacy (Panel B) by gender and child cohort. See Section~\ref{subsec:child-education}
for detail.}{\small\par}
\end{figure}
\pagebreak{}
\begin{figure}[H]
\begin{centering}
\caption{\label{fig:white-yearseduc}Gaps in perceived race and educational
attainment as an adult, by perceived child race and gender}
\par\end{centering}
\begin{centering}
\subfloat[\label{subfig:white-plus20}Classification as White in \emph{t+20}]{
\centering{}\includegraphics[width=0.8\textwidth]{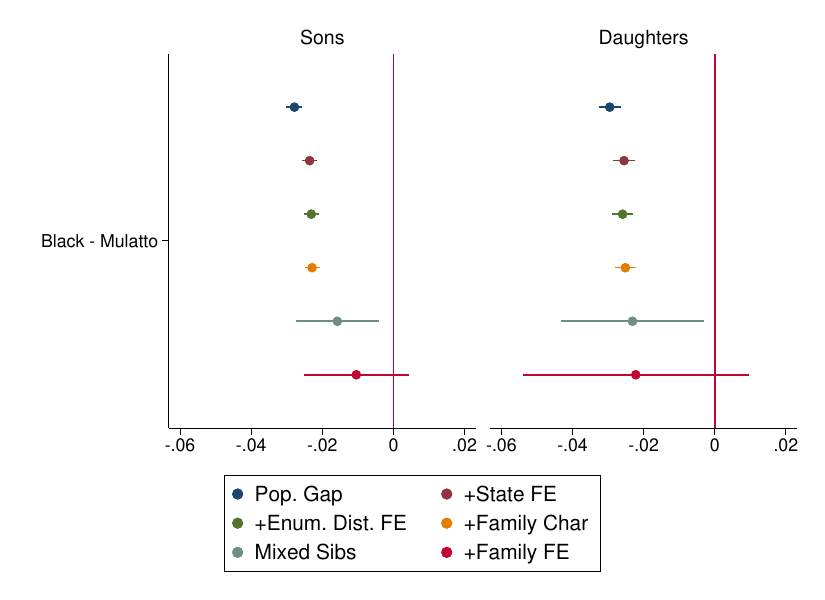}}
\par\end{centering}
\begin{centering}
\subfloat[\label{subfig:years-educ-1940}Years of education by 1940]{
\centering{}\includegraphics[width=0.8\textwidth]{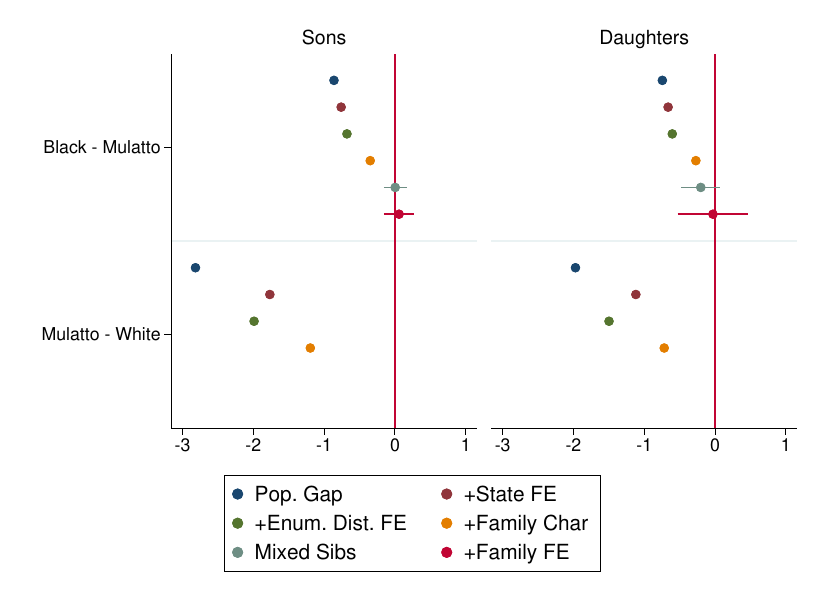}}
\par\end{centering}
{\small Note: Figure shows regression coefficients and 95\% confidence
intervals documenting gaps by perceived childhood race and gender
in an indicator for classification as a White adult in }{\small\emph{t+20}}{\small{}
(Panel A) and for years of education attained by 1940 (Panel B). See
Sections~\ref{subsec:white-adult} and \ref{subsec:adult_educ_married}
for detail, and see Tables~\ref{tab:white-plus20} and \ref{tab:years-educ-1940}
for corresponding regression tables.}{\small\par}
\end{figure}

\pagebreak{}
\begin{figure}[H]
\begin{centering}
\caption{\label{fig:married-Syearseduc}Gaps in marital outcomes as an adult,
by perceived child race and gender}
\par\end{centering}
\begin{centering}
\subfloat[\label{subfig:married-plus20}Married in \emph{t+20}]{
\centering{}\includegraphics[width=0.8\textwidth]{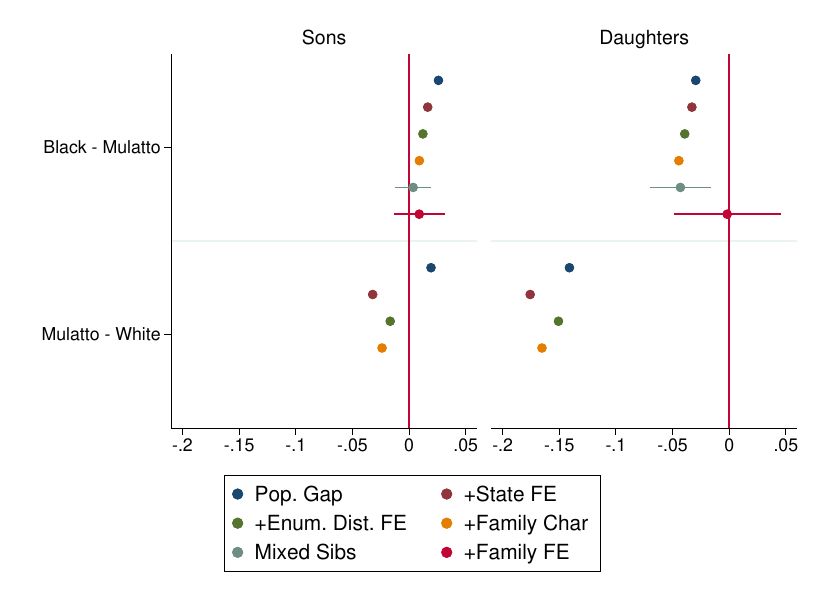}}
\par\end{centering}
\begin{centering}
\subfloat[\label{subfig:Syears-educ-1940}Spouses' years of education by 1940]{
\centering{}\includegraphics[width=0.8\textwidth]{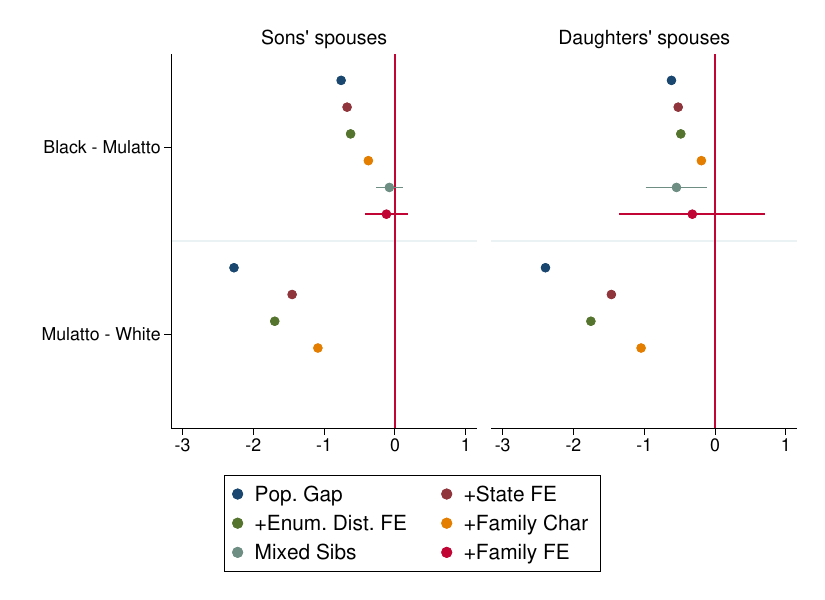}}
\par\end{centering}
{\small Note: Figure shows regression coefficients and 95\% confidence
intervals documenting gaps by perceived childhood race and gender
in an indicator for being married twenty years later (Panel A) and
in the years of education attained by one's }{\small\emph{spouse}}{\small{}
by 1940 (Panel B). See Section~\ref{subsec:adult_educ_married} for
detail, and see Tables~\ref{tab:married-plus20} and \ref{tab:Syears-educ-1940}
for corresponding regression tables.}{\small\par}
\end{figure}

\pagebreak{}
\begin{figure}[H]
\begin{centering}
\caption{\label{fig:income}Gaps in income and income scores, by perceived
child race and gender}
\par\end{centering}
\begin{centering}
\subfloat[\label{subfig:log-inc-1940}Log wage income in 1940]{
\centering{}\includegraphics[width=0.8\textwidth]{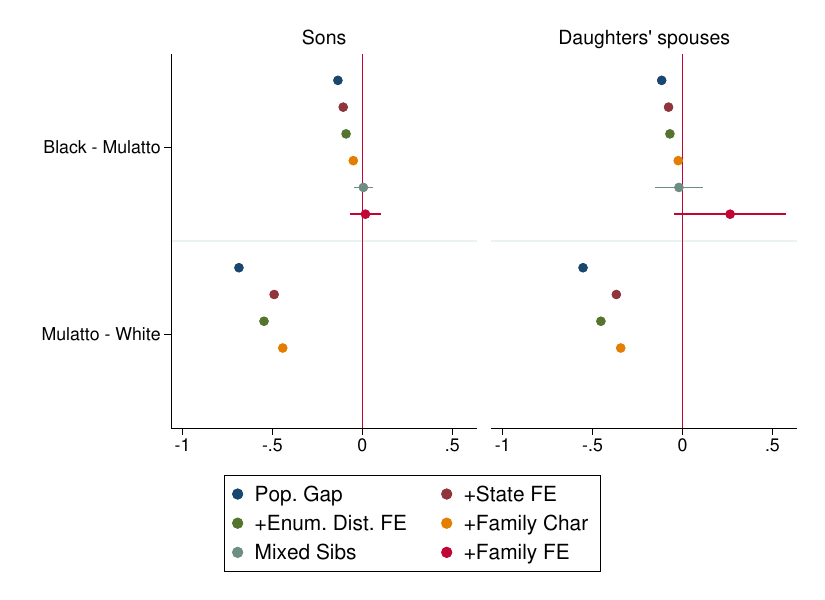}}
\par\end{centering}
\begin{centering}
\subfloat[\label{subfig:log-inc-score-plus20}Log wage income score in \emph{t+20}]{
\centering{}\includegraphics[width=0.8\textwidth]{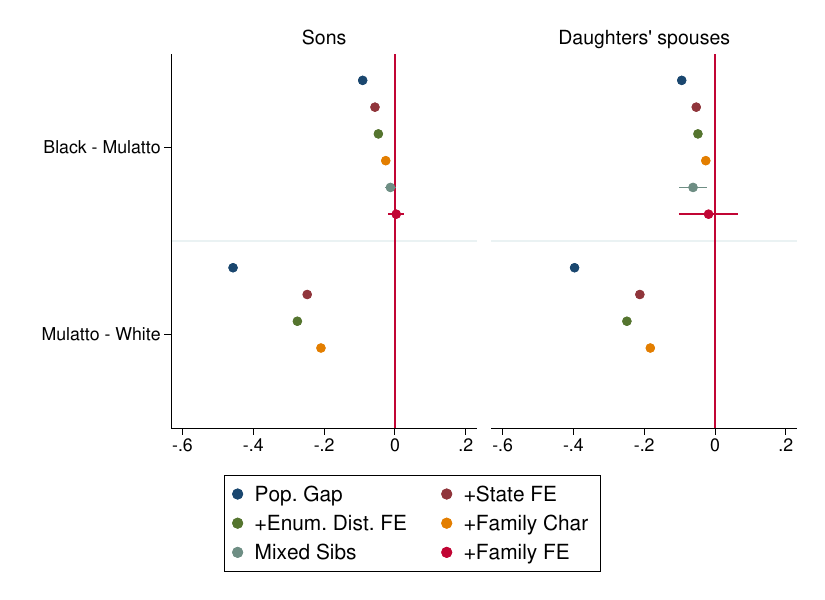}}
\par\end{centering}
{\small Note: Figure shows regression coefficients and 95\% confidence
intervals documenting gaps by perceived childhood race and gender
in actual log wage income in 1940 and in log wage income scores in
}{\small\emph{t+20}}{\small . See Section~\ref{subsec:adult_income}
for detail and Tables~\ref{tab:loginc-1940} and \ref{tab:logincscore-plus20}
for corresponding regression tables.}{\small\par}
\end{figure}

%% file: tables.tex
\begin{table}[H]
\caption{\label{tab:bm-couple-stats}Parents' perceived races in the population
and among differently-classified families}

\begin{centering}
\subfloat[\label{tab:bm-couples}Attributes of African American couples, by
perceived race]{
\centering{}\input{input/Tables/BM_couples.tex}}
\par\end{centering}
\begin{centering}
\subfloat[\label{tab:bms-parent-race}Parent perceived race frequencies among
families with differently-classified Black and Mulatto sons (BMS sample)]{
\centering{}\input{input/Tables/BMS_parent_race.tex}}
\par\end{centering}
\begin{centering}
\subfloat[\label{tab:bmd-parent-race}Parent perceived race frequencies among
families with differently-classified Black and Mulatto daughters (BMD
sample)]{
\centering{}\input{input/Tables/BMD_parent_race.tex}}
\par\end{centering}
{\small Note: Panel A shows differences in perceived race frequency
and characteristics among African-American couples in the general
population in years 1870, 1880, 1910, and 1920. African-American couples
are defined as two spouses who were both classified as Black/Mulatto.
Panel B shows parent perceived race frequencies among families containing
at least one Black-coded and at least one Mulatto-coded son (BMS sample)
both age 3--18, while Panel C shows the analogous panel for differently-classified
Black and Mulatto daughters (BMD sample). These panels again cover
the 1870, 1880, 1910, and 1920 census years.}{\small\par}
\end{table}

\pagebreak{}
\begin{table}[H]
\begin{centering}
\caption{\label{tab:attribs}Attributes of families with same vs.\ differently-classified
siblings}
\par\end{centering}
\begin{centering}
\subfloat[\label{tab:attribs-sons}Sons]{
\centering{}\input{input/Tables/fam_attributes_sons.tex}}
\par\end{centering}
\begin{centering}
\subfloat[\label{tab:attribs-daughters}Daughters]{
\centering{}\input{input/Tables/fam_attributes_daughters.tex}}
\par\end{centering}
{\small Note: Panel A shows mean characteristics for the sample of
sons in households with at least two sons age 3--18 all Black-coded
or all Mulatto-coded (columns 1--2) and for sons in BMS households
with at least one Black- and one Mulatto-coded son age 3--18 (column
3). Panel B shows analogous results for daughters.}{\small\par}
\end{table}

\pagebreak{}
\begin{sidewaystable}[H]
\begin{centering}
\caption{\label{tab:match-characteristics}Characteristics of linked vs.\ all
searched sons}
\par\end{centering}
\begin{centering}
\resizebox{\textwidth}{!}{\input{input/Tables/match_characteristics_plus20_bw_combined_byuautoabeexcon.tex}}
\par\end{centering}
{\small Note: Table shows differences in mean characteristics for our
initial and linked samples of sons. Column (1) reports counts and
mean statistics among our initial sample prior to linking, including
children from 1870, 1880, 1910, and 1920. Columns (2) and (4) denote
mean characteristics under our Census Tree links when linking to }{\small\emph{t+20}}{\small ,
without and with inverse probability link weights respectively (see
Section~\ref{subsec:Matching-across-censuses} for detail on linking
and weights). Columns (3) and (5) report differences between our initial
and linked means, without and with weights respectively. Columns (6)--(9)
repeats the above for the ABE Exact Conservative linking algorithm.
Standard errors are clustered by (enumeration district)$\times$(year).
See Appendix Table~\ref{tab:match-characteristics-daughters} for
an analogous table comparing the characteristics of linked vs.\ all
searched daughters.}{\small\par}
\end{sidewaystable}

\pagebreak{}
\begin{table}[H]
\begin{centering}
\caption{\label{tab:descriptive-means}Average characteristics, by perceived
child race and gender}
\subfloat[\label{tab:descriptive-means-sons}Sons]{
\centering{}\resizebox{!}{0.24\textheight}{\input{input/Tables/descriptive_means_sons_byuauto.tex}}}
\par\end{centering}
\begin{centering}
\subfloat[\label{tab:descriptive-means-daughters}Daughters]{
\centering{}\resizebox{!}{0.24\textheight}{\input{input/Tables/descriptive_means_daughters_byuauto.tex}}}
\par\end{centering}
{\small Note: Table shows counts and mean characteristics by perceived
child race and gender in 1870, 1880, 1910, and 1920. The bottom portion
of each panel includes adult outcomes identified by the Census Tree
links. Panels A and B analyze means among sons and daughters respectively.}{\small\par}
\end{table}

\pagebreak{}
\begin{table}[H]
\begin{centering}
\caption{\label{tab:inschool}Gaps in childhood school attendance, by perceived
child race and gender}
\par\end{centering}
\begin{centering}
\subfloat[\label{tab:inschool-sons}Sons]{
\centering{}\resizebox{0.9\textwidth}{!}{\input{input/Tables/regressions_inschool_combined_sons_black_mulatto2_wfamfe.tex}}}
\par\end{centering}
\begin{centering}
\subfloat[\label{tab:inschool-daughters}Daughters]{
\centering{}\resizebox{0.9\textwidth}{!}{\input{input/Tables/regressions_inschool_combined_daughters_black_mulatto2_wfamfe.tex}}}
\par\end{centering}
{\small Note: Table shows regressions measuring gaps in an indicator
for childhood school attendance, by perceived child race among children
age 5--18 in 1870, 1880, 1910, and 1920. Columns (1)--(4) follow
Equation~\ref{eq:naive-discrimination-estimation}, starting with
age and birth order fixed effects and adding controls for state fixed
effects, enumeration district fixed effects, and family characteristics
in columns (2)--(4), respectively. See Footnote~\ref{fn:family-characteristics}
for detail regarding the set of family characteristics included as
controls. Column (5) implements Equation~\ref{eq:within-BMS-estimation}
by further adding BMS\slash BMD race interaction terms and weights
such that our }{\small\emph{Black--Mulatto}}{\small{} coefficient measures
differences among differently-classified siblings with approximately
equal weighted distributions of family characteristics. Column (6)
adds family fixed effects to the specification in column (4). Panels
A and B analyze differences among sons and daughters respectively.
The bottom part of each panel presents observation counts and mean
outcome variables by perceived child race. All non-race controls are
interacted with cohort fixed effects. Standard errors are clustered
by (enumeration district)$\times$(year).}{\small\par}
\end{table}

\pagebreak{}
\begin{table}[H]
\begin{centering}
\caption{\label{tab:literacy}Gaps in childhood literacy, by perceived child
race and gender}
\par\end{centering}
\begin{centering}
\subfloat[\label{tab:literacy-sons}Sons]{
\centering{}\resizebox{0.9\textwidth}{!}{\input{input/Tables/regressions_literate_combined_sons_black_mulatto2_wfamfe.tex}}}
\par\end{centering}
\begin{centering}
\subfloat[\label{tab:literacy-daughters}Daughters]{
\centering{}\resizebox{0.9\textwidth}{!}{\input{input/Tables/regressions_literate_combined_daughters_black_mulatto2_wfamfe.tex}}}
\par\end{centering}
{\small Note: Table shows regressions measuring gaps in an indicator
for a child's ability to read and write, by perceived child race among
children age 10-18 in 1870, 1880, 1910, and 1920. Columns (1)--(4)
follow Equation~\ref{eq:naive-discrimination-estimation}, starting
with age and birth order fixed effects and adding controls for state
fixed effects, enumeration district fixed effects, and family characteristics
in columns (2)--(4) respectively. See Footnote~\ref{fn:family-characteristics}
for detail regarding the set of family characteristics included as
controls. Column (5) implements Equation~\ref{eq:within-BMS-estimation}
by further adding BMS\slash BMD race interaction terms and weights
such that our }{\small\emph{Black--Mulatto}}{\small{} coefficient measures
differences among differently-classified siblings with approximately
equal weighted distributions of family characteristics. Column (6)
adds family fixed effects to the specification in column (4). Panels
A and B analyze differences among sons and daughters respectively.
The bottom part of each panel presents observation counts and mean
outcome variables by perceived child race. All non-race controls are
interacted with cohort fixed effects. Standard errors are clustered
by (enumeration district)$\times$(year).}{\small\par}
\end{table}

\pagebreak{}
\begin{table}[H]
\begin{centering}
\caption{\label{tab:white-plus20}Gaps in classification as White adult in
\emph{t+20}, by perceived child race and gender}
\par\end{centering}
\begin{centering}
\subfloat[\label{tab:white-plus20-sons}Sons]{
\centering{}\resizebox{0.9\textwidth}{!}{\input{input/Tables/regressions_white_plus20_byuauto_combined_black_mulatto2_wgtd_wfamfe.tex}}}
\par\end{centering}
\begin{centering}
\subfloat[\label{tab:white-plus20-daughters}Daughters]{
\centering{}\resizebox{0.9\textwidth}{!}{\input{input/Tables/regressions_women_white_plus20_byuauto_combined_black_mulatto2_wgtd_wfamfe.tex}}}
\par\end{centering}
{\small Note: Table shows regressions measuring gaps in an indicator
for future classification as White two decades later, by perceived
child race among children age 3--18 in 1870, 1880, 1910, and 1920.
Columns (1)--(4) follow Equation~\ref{eq:naive-discrimination-estimation},
starting with age and birth order fixed effects and adding controls
for state fixed effects, enumeration district fixed effects, and family
characteristics in columns (2)--(4) respectively. See Footnote~\ref{fn:family-characteristics}
for detail regarding the set of family characteristics included as
controls. Column (5) implements Equation~\ref{eq:within-BMS-estimation}
by further adding BMS\slash BMD race interaction terms and weights
such that our }{\small\emph{Black--Mulatto}}{\small{} coefficient measures
differences among differently-classified siblings with approximately
equal weighted distributions of family characteristics. Panels A and
B analyze differences among sons and daughters respectively. The bottom
part of each panel presents observation counts and mean outcome variables
by perceived child race. All non-race controls are interacted with
cohort fixed effects. Observations are weighted by their inverse link
probabilities as described in Section~\ref{subsec:Matching-across-censuses}.
Standard errors are clustered by (enumeration district)$\times$(year).
Observations in which a child was identified as a race other than
White or Black as an adult are dropped. All controls come from the
child's census year, rather than from their linked adult self.}{\small\par}
\end{table}

\pagebreak{}
\begin{table}[H]
\begin{centering}
\caption{\label{tab:years-educ-1940}Gaps in years of educational attainment
by 1940, by perceived child race and gender}
\par\end{centering}
\begin{centering}
\subfloat[\label{tab:years-educ-1940-sons}Sons]{
\centering{}\resizebox{0.9\textwidth}{!}{\input{input/Tables/regressions_years_educ_1940_byuauto_combined_black_mulatto2_wgtd_wfamfe.tex}}}
\par\end{centering}
\begin{centering}
\subfloat[\label{tab:years-educ-1940-daughters}Daughters]{
\centering{}\resizebox{0.9\textwidth}{!}{\input{input/Tables/regressions_women_years_educ_1940_byuauto_combined_black_mulatto2_wgtd_wfamfe.tex}}}
\par\end{centering}
{\small Note: Table shows regressions measuring gaps in a child's years
of education by 1940, by perceived child race among children age 3--18
in 1910 and 1920. Columns (1)--(4) follow Equation~\ref{eq:naive-discrimination-estimation},
starting with age and birth order fixed effects and adding controls
for state fixed effects, enumeration district fixed effects, and family
characteristics in columns (2)--(4) respectively. See Footnote~\ref{fn:family-characteristics}
for detail regarding the set of family characteristics included as
controls. Column (5) implements Equation~\ref{eq:within-BMS-estimation}
by further adding BMS\slash BMD race interaction terms and weights
such that our }{\small\emph{Black--Mulatto}}{\small{} coefficient measures
differences among differently-classified siblings with approximately
equal weighted distributions of family characteristics. Panels A and
B analyze differences among sons and daughters respectively. The bottom
part of each panel presents observation counts and mean outcome variables
by perceived child race. All non-race controls are interacted with
cohort fixed effects. Observations are weighted by their inverse link
probabilities as described in Section~\ref{subsec:Matching-across-censuses}.
Standard errors are clustered by (enumeration district)$\times$(year).
All controls come from the child's census year, rather than from their
linked adult self.}{\small\par}
\end{table}

\pagebreak{}
\begin{table}[H]
\begin{centering}
\caption{\label{tab:married-plus20}Gaps in marriage rates in \emph{t+20},
by perceived child race and gender}
\par\end{centering}
\begin{centering}
\subfloat[\label{tab:married-plus20-sons}Sons]{
\centering{}\resizebox{0.9\textwidth}{!}{\input{input/Tables/regressions_married_plus20_byuauto_combined_black_mulatto2_wgtd_wfamfe.tex}}}
\par\end{centering}
\begin{centering}
\subfloat[\label{tab:married-plus20-daughters}Daughters]{
\centering{}\resizebox{0.9\textwidth}{!}{\input{input/Tables/regressions_women_married_plus20_byuauto_combined_black_mulatto2_wgtd_wfamfe.tex}}}
\par\end{centering}
{\small Note: Table shows regressions measuring gaps in an indicator
for being married two decades later (three decades for the 1870 cohort),
by perceived child race among children age 3--18 in 1870, 1880, 1910,
and 1920. Columns (1)--(4) follow Equation~\ref{eq:naive-discrimination-estimation},
starting with age and birth order fixed effects and adding controls
for state fixed effects, enumeration district fixed effects, and family
characteristics in columns (2)--(4) respectively. See Footnote~\ref{fn:family-characteristics}
for detail regarding the set of family characteristics included as
controls. Column (5) implements Equation~\ref{eq:within-BMS-estimation}
by further adding BMS\slash BMD race interaction terms and weights
such that our }{\small\emph{Black--Mulatto}}{\small{} coefficient measures
differences among differently-classified siblings with approximately
equal weighted distributions of family characteristics. Panels A and
B analyze differences among sons and daughters respectively. The bottom
part of each panel presents observation counts and mean outcome variables
by perceived child race. All non-race controls are interacted with
cohort fixed effects. Observations are weighted by their inverse link
probabilities as described in Section~\ref{subsec:Matching-across-censuses}.
Standard errors are clustered by (enumeration district)$\times$(year).
All controls come from the child's census year, rather than from their
linked adult self.}{\small\par}
\end{table}
\pagebreak{}
\begin{table}[H]
\begin{centering}
\caption{\label{tab:Syears-educ-1940}Gaps in spouses' years of educational
attainment by 1940, by perceived child race and gender}
\par\end{centering}
\begin{centering}
\subfloat[\label{tab:Syears-educ-1940-sons}Sons]{
\centering{}\resizebox{0.9\textwidth}{!}{\input{input/Tables/regressions_Syears_educ_1940_byuauto_combined_black_mulatto2_wgtd_wfamfe.tex}}}
\par\end{centering}
\begin{centering}
\subfloat[\label{tab:Syears-educ-1940-daughters}Daughters]{
\centering{}\resizebox{0.9\textwidth}{!}{\input{input/Tables/regressions_women_Syears_educ_1940_byuauto_combined_black_mulatto2_wgtd_wfamfe.tex}}}
\par\end{centering}
{\small Note: Table shows regressions measuring gaps in the years of
education a child's future spouse had attained by 1940, by perceived
child race among children age 3--18 in 1910 and 1920. Columns (1)--(4)
follow Equation~\ref{eq:naive-discrimination-estimation}, starting
with age and birth order fixed effects and adding controls for state
fixed effects, enumeration district fixed effects, and family characteristics
in columns (2)--(4) respectively. See Footnote~\ref{fn:family-characteristics}
for detail regarding the set of family characteristics included as
controls. Column (5) implements Equation~\ref{eq:within-BMS-estimation}
by further adding BMS\slash BMD race interaction terms and weights
such that our }{\small\emph{Black--Mulatto}}{\small{} coefficient measures
differences among differently-classified siblings with approximately
equal weighted distributions of family characteristics. Panels A and
B analyze differences among sons and daughters respectively. The bottom
part of each panel presents observation counts and mean outcome variables
by perceived child race. All non-race controls are interacted with
cohort fixed effects. Observations are weighted by their inverse link
probabilities as described in Section~\ref{subsec:Matching-across-censuses}.
Standard errors are clustered by (enumeration district)$\times$(year).
All controls come from the child's census year, rather than from their
linked adult self. Restricts to individuals married in 1940.}{\small\par}
\end{table}

\pagebreak{}
\begin{table}[H]
\begin{centering}
\caption{\label{tab:loginc-1940}Gaps in log wage income in 1940, by perceived
child race and gender}
\par\end{centering}
\begin{centering}
\subfloat[\label{tab:loginc-1940-sons}Sons]{
\centering{}\resizebox{0.9\textwidth}{!}{\input{input/Tables/regressions_log_inc_1940_byuauto_combined_black_mulatto2_wgtd_wfamfe.tex}}}
\par\end{centering}
\begin{centering}
\subfloat[\label{tab:loginc-1940-daughters-spouse}Daughters' spouses]{
\centering{}\resizebox{0.9\textwidth}{!}{\input{input/Tables/regressions_women_Slog_inc_1940_byuauto_combined_black_mulatto2_wgtd_wfamfe.tex}}}
\par\end{centering}
{\small Note: Table shows regressions measuring gaps in the log wage
income of a son or of a daughter's spouse as an adult in 1940, by
perceived child race among children age 3--18 in 1910 and 1920. Columns
(1)--(4) follow Equation~\ref{eq:naive-discrimination-estimation},
starting with age and birth order fixed effects and adding controls
for state fixed effects, enumeration district fixed effects, and family
characteristics in columns (2)--(4) respectively. See Footnote~\ref{fn:family-characteristics}
for detail regarding the set of family characteristics included as
controls. Column (5) implements Equation~\ref{eq:within-BMS-estimation}
by further adding BMS\slash BMD race interaction terms and weights
such that our }{\small\emph{Black--Mulatto}}{\small{} coefficient measures
differences among differently-classified siblings with approximately
equal weighted distributions of family characteristics. Panels A and
B analyze differences among sons and among daughters' spouses respectively.
The bottom part of each panel presents observation counts and mean
outcome variables by perceived child race. All non-race controls are
interacted with cohort fixed effects. Observations are weighted by
their inverse link probabilities as described in Section~\ref{subsec:Matching-across-censuses}.
Standard errors are clustered by (enumeration district)$\times$(year).
All controls come from the child's census year, rather than from their
linked adult self. Panel B restricts to daughters married in 1940.}{\small\par}
\end{table}

\pagebreak{}
\begin{table}[H]
\begin{centering}
\caption{\label{tab:logincscore-plus20}Gaps in predicted log wage in \emph{t+20},
by perceived child race and gender}
\par\end{centering}
\begin{centering}
\subfloat[\label{tab:logincscore-plus20-sons}Sons]{
\centering{}\resizebox{0.9\textwidth}{!}{\input{input/Tables/regressions_scorewgCWadj_plus20_byuauto_combined_black_mulatto2_wgtd_wfamfe.tex}}}
\par\end{centering}
\begin{centering}
\subfloat[\label{tab:logincscore-plus20-daughters-spouse}Daughters' spouses]{
\centering{}\resizebox{0.9\textwidth}{!}{\input{input/Tables/regressions_women_SscorewgCWadj_plus20_byuauto_combined_black_mulatto2_wgtd_wfamfe.tex}}}
\par\end{centering}
{\small Note: Table shows regressions measuring gaps in the predicted
log wage income score of a son or a daughter's spouse, by perceived
child race among children age 3--18 in 1870, 1880, 1910, and 1920.
Columns (1)--(4) follow Equation~\ref{eq:naive-discrimination-estimation},
starting with age and birth order fixed effects and adding controls
for state fixed effects, enumeration district fixed effects, and family
characteristics in columns (2)--(4) respectively. See Footnote~\ref{fn:family-characteristics}
for detail regarding the set of family characteristics included as
controls. Column (5) implements Equation~\ref{eq:within-BMS-estimation}
by further adding BMS\slash BMD race interaction terms and weights
such that our }{\small\emph{Black--Mulatto}}{\small{} coefficient measures
differences among differently-classified siblings with approximately
equal weighted distributions of family characteristics. Panels A and
B analyze differences among sons and daughters' spouses respectively.
The bottom part of each panel presents observation counts and mean
outcome variables by perceived child race. All non-race controls are
interacted with cohort fixed effects. Observations are weighted by
their inverse link probabilities as described in Section~\ref{subsec:Matching-across-censuses}.
Standard errors are clustered by (enumeration district)$\times$(year).
Predictions made based on adult occupation, geography, and age (see
Section~\ref{subsec:income-prediction}). All controls come from
the child's census year, rather than from their linked adult self.
Panel B restricts to married daughters.}{\small\par}
\end{table}

%% file: input/Tables/BM_couples.tex
\begin{tabular}{ll*{4}{c}} \toprule \multicolumn{2}{c}{Parents' Race} & & & & \\ \cmidrule(lr){1-2} 
Husband & Wife & \% of Couples & N Children & \% Black & \% Female \\ \midrule
Black & Black & 78.6\% & 2.22 & 98.8\% & 49.7\% \\
Black & Mulatto & 6.6\% & 2.23 & 33.8\% & 49.6\% \\
Mulatto & Black & 3.5\% & 2.37 & 53.3\% & 49.2\% \\
Mulatto & Mulatto & 11.3\% & 2.17 & 3.0\% & 49.5\% \\
\bottomrule \end{tabular}

%% file: input/Tables/BMS_parent_race.tex
\begin{tabular}{l*{5}{c}} \toprule & \multicolumn{4}{c}{Mother} & \\ \cmidrule(lr){2-5} 
Father & Black & Mulatto & White & Other & Total \\ \midrule
Black & 29.2\% & 38.2\% & 0.2\% & - & 67.6\% \\
Mulatto & 20.9\% & 10.8\% & 0.0\% & - & 31.7\% \\
White & 0.3\% & 0.1\% & 0.3\% & - & 0.7\% \\
Other & 0.0\% & - & - & - & 0.0\% \\ \midrule
Total & 50.4\% & 49.1\% & 0.5\% & - & 100.0\% \\
\bottomrule \end{tabular}

%% file: input/Tables/BMD_parent_race.tex
\begin{tabular}{l*{5}{c}} \toprule & \multicolumn{4}{c}{Mother} & \\ \cmidrule(lr){2-5} 
Father & Black & Mulatto & White & Other & Total \\ \midrule
Black & 35.3\% & 35.7\% & 0.2\% & - & 71.2\% \\
Mulatto & 19.1\% & 8.9\% & 0.0\% & - & 28.1\% \\
White & 0.3\% & 0.1\% & 0.3\% & - & 0.7\% \\
Other & - & - & - & - & - \\ \midrule
Total & 54.7\% & 44.8\% & 0.6\% & - & 100.0\% \\
\bottomrule \end{tabular}

%% file: input/Tables/fam_attributes_sons.tex
\begin{tabular}{l*{4}{c}} \toprule Sample: & \multicolumn{2}{c}{Single-Classification Families} & & BMS \\ Household classification: & Black & Mulatto & & Both \\ \cmidrule(lr){2-3} \cmidrule(lr){5-5} 
& (1) & (2) & & (3) \\ \midrule
Sons in family & 3.35 & 3.34 & $<$ & 3.60 \\
Age & 9.52 & 9.51 & $<$ & 9.72 \\
Able to write & 0.54 & 0.70 & $>$ & 0.49 \\
Head is literate & 0.45 & 0.60 & $>$ & 0.39 \\ Head's occupation: & & & & \\
- Farm laborer & 0.19 & 0.11 & $<$ & 0.20 \\
- Farmer & 0.55 & 0.53 & $\ni$ & 0.54 \\
- Laborer (not farm) & 0.15 & 0.14 & $>$ & 0.14 \\
- In home services & 0.00 & 0.00 & $<$ & 0.00 \\
County urban & 0.15 & 0.21 & $>$ & 0.13 \\
County Afr. Americans & 0.47 & 0.40 & $\ni$ & 0.46 \\
Observations & 2,252,690 & 438,543 & & 50,333 \\
\bottomrule \end{tabular}

%% file: input/Tables/fam_attributes_daughters.tex
\begin{tabular}{l*{4}{c}} \toprule Sample: & \multicolumn{2}{c}{Single-Classification Families} & & BMD \\ Household classification: & Black & Mulatto & & Both \\ \cmidrule(lr){2-3} \cmidrule(lr){5-5} 
& (1) & (2) & & (3) \\ \midrule
Daughters in family & 3.32 & 3.32 & $<$ & 3.57 \\
Age & 9.47 & 9.47 & $<$ & 9.77 \\
Able to write & 0.60 & 0.76 & $>$ & 0.55 \\
Head is literate & 0.45 & 0.61 & $>$ & 0.39 \\ Head's occupation: & & & & \\
- Farm laborer & 0.19 & 0.11 & $<$ & 0.20 \\
- Farmer & 0.54 & 0.52 & $>$ & 0.50 \\
- Laborer (not farm) & 0.16 & 0.15 & $<$ & 0.16 \\
- In home services & 0.00 & 0.00 & $<$ & 0.00 \\
County urban & 0.16 & 0.21 & $>$ & 0.13 \\
County Afr. Americans & 0.47 & 0.40 & $\ni$ & 0.46 \\
Observations & 2,212,548 & 449,556 & & 53,762 \\
\bottomrule \end{tabular}

%% file: input/Tables/match_characteristics_plus20_bw_combined_byuautoabeexcon.tex
\begin{tabular}{l|c|cccc|cccc} \toprule  & All obs & \multicolumn{4}{|c}{Census Tree} & \multicolumn{4}{|c}{ABE Exact Conservative} \\ \midrule Observations (count) & \multicolumn{1}{|c}{49,609,678} & \multicolumn{4}{|c}{25,842,326} & \multicolumn{4}{|c}{8,558,226}  \\ Black+Mulatto in county & \multicolumn{1}{|c}{6,240,968} & \multicolumn{4}{|c}{1,569,064} & \multicolumn{4}{|c}{426,190}  \\ Black+Mulatto in family & \multicolumn{1}{|c}{50,333} & \multicolumn{4}{|c}{6,104} & \multicolumn{4}{|c}{652} \\ \midrule   &  & \multicolumn{2}{|c}{Unweighted} & \multicolumn{2}{c}{Weighted} & \multicolumn{2}{|c}{Unweighted} & \multicolumn{2}{c}{Weighted} \\  \cmidrule(lr){3-4} \cmidrule(lr){5-6} \cmidrule(lr){7-8} \cmidrule(lr){9-10} Child characteristics & Mean & Mean & Difference & Mean & Difference & Mean & Difference & Mean & Difference \\ \midrule  
& (1) & (2) & (3) & (4) & (5) & (6) & (7) & (8) & (9) \\ \midrule
Black & .105 & .0489 & -.056*** & .0994 & -.00542*** & .0403 & -.0646*** & .101 & -.00348*** \\
& & & (.000365) & & (.000845) & & (.000325) & & (.000922) \\
Mulatto & .0212 & .0122 & -.00904*** & .023 & .00181*** & .0105 & -.0107*** & .0236 & .00243*** \\
& & & (.000105) & & (.000213) & & (.0000942) & & (.000226) \\
Age & 10.1 & 10.1 & -.0438*** & 10.1 & -.0102*** & 10.1 & -.0212*** & 10.1 & -.00417 \\
& & & (.00177) & & (.00218) & & (.00224) & & (.00275) \\
Able to write & .888 & .93 & .0421*** & .896 & .00801*** & .947 & .0588*** & .899 & .0108*** \\
& & & (.000565) & & (.000941) & & (.000487) & & (.001) \\
Sons in household & 2.18 & 2.25 & .0766*** & 2.18 & .00839*** & 2.21 & .0392*** & 2.18 & .00877*** \\
& & & (.00111) & & (.00129) & & (.00116) & & (.00139) \\
Head is literate & .852 & .901 & .0489*** & .854 & .0021** & .922 & .0703*** & .852 & .000236 \\
& & & (.000468) & & (.000837) & & (.000422) & & (.000925) \\ Head's occupation: & & & & & & & & & \\
- Farm laborer & .0481 & .0344 & -.0138*** & .0467 & -.0014*** & .0317 & -.0164*** & .0476 & -.000483 \\
& & & (.000248) & & (.000491) & & (.000235) & & (.000561) \\
- Farmer & .387 & .406 & .0194*** & .387 & .000317 & .392 & .0053*** & .387 & .000234 \\
& & & (.00139) & & (.00179) & & (.00137) & & (.00181) \\
- Laborer (not farm) & .101 & .0873 & -.0137*** & .101 & .0000115 & .0798 & -.0213*** & .101 & .000106 \\
& & & (.00031) & & (.000438) & & (.000277) & & (.000442) \\
County urban & .37 & .379 & .00954*** & .365 & -.00464* & .395 & .0251*** & .364 & -.00576** \\
& & & (.00177) & & (.00257) & & (.00176) & & (.00263) \\
Census Year & 1902 & 1904 & 2.64*** & 1902 & .168 & 1906 & 4.02*** & 1902 & .0512 \\
& & & (.135) & & (.183) & & (.137) & & (.186) \\
\bottomrule \end{tabular}

%% file: input/Tables/descriptive_means_sons_byuauto.tex
\begin{tabular}{l*{3}{c}} \toprule Perceived Race: & White & Black & Mulatto \\ 
& (1) & (2) & (3) \\ \midrule
Observations & 43,355,734 & 5,202,057 & 1,051,887 \\ \midrule
Age & 10.14 & 10.02 & 9.95 \\
In school & 0.67 & 0.40 & 0.50 \\
Literate & 0.93 & 0.54 & 0.68 \\
In South & 0.22 & 0.88 & 0.82 \\
Father present & 0.87 & 0.72 & 0.73 \\
Mother present & 0.91 & 0.83 & 0.85 \\
Sons in family & 2.19 & 2.07 & 2.07 \\
Head is literate & 0.91 & 0.45 & 0.58 \\ Head's occupation: & & & \\
- Farm laborer & 0.03 & 0.18 & 0.11 \\
- Farmer & 0.38 & 0.47 & 0.44 \\
- Laborer (not farm) & 0.09 & 0.15 & 0.14 \\
- In home services & 0.02 & 0.02 & 0.03 \\
County urban & 0.40 & 0.18 & 0.23 \\
County Afr. Americans & 0.08 & 0.46 & 0.40 \\ \midrule
Is White in t+20 & 0.99 & 0.12 & 0.15 \\
Years of education in 1940 & 9.59 & 5.98 & 6.81 \\
Married in t+20 & 0.68 & 0.71 & 0.68 \\
Spouse years of educ. in 1940 & 9.80 & 6.81 & 7.55 \\
Is employed in 1940 & 0.94 & 0.92 & 0.92 \\
Log wage income in 1940 & 6.85 & 6.03 & 6.19 \\
Log wage income score in t+20 & 6.75 & 6.19 & 6.28 \\
\bottomrule \end{tabular}

%% file: input/Tables/descriptive_means_daughters_byuauto.tex
\begin{tabular}{l*{3}{c}} \toprule Perceived Race: & White & Black & Mulatto \\ 
& (1) & (2) & (3) \\ \midrule
Observations & 42,688,098 & 5,213,710 & 1,132,525 \\ \midrule
Age & 10.19 & 10.09 & 10.18 \\
In school & 0.67 & 0.43 & 0.53 \\
Literate & 0.95 & 0.61 & 0.74 \\
In South & 0.22 & 0.88 & 0.82 \\
Father present & 0.85 & 0.71 & 0.70 \\
Mother present & 0.90 & 0.82 & 0.82 \\
Daughters in family & 2.13 & 2.04 & 2.02 \\
Head is literate & 0.91 & 0.46 & 0.59 \\ Head's occupation: & & & \\
- Farm laborer & 0.03 & 0.18 & 0.11 \\
- Farmer & 0.36 & 0.45 & 0.41 \\
- Laborer (not farm) & 0.10 & 0.15 & 0.15 \\
- In home services & 0.02 & 0.02 & 0.03 \\
County urban & 0.40 & 0.19 & 0.24 \\
County Afr. Americans & 0.08 & 0.46 & 0.40 \\ \midrule
Is White in t+20 & 0.99 & 0.17 & 0.20 \\
Years of education in 1940 & 9.91 & 7.27 & 7.96 \\
Married in t+20 & 0.67 & 0.48 & 0.51 \\
Spouse years of educ. in 1940 & 9.15 & 6.12 & 6.71 \\
Spouse is employed in 1940 & 0.96 & 0.95 & 0.95 \\
Spouse log wg. inc. in 1940 & 6.95 & 6.28 & 6.41 \\
Spouse log wg. inc. score in t+20 & 6.83 & 6.35 & 6.45 \\
\bottomrule \end{tabular}

%% file: input/Tables/regressions_inschool_combined_sons_black_mulatto2_wfamfe.tex
{
\def\sym#1{\ifmmode^{#1}\else\(^{#1}\)\fi}
\begin{tabular}{l*{6}{c}}
\toprule
                &\multicolumn{1}{c}{(1)}         &\multicolumn{1}{c}{(2)}         &\multicolumn{1}{c}{(3)}         &\multicolumn{1}{c}{(4)}         &\multicolumn{1}{c}{(5)}         &\multicolumn{1}{c}{(6)}         \\
\midrule
Black - Mulatto &   -0.078\sym{***}&   -0.056\sym{***}&   -0.050\sym{***}&   -0.031\sym{***}&   -0.006\sym{*}  &   -0.005         \\
                &  (0.002)         &  (0.001)         &  (0.001)         &  (0.001)         &  (0.004)         &  (0.004)         \\
Mulatto - White &   -0.186\sym{***}&   -0.102\sym{***}&   -0.107\sym{***}&   -0.054\sym{***}&                  &                  \\
                &  (0.002)         &  (0.001)         &  (0.001)         &  (0.001)         &                  &                  \\
(Year)X(Age+Birth Order) FE &      Yes         &      Yes         &      Yes         &      Yes         &      Yes         &      Yes         \\
(Year)X(State) FE &       No         &      Yes         &       No         &       No         &       No         &       No         \\
(Year)X(Enum. Dist.) FE &       No         &       No         &      Yes         &      Yes         &      Yes         &       No         \\
(Year)X(Fam. Char.) &       No         &       No         &       No         &      Yes         &      Yes         &       No         \\
BMS Interactions &       No         &       No         &       No         &       No         &      Yes         &       No         \\
(Year)X(Fam.) FE &       No         &       No         &       No         &       No         &       No         &      Yes         \\
\midrule
Observations    &42,583,862         &42,583,862         &42,583,669         &42,583,669         &42,583,669         &20,992,325         \\
R-Squared       &    0.299         &    0.333         &    0.393         &    0.404         &    0.404         &    0.734         \\
Count Black     &4,462,974         &4,462,974         &4,462,971         &4,462,971         &4,462,971         &1,813,693         \\
Count Mulatto   &  897,591         &  897,591         &  897,590         &  897,590         &  897,590         &  367,455         \\
Count White     &37,223,297         &37,223,297         &37,223,108         &37,223,108         &37,223,108         &18,811,177         \\
Mean[Outcome $ | $ Black]&    0.399         &    0.399         &    0.399         &    0.399         &    0.399         &    0.432         \\
Mean[Outcome $ | $ Mulatto]&    0.498         &    0.498         &    0.498         &    0.498         &    0.498         &    0.545         \\
Mean[Outcome $ | $ White]&    0.673         &    0.673         &    0.673         &    0.673         &    0.673         &    0.710         \\
Count Mixed Siblings&   43,221         &   43,221         &   43,221         &   43,221         &   43,221         &   38,647         \\
\bottomrule
\end{tabular}
}

%% file: input/Tables/regressions_inschool_combined_daughters_black_mulatto2_wfamfe.tex
{
\def\sym#1{\ifmmode^{#1}\else\(^{#1}\)\fi}
\begin{tabular}{l*{6}{c}}
\toprule
                &\multicolumn{1}{c}{(1)}         &\multicolumn{1}{c}{(2)}         &\multicolumn{1}{c}{(3)}         &\multicolumn{1}{c}{(4)}         &\multicolumn{1}{c}{(5)}         &\multicolumn{1}{c}{(6)}         \\
\midrule
Black - Mulatto &   -0.080\sym{***}&   -0.059\sym{***}&   -0.051\sym{***}&   -0.034\sym{***}&   -0.022\sym{***}&   -0.020\sym{***}\\
                &  (0.002)         &  (0.001)         &  (0.001)         &  (0.001)         &  (0.003)         &  (0.003)         \\
Mulatto - White &   -0.149\sym{***}&   -0.072\sym{***}&   -0.085\sym{***}&   -0.030\sym{***}&                  &                  \\
                &  (0.002)         &  (0.001)         &  (0.001)         &  (0.001)         &                  &                  \\
(Year)X(Age+Birth Order) FE &      Yes         &      Yes         &      Yes         &      Yes         &      Yes         &      Yes         \\
(Year)X(State) FE &       No         &      Yes         &       No         &       No         &       No         &       No         \\
(Year)X(Enum. Dist.) FE &       No         &       No         &      Yes         &      Yes         &      Yes         &       No         \\
(Year)X(Fam. Char.) &       No         &       No         &       No         &      Yes         &      Yes         &       No         \\
BMD Interactions &       No         &       No         &       No         &       No         &      Yes         &       No         \\
(Year)X(Fam.) FE &       No         &       No         &       No         &       No         &       No         &      Yes         \\
\midrule
Observations    &42,154,560         &42,154,560         &42,154,322         &42,154,322         &42,154,322         &19,997,967         \\
R-Squared       &    0.303         &    0.334         &    0.392         &    0.405         &    0.405         &    0.733         \\
Count Black     &4,476,458         &4,476,458         &4,476,443         &4,476,443         &4,476,443         &1,773,183         \\
Count Mulatto   &  972,931         &  972,931         &  972,930         &  972,930         &  972,930         &  376,913         \\
Count White     &36,705,171         &36,705,171         &36,704,949         &36,704,949         &36,704,949         &17,847,871         \\
Mean[Outcome $ | $ Black]&    0.429         &    0.429         &    0.429         &    0.429         &    0.429         &    0.467         \\
Mean[Outcome $ | $ Mulatto]&    0.526         &    0.526         &    0.526         &    0.526         &    0.526         &    0.584         \\
Mean[Outcome $ | $ White]&    0.671         &    0.671         &    0.671         &    0.671         &    0.671         &    0.718         \\
Count Mixed Siblings&   46,214         &   46,214         &   46,214         &   46,214         &   46,214         &   41,239         \\
\bottomrule
\multicolumn{7}{l}{\footnotesize Standard errors in parentheses}\\
\multicolumn{7}{l}{\footnotesize \sym{*} \(p<0.10\), \sym{**} \(p<0.05\), \sym{***} \(p<0.01\)}\\
\end{tabular}
}

%% file: input/Tables/regressions_literate_combined_sons_black_mulatto2_wfamfe.tex
{
\def\sym#1{\ifmmode^{#1}\else\(^{#1}\)\fi}
\begin{tabular}{l*{6}{c}}
\toprule
                &\multicolumn{1}{c}{(1)}         &\multicolumn{1}{c}{(2)}         &\multicolumn{1}{c}{(3)}         &\multicolumn{1}{c}{(4)}         &\multicolumn{1}{c}{(5)}         &\multicolumn{1}{c}{(6)}         \\
\midrule
Black - Mulatto &   -0.126\sym{***}&   -0.102\sym{***}&   -0.095\sym{***}&   -0.064\sym{***}&   -0.003         &   -0.009\sym{*}  \\
                &  (0.002)         &  (0.002)         &  (0.001)         &  (0.001)         &  (0.005)         &  (0.005)         \\
Mulatto - White &   -0.257\sym{***}&   -0.162\sym{***}&   -0.184\sym{***}&   -0.097\sym{***}&                  &                  \\
                &  (0.002)         &  (0.002)         &  (0.002)         &  (0.001)         &                  &                  \\
(Year)X(Age+Birth Order) FE &      Yes         &      Yes         &      Yes         &      Yes         &      Yes         &      Yes         \\
(Year)X(State) FE &       No         &      Yes         &       No         &       No         &       No         &       No         \\
(Year)X(Enum. Dist.) FE &       No         &       No         &      Yes         &      Yes         &      Yes         &       No         \\
(Year)X(Fam. Char.) &       No         &       No         &       No         &      Yes         &      Yes         &       No         \\
BMS Interactions &       No         &       No         &       No         &       No         &      Yes         &       No         \\
(Year)X(Fam.) FE &       No         &       No         &       No         &       No         &       No         &      Yes         \\
\midrule
Observations    &26,230,967         &26,230,967         &26,230,691         &26,230,691         &26,230,691         &9,959,614         \\
R-Squared       &    0.210         &    0.298         &    0.391         &    0.449         &    0.449         &    0.835         \\
Count Black     &2,715,065         &2,715,065         &2,715,059         &2,715,059         &2,715,059         &  819,673         \\
Count Mulatto   &  540,816         &  540,816         &  540,815         &  540,815         &  540,815         &  166,382         \\
Count White     &22,975,086         &22,975,086         &22,974,817         &22,974,817         &22,974,817         &8,973,559         \\
Mean[Outcome $ | $ Black]&    0.543         &    0.543         &    0.543         &    0.543         &    0.543         &    0.554         \\
Mean[Outcome $ | $ Mulatto]&    0.678         &    0.678         &    0.678         &    0.678         &    0.678         &    0.700         \\
Mean[Outcome $ | $ White]&    0.934         &    0.934         &    0.934         &    0.934         &    0.934         &    0.934         \\
Count Mixed Siblings&   25,015         &   25,015         &   25,015         &   25,015         &   25,015         &   15,434         \\
\bottomrule
\end{tabular}
}

%% file: input/Tables/regressions_literate_combined_daughters_black_mulatto2_wfamfe.tex
{
\def\sym#1{\ifmmode^{#1}\else\(^{#1}\)\fi}
\begin{tabular}{l*{6}{c}}
\toprule
                &\multicolumn{1}{c}{(1)}         &\multicolumn{1}{c}{(2)}         &\multicolumn{1}{c}{(3)}         &\multicolumn{1}{c}{(4)}         &\multicolumn{1}{c}{(5)}         &\multicolumn{1}{c}{(6)}         \\
\midrule
Black - Mulatto &   -0.126\sym{***}&   -0.103\sym{***}&   -0.094\sym{***}&   -0.067\sym{***}&   -0.024\sym{***}&   -0.029\sym{***}\\
                &  (0.002)         &  (0.002)         &  (0.001)         &  (0.001)         &  (0.004)         &  (0.005)         \\
Mulatto - White &   -0.207\sym{***}&   -0.132\sym{***}&   -0.161\sym{***}&   -0.080\sym{***}&                  &                  \\
                &  (0.002)         &  (0.002)         &  (0.002)         &  (0.001)         &                  &                  \\
(Year)X(Age+Birth Order) FE &      Yes         &      Yes         &      Yes         &      Yes         &      Yes         &      Yes         \\
(Year)X(State) FE &       No         &      Yes         &       No         &       No         &       No         &       No         \\
(Year)X(Enum. Dist.) FE &       No         &       No         &      Yes         &      Yes         &      Yes         &       No         \\
(Year)X(Fam. Char.) &       No         &       No         &       No         &      Yes         &      Yes         &       No         \\
BMD Interactions &       No         &       No         &       No         &       No         &      Yes         &       No         \\
(Year)X(Fam.) FE &       No         &       No         &       No         &       No         &       No         &      Yes         \\
\midrule
Observations    &26,103,668         &26,103,668         &26,103,350         &26,103,350         &26,103,350         &9,283,820         \\
R-Squared       &    0.200         &    0.294         &    0.385         &    0.447         &    0.447         &    0.841         \\
Count Black     &2,735,431         &2,735,431         &2,735,413         &2,735,413         &2,735,413         &  784,972         \\
Count Mulatto   &  600,579         &  600,579         &  600,578         &  600,578         &  600,578         &  168,842         \\
Count White     &22,767,658         &22,767,658         &22,767,359         &22,767,359         &22,767,359         &8,330,006         \\
Mean[Outcome $ | $ Black]&    0.606         &    0.606         &    0.606         &    0.606         &    0.606         &    0.620         \\
Mean[Outcome $ | $ Mulatto]&    0.742         &    0.742         &    0.742         &    0.742         &    0.742         &    0.767         \\
Mean[Outcome $ | $ White]&    0.946         &    0.946         &    0.946         &    0.946         &    0.946         &    0.948         \\
Count Mixed Siblings&   27,033         &   27,033         &   27,033         &   27,033         &   27,033         &   16,378         \\
\bottomrule
\multicolumn{7}{l}{\footnotesize Standard errors in parentheses}\\
\multicolumn{7}{l}{\footnotesize \sym{*} \(p<0.10\), \sym{**} \(p<0.05\), \sym{***} \(p<0.01\)}\\
\end{tabular}
}

%% file: input/Tables/regressions_white_plus20_byuauto_combined_black_mulatto2_wgtd_wfamfe.tex
{
\def\sym#1{\ifmmode^{#1}\else\(^{#1}\)\fi}
\begin{tabular}{l*{6}{c}}
\toprule
                &\multicolumn{1}{c}{(1)}         &\multicolumn{1}{c}{(2)}         &\multicolumn{1}{c}{(3)}         &\multicolumn{1}{c}{(4)}         &\multicolumn{1}{c}{(5)}         &\multicolumn{1}{c}{(6)}         \\
\midrule
Black - Mulatto &   -0.028\sym{***}&   -0.024\sym{***}&   -0.023\sym{***}&   -0.023\sym{***}&   -0.016\sym{***}&   -0.010         \\
                &  (0.001)         &  (0.001)         &  (0.001)         &  (0.001)         &  (0.006)         &  (0.008)         \\
Mulatto - White &   -0.840\sym{***}&   -0.816\sym{***}&   -0.804\sym{***}&   -0.802\sym{***}&                  &                  \\
                &  (0.001)         &  (0.001)         &  (0.001)         &  (0.001)         &                  &                  \\
(Year)X(Age+Birth Order) FE &      Yes         &      Yes         &      Yes         &      Yes         &      Yes         &      Yes         \\
(Year)X(State) FE &       No         &      Yes         &       No         &       No         &       No         &       No         \\
(Year)X(Enum. Dist.) FE &       No         &       No         &      Yes         &      Yes         &      Yes         &       No         \\
(Year)X(Fam. Char.) &       No         &       No         &       No         &      Yes         &      Yes         &       No         \\
BMS Interactions &       No         &       No         &       No         &       No         &      Yes         &       No         \\
(Year)X(Fam.) FE &       No         &       No         &       No         &       No         &       No         &      Yes         \\
\midrule
Observations    &25,832,887         &25,832,887         &25,832,499         &25,832,499         &25,832,499         &11,492,652         \\
R-Squared       &    0.795         &    0.797         &    0.800         &    0.801         &    0.801         &    0.929         \\
Count Black     &1,263,393         &1,263,393         &1,263,381         &1,263,381         &1,263,381         &  359,787         \\
Count Mulatto   &  313,482         &  313,482         &  313,477         &  313,477         &  313,477         &   99,593         \\
Count White     &24,256,012         &24,256,012         &24,255,641         &24,255,641         &24,255,641         &11,033,272         \\
Mean[Outcome $ | $ Black]&    0.125         &    0.125         &    0.125         &    0.125         &    0.125         &    0.082         \\
Mean[Outcome $ | $ Mulatto]&    0.153         &    0.153         &    0.153         &    0.153         &    0.153         &    0.126         \\
Mean[Outcome $ | $ White]&    0.992         &    0.992         &    0.992         &    0.992         &    0.992         &    0.994         \\
Count Mixed Siblings&   13,969         &   13,969         &   13,969         &   13,969         &   13,969         &    6,098         \\
\bottomrule
\end{tabular}
}

%% file: input/Tables/regressions_women_white_plus20_byuauto_combined_black_mulatto2_wgtd_wfamfe.tex
{
\def\sym#1{\ifmmode^{#1}\else\(^{#1}\)\fi}
\begin{tabular}{l*{6}{c}}
\toprule
                &\multicolumn{1}{c}{(1)}         &\multicolumn{1}{c}{(2)}         &\multicolumn{1}{c}{(3)}         &\multicolumn{1}{c}{(4)}         &\multicolumn{1}{c}{(5)}         &\multicolumn{1}{c}{(6)}         \\
\midrule
Black - Mulatto &   -0.030\sym{***}&   -0.026\sym{***}&   -0.026\sym{***}&   -0.025\sym{***}&   -0.023\sym{**} &   -0.022         \\
                &  (0.002)         &  (0.002)         &  (0.001)         &  (0.001)         &  (0.010)         &  (0.016)         \\
Mulatto - White &   -0.791\sym{***}&   -0.768\sym{***}&   -0.760\sym{***}&   -0.757\sym{***}&                  &                  \\
                &  (0.002)         &  (0.002)         &  (0.002)         &  (0.002)         &                  &                  \\
(Year)X(Age+Birth Order) FE &      Yes         &      Yes         &      Yes         &      Yes         &      Yes         &      Yes         \\
(Year)X(State) FE &       No         &      Yes         &       No         &       No         &       No         &       No         \\
(Year)X(Enum. Dist.) FE &       No         &       No         &      Yes         &      Yes         &      Yes         &       No         \\
(Year)X(Fam. Char.) &       No         &       No         &       No         &      Yes         &      Yes         &       No         \\
BMD Interactions &       No         &       No         &       No         &       No         &      Yes         &       No         \\
(Year)X(Fam.) FE &       No         &       No         &       No         &       No         &       No         &      Yes         \\
\midrule
Observations    &17,034,764         &17,034,764         &17,034,221         &17,034,220         &17,034,220         &6,275,046         \\
R-Squared       &    0.743         &    0.745         &    0.754         &    0.754         &    0.754         &    0.918         \\
Count Black     &  537,871         &  537,871         &  537,850         &  537,850         &  537,850         &   90,869         \\
Count Mulatto   &  152,555         &  152,555         &  152,549         &  152,549         &  152,549         &   30,733         \\
Count White     &16,344,338         &16,344,338         &16,343,822         &16,343,821         &16,343,821         &6,153,444         \\
Mean[Outcome $ | $ Black]&    0.171         &    0.171         &    0.171         &    0.171         &    0.172         &    0.110         \\
Mean[Outcome $ | $ Mulatto]&    0.201         &    0.201         &    0.201         &    0.201         &    0.201         &    0.181         \\
Mean[Outcome $ | $ White]&    0.993         &    0.993         &    0.993         &    0.993         &    0.993         &    0.995         \\
Count Mixed Siblings&    5,779         &    5,779         &    5,779         &    5,779         &    5,779         &    1,440         \\
\bottomrule
\multicolumn{7}{l}{\footnotesize Standard errors in parentheses}\\
\multicolumn{7}{l}{\footnotesize \sym{*} \(p<0.10\), \sym{**} \(p<0.05\), \sym{***} \(p<0.01\)}\\
\end{tabular}
}

%% file: input/Tables/regressions_years_educ_1940_byuauto_combined_black_mulatto2_wgtd_wfamfe.tex
{
\def\sym#1{\ifmmode^{#1}\else\(^{#1}\)\fi}
\begin{tabular}{l*{6}{c}}
\toprule
                &\multicolumn{1}{c}{(1)}         &\multicolumn{1}{c}{(2)}         &\multicolumn{1}{c}{(3)}         &\multicolumn{1}{c}{(4)}         &\multicolumn{1}{c}{(5)}         &\multicolumn{1}{c}{(6)}         \\
\midrule
Black - Mulatto &   -0.861\sym{***}&   -0.760\sym{***}&   -0.678\sym{***}&   -0.349\sym{***}&    0.004         &    0.058         \\
                &  (0.018)         &  (0.015)         &  (0.012)         &  (0.011)         &  (0.081)         &  (0.110)         \\
Mulatto - White &   -2.817\sym{***}&   -1.767\sym{***}&   -1.989\sym{***}&   -1.195\sym{***}&                  &                  \\
                &  (0.018)         &  (0.016)         &  (0.013)         &  (0.011)         &                  &                  \\
(Year)X(Age+Birth Order) FE &      Yes         &      Yes         &      Yes         &      Yes         &      Yes         &      Yes         \\
(Year)X(State) FE &       No         &      Yes         &       No         &       No         &       No         &       No         \\
(Year)X(Enum. Dist.) FE &       No         &       No         &      Yes         &      Yes         &      Yes         &       No         \\
(Year)X(Fam. Char.) &       No         &       No         &       No         &      Yes         &      Yes         &       No         \\
BMS Interactions &       No         &       No         &       No         &       No         &      Yes         &       No         \\
(Year)X(Fam.) FE &       No         &       No         &       No         &       No         &       No         &      Yes         \\
\midrule
Observations    &17,254,081         &17,254,081         &17,253,723         &17,253,723         &17,253,723         &7,377,281         \\
R-Squared       &    0.129         &    0.182         &    0.291         &    0.354         &    0.354         &    0.769         \\
Count Black     &  791,274         &  791,274         &  791,262         &  791,262         &  791,262         &  223,812         \\
Count Mulatto   &  212,622         &  212,622         &  212,618         &  212,618         &  212,618         &   66,325         \\
Count White     &16,250,185         &16,250,185         &16,249,843         &16,249,843         &16,249,843         &7,087,144         \\
Mean[Outcome $ | $ Black]&    5.841         &    5.841         &    5.841         &    5.841         &    5.841         &    5.886         \\
Mean[Outcome $ | $ Mulatto]&    6.643         &    6.643         &    6.643         &    6.643         &    6.644         &    6.739         \\
Mean[Outcome $ | $ White]&    9.517         &    9.517         &    9.517         &    9.517         &    9.517         &    9.389         \\
Count Mixed Siblings&    6,787         &    6,787         &    6,787         &    6,787         &    6,787         &    3,041         \\
\bottomrule
\end{tabular}
}

%% file: input/Tables/regressions_women_years_educ_1940_byuauto_combined_black_mulatto2_wgtd_wfamfe.tex
{
\def\sym#1{\ifmmode^{#1}\else\(^{#1}\)\fi}
\begin{tabular}{l*{6}{c}}
\toprule
                &\multicolumn{1}{c}{(1)}         &\multicolumn{1}{c}{(2)}         &\multicolumn{1}{c}{(3)}         &\multicolumn{1}{c}{(4)}         &\multicolumn{1}{c}{(5)}         &\multicolumn{1}{c}{(6)}         \\
\midrule
Black - Mulatto &   -0.744\sym{***}&   -0.663\sym{***}&   -0.604\sym{***}&   -0.271\sym{***}&   -0.201         &   -0.029         \\
                &  (0.020)         &  (0.018)         &  (0.016)         &  (0.015)         &  (0.139)         &  (0.254)         \\
Mulatto - White &   -1.971\sym{***}&   -1.118\sym{***}&   -1.497\sym{***}&   -0.718\sym{***}&                  &                  \\
                &  (0.020)         &  (0.019)         &  (0.016)         &  (0.015)         &                  &                  \\
(Year)X(Age+Birth Order) FE &      Yes         &      Yes         &      Yes         &      Yes         &      Yes         &      Yes         \\
(Year)X(State) FE &       No         &      Yes         &       No         &       No         &       No         &       No         \\
(Year)X(Enum. Dist.) FE &       No         &       No         &      Yes         &      Yes         &      Yes         &       No         \\
(Year)X(Fam. Char.) &       No         &       No         &       No         &      Yes         &      Yes         &       No         \\
BMD Interactions &       No         &       No         &       No         &       No         &      Yes         &       No         \\
(Year)X(Fam.) FE &       No         &       No         &       No         &       No         &       No         &      Yes         \\
\midrule
Observations    &10,982,176         &10,982,176         &10,981,682         &10,981,682         &10,981,682         &3,647,967         \\
R-Squared       &    0.093         &    0.141         &    0.257         &    0.322         &    0.322         &    0.776         \\
Count Black     &  349,451         &  349,451         &  349,431         &  349,431         &  349,431         &   54,002         \\
Count Mulatto   &  101,792         &  101,792         &  101,786         &  101,786         &  101,786         &   18,447         \\
Count White     &10,530,933         &10,530,933         &10,530,465         &10,530,465         &10,530,465         &3,575,518         \\
Mean[Outcome $ | $ Black]&    7.122         &    7.122         &    7.122         &    7.122         &    7.122         &    7.584         \\
Mean[Outcome $ | $ Mulatto]&    7.785         &    7.785         &    7.785         &    7.785         &    7.786         &    8.325         \\
Mean[Outcome $ | $ White]&    9.825         &    9.825         &    9.825         &    9.825         &    9.825         &    9.816         \\
Count Mixed Siblings&    2,855         &    2,855         &    2,855         &    2,855         &    2,855         &      613         \\
\bottomrule
\multicolumn{7}{l}{\footnotesize Standard errors in parentheses}\\
\multicolumn{7}{l}{\footnotesize \sym{*} \(p<0.10\), \sym{**} \(p<0.05\), \sym{***} \(p<0.01\)}\\
\end{tabular}
}

%% file: input/Tables/regressions_married_plus20_byuauto_combined_black_mulatto2_wgtd_wfamfe.tex
{
\def\sym#1{\ifmmode^{#1}\else\(^{#1}\)\fi}
\begin{tabular}{l*{6}{c}}
\toprule
                &\multicolumn{1}{c}{(1)}         &\multicolumn{1}{c}{(2)}         &\multicolumn{1}{c}{(3)}         &\multicolumn{1}{c}{(4)}         &\multicolumn{1}{c}{(5)}         &\multicolumn{1}{c}{(6)}         \\
\midrule
Black - Mulatto &    0.026\sym{***}&    0.016\sym{***}&    0.012\sym{***}&    0.009\sym{***}&    0.004         &    0.009         \\
                &  (0.001)         &  (0.001)         &  (0.001)         &  (0.001)         &  (0.008)         &  (0.012)         \\
Mulatto - White &    0.019\sym{***}&   -0.032\sym{***}&   -0.017\sym{***}&   -0.024\sym{***}&                  &                  \\
                &  (0.001)         &  (0.001)         &  (0.001)         &  (0.001)         &                  &                  \\
(Year)X(Age+Birth Order) FE &      Yes         &      Yes         &      Yes         &      Yes         &      Yes         &      Yes         \\
(Year)X(State) FE &       No         &      Yes         &       No         &       No         &       No         &       No         \\
(Year)X(Enum. Dist.) FE &       No         &       No         &      Yes         &      Yes         &      Yes         &       No         \\
(Year)X(Fam. Char.) &       No         &       No         &       No         &      Yes         &      Yes         &       No         \\
BMS Interactions &       No         &       No         &       No         &       No         &      Yes         &       No         \\
(Year)X(Fam.) FE &       No         &       No         &       No         &       No         &       No         &      Yes         \\
\midrule
Observations    &25,842,325         &25,842,325         &25,841,935         &25,841,935         &25,841,935         &11,497,769         \\
R-Squared       &    0.116         &    0.129         &    0.149         &    0.150         &    0.150         &    0.594         \\
Count Black     &1,263,923         &1,263,923         &1,263,911         &1,263,911         &1,263,911         &  360,022         \\
Count Mulatto   &  314,335         &  314,335         &  314,330         &  314,330         &  314,330         &  100,041         \\
Count White     &24,264,067         &24,264,067         &24,263,694         &24,263,694         &24,263,694         &11,037,706         \\
Mean[Outcome $ | $ Black]&    0.722         &    0.722         &    0.722         &    0.722         &    0.722         &    0.686         \\
Mean[Outcome $ | $ Mulatto]&    0.693         &    0.693         &    0.693         &    0.693         &    0.693         &    0.654         \\
Mean[Outcome $ | $ White]&    0.689         &    0.689         &    0.689         &    0.689         &    0.689         &    0.675         \\
Count Mixed Siblings&   13,974         &   13,974         &   13,974         &   13,974         &   13,974         &    6,104         \\
\bottomrule
\end{tabular}
}

%% file: input/Tables/regressions_women_married_plus20_byuauto_combined_black_mulatto2_wgtd_wfamfe.tex
{
\def\sym#1{\ifmmode^{#1}\else\(^{#1}\)\fi}
\begin{tabular}{l*{6}{c}}
\toprule
                &\multicolumn{1}{c}{(1)}         &\multicolumn{1}{c}{(2)}         &\multicolumn{1}{c}{(3)}         &\multicolumn{1}{c}{(4)}         &\multicolumn{1}{c}{(5)}         &\multicolumn{1}{c}{(6)}         \\
\midrule
Black - Mulatto &   -0.030\sym{***}&   -0.033\sym{***}&   -0.039\sym{***}&   -0.044\sym{***}&   -0.043\sym{***}&   -0.002         \\
                &  (0.002)         &  (0.002)         &  (0.002)         &  (0.002)         &  (0.014)         &  (0.024)         \\
Mulatto - White &   -0.141\sym{***}&   -0.176\sym{***}&   -0.151\sym{***}&   -0.165\sym{***}&                  &                  \\
                &  (0.002)         &  (0.002)         &  (0.002)         &  (0.002)         &                  &                  \\
(Year)X(Age+Birth Order) FE &      Yes         &      Yes         &      Yes         &      Yes         &      Yes         &      Yes         \\
(Year)X(State) FE &       No         &      Yes         &       No         &       No         &       No         &       No         \\
(Year)X(Enum. Dist.) FE &       No         &       No         &      Yes         &      Yes         &      Yes         &       No         \\
(Year)X(Fam. Char.) &       No         &       No         &       No         &      Yes         &      Yes         &       No         \\
BMD Interactions &       No         &       No         &       No         &       No         &      Yes         &       No         \\
(Year)X(Fam.) FE &       No         &       No         &       No         &       No         &       No         &      Yes         \\
\midrule
Observations    &17,040,433         &17,040,433         &17,039,884         &17,039,883         &17,039,883         &6,277,543         \\
R-Squared       &    0.077         &    0.108         &    0.150         &    0.155         &    0.155         &    0.648         \\
Count Black     &  538,194         &  538,194         &  538,173         &  538,173         &  538,173         &   90,979         \\
Count Mulatto   &  153,073         &  153,073         &  153,066         &  153,066         &  153,066         &   30,905         \\
Count White     &16,349,166         &16,349,166         &16,348,645         &16,348,644         &16,348,644         &6,155,659         \\
Mean[Outcome $ | $ Black]&    0.482         &    0.482         &    0.482         &    0.482         &    0.482         &    0.417         \\
Mean[Outcome $ | $ Mulatto]&    0.515         &    0.515         &    0.515         &    0.515         &    0.515         &    0.480         \\
Mean[Outcome $ | $ White]&    0.676         &    0.676         &    0.676         &    0.676         &    0.676         &    0.671         \\
Count Mixed Siblings&    5,781         &    5,781         &    5,781         &    5,781         &    5,781         &    1,440         \\
\bottomrule
\multicolumn{7}{l}{\footnotesize Standard errors in parentheses}\\
\multicolumn{7}{l}{\footnotesize \sym{*} \(p<0.10\), \sym{**} \(p<0.05\), \sym{***} \(p<0.01\)}\\
\end{tabular}
}

%% file: input/Tables/regressions_Syears_educ_1940_byuauto_combined_black_mulatto2_wgtd_wfamfe.tex
{
\def\sym#1{\ifmmode^{#1}\else\(^{#1}\)\fi}
\begin{tabular}{l*{6}{c}}
\toprule
                &\multicolumn{1}{c}{(1)}         &\multicolumn{1}{c}{(2)}         &\multicolumn{1}{c}{(3)}         &\multicolumn{1}{c}{(4)}         &\multicolumn{1}{c}{(5)}         &\multicolumn{1}{c}{(6)}         \\
\midrule
Black - Mulatto &   -0.760\sym{***}&   -0.676\sym{***}&   -0.626\sym{***}&   -0.375\sym{***}&   -0.079         &   -0.120         \\
                &  (0.017)         &  (0.015)         &  (0.013)         &  (0.012)         &  (0.097)         &  (0.155)         \\
Mulatto - White &   -2.273\sym{***}&   -1.452\sym{***}&   -1.697\sym{***}&   -1.088\sym{***}&                  &                  \\
                &  (0.017)         &  (0.016)         &  (0.014)         &  (0.012)         &                  &                  \\
(Year)X(Age+Birth Order) FE &      Yes         &      Yes         &      Yes         &      Yes         &      Yes         &      Yes         \\
(Year)X(State) FE &       No         &      Yes         &       No         &       No         &       No         &       No         \\
(Year)X(Enum. Dist.) FE &       No         &       No         &      Yes         &      Yes         &      Yes         &       No         \\
(Year)X(Fam. Char.) &       No         &       No         &       No         &      Yes         &      Yes         &       No         \\
BMS Interactions &       No         &       No         &       No         &       No         &      Yes         &       No         \\
(Year)X(Fam.) FE &       No         &       No         &       No         &       No         &       No         &      Yes         \\
\midrule
Observations    &12,489,464         &12,489,464         &12,488,977         &12,488,977         &12,488,977         &4,470,376         \\
R-Squared       &    0.104         &    0.154         &    0.251         &    0.292         &    0.292         &    0.696         \\
Count Black     &  542,386         &  542,386         &  542,365         &  542,365         &  542,365         &  116,883         \\
Count Mulatto   &  144,406         &  144,406         &  144,404         &  144,404         &  144,404         &   34,484         \\
Count White     &11,802,672         &11,802,672         &11,802,208         &11,802,208         &11,802,208         &4,319,009         \\
Mean[Outcome $ | $ Black]&    6.708         &    6.708         &    6.708         &    6.708         &    6.708         &    6.598         \\
Mean[Outcome $ | $ Mulatto]&    7.420         &    7.420         &    7.420         &    7.420         &    7.421         &    7.367         \\
Mean[Outcome $ | $ White]&    9.729         &    9.729         &    9.729         &    9.729         &    9.729         &    9.626         \\
Count Mixed Siblings&    4,687         &    4,687         &    4,687         &    4,687         &    4,687         &    1,484         \\
\bottomrule
\end{tabular}
}

%% file: input/Tables/regressions_women_Syears_educ_1940_byuauto_combined_black_mulatto2_wgtd_wfamfe.tex
{
\def\sym#1{\ifmmode^{#1}\else\(^{#1}\)\fi}
\begin{tabular}{l*{6}{c}}
\toprule
                &\multicolumn{1}{c}{(1)}         &\multicolumn{1}{c}{(2)}         &\multicolumn{1}{c}{(3)}         &\multicolumn{1}{c}{(4)}         &\multicolumn{1}{c}{(5)}         &\multicolumn{1}{c}{(6)}         \\
\midrule
Black - Mulatto &   -0.615\sym{***}&   -0.520\sym{***}&   -0.484\sym{***}&   -0.193\sym{***}&   -0.545\sym{**} &   -0.321         \\
                &  (0.026)         &  (0.023)         &  (0.022)         &  (0.021)         &  (0.222)         &  (0.527)         \\
Mulatto - White &   -2.395\sym{***}&   -1.464\sym{***}&   -1.753\sym{***}&   -1.044\sym{***}&                  &                  \\
                &  (0.024)         &  (0.022)         &  (0.021)         &  (0.020)         &                  &                  \\
(Year)X(Age+Birth Order) FE &      Yes         &      Yes         &      Yes         &      Yes         &      Yes         &      Yes         \\
(Year)X(State) FE &       No         &      Yes         &       No         &       No         &       No         &       No         \\
(Year)X(Enum. Dist.) FE &       No         &       No         &      Yes         &      Yes         &      Yes         &       No         \\
(Year)X(Fam. Char.) &       No         &       No         &       No         &      Yes         &      Yes         &       No         \\
BMD Interactions &       No         &       No         &       No         &       No         &      Yes         &       No         \\
(Year)X(Fam.) FE &       No         &       No         &       No         &       No         &       No         &      Yes         \\
\midrule
Observations    &7,372,514         &7,372,514         &7,371,725         &7,371,725         &7,371,725         &2,072,639         \\
R-Squared       &    0.067         &    0.122         &    0.233         &    0.279         &    0.279         &    0.694         \\
Count Black     &  160,958         &  160,958         &  160,925         &  160,925         &  160,925         &   11,990         \\
Count Mulatto   &   50,526         &   50,526         &   50,513         &   50,513         &   50,513         &    5,769         \\
Count White     &7,161,030         &7,161,030         &7,160,287         &7,160,287         &7,160,287         &2,054,880         \\
Mean[Outcome $ | $ Black]&    6.072         &    6.072         &    6.072         &    6.072         &    6.072         &    5.908         \\
Mean[Outcome $ | $ Mulatto]&    6.633         &    6.633         &    6.634         &    6.634         &    6.635         &    6.791         \\
Mean[Outcome $ | $ White]&    9.101         &    9.101         &    9.101         &    9.101         &    9.101         &    8.932         \\
Count Mixed Siblings&    1,316         &    1,316         &    1,316         &    1,316         &    1,316         &      175         \\
\bottomrule
\multicolumn{7}{l}{\footnotesize Standard errors in parentheses}\\
\multicolumn{7}{l}{\footnotesize \sym{*} \(p<0.10\), \sym{**} \(p<0.05\), \sym{***} \(p<0.01\)}\\
\end{tabular}
}

%% file: input/Tables/regressions_log_inc_1940_byuauto_combined_black_mulatto2_wgtd_wfamfe.tex
{
\def\sym#1{\ifmmode^{#1}\else\(^{#1}\)\fi}
\begin{tabular}{l*{6}{c}}
\toprule
                &\multicolumn{1}{c}{(1)}         &\multicolumn{1}{c}{(2)}         &\multicolumn{1}{c}{(3)}         &\multicolumn{1}{c}{(4)}         &\multicolumn{1}{c}{(5)}         &\multicolumn{1}{c}{(6)}         \\
\midrule
Black - Mulatto &   -0.136\sym{***}&   -0.107\sym{***}&   -0.091\sym{***}&   -0.051\sym{***}&    0.006         &    0.017         \\
                &  (0.004)         &  (0.003)         &  (0.003)         &  (0.003)         &  (0.027)         &  (0.044)         \\
Mulatto - White &   -0.686\sym{***}&   -0.490\sym{***}&   -0.547\sym{***}&   -0.443\sym{***}&                  &                  \\
                &  (0.004)         &  (0.004)         &  (0.003)         &  (0.003)         &                  &                  \\
(Year)X(Age+Birth Order) FE &      Yes         &      Yes         &      Yes         &      Yes         &      Yes         &      Yes         \\
(Year)X(State) FE &       No         &      Yes         &       No         &       No         &       No         &       No         \\
(Year)X(Enum. Dist.) FE &       No         &       No         &      Yes         &      Yes         &      Yes         &       No         \\
(Year)X(Fam. Char.) &       No         &       No         &       No         &      Yes         &      Yes         &       No         \\
BMS Interactions &       No         &       No         &       No         &       No         &      Yes         &       No         \\
(Year)X(Fam.) FE &       No         &       No         &       No         &       No         &       No         &      Yes         \\
\midrule
Observations    &12,690,326         &12,690,326         &12,689,868         &12,689,868         &12,689,868         &4,522,234         \\
R-Squared       &    0.111         &    0.157         &    0.237         &    0.251         &    0.252         &    0.670         \\
Count Black     &  549,575         &  549,575         &  549,560         &  549,560         &  549,560         &  121,309         \\
Count Mulatto   &  150,040         &  150,040         &  150,036         &  150,036         &  150,036         &   36,895         \\
Count White     &11,990,711         &11,990,711         &11,990,272         &11,990,272         &11,990,272         &4,364,030         \\
Mean[Outcome $ | $ Black]&    6.034         &    6.034         &    6.034         &    6.034         &    6.034         &    5.994         \\
Mean[Outcome $ | $ Mulatto]&    6.189         &    6.189         &    6.189         &    6.189         &    6.189         &    6.168         \\
Mean[Outcome $ | $ White]&    6.851         &    6.851         &    6.851         &    6.851         &    6.851         &    6.823         \\
Count Mixed Siblings&    4,611         &    4,611         &    4,611         &    4,611         &    4,611         &    1,483         \\
\bottomrule
\end{tabular}
}

%% file: input/Tables/regressions_women_Slog_inc_1940_byuauto_combined_black_mulatto2_wgtd_wfamfe.tex
{
\def\sym#1{\ifmmode^{#1}\else\(^{#1}\)\fi}
\begin{tabular}{l*{6}{c}}
\toprule
                &\multicolumn{1}{c}{(1)}         &\multicolumn{1}{c}{(2)}         &\multicolumn{1}{c}{(3)}         &\multicolumn{1}{c}{(4)}         &\multicolumn{1}{c}{(5)}         &\multicolumn{1}{c}{(6)}         \\
\midrule
Black - Mulatto &   -0.116\sym{***}&   -0.077\sym{***}&   -0.070\sym{***}&   -0.024\sym{***}&   -0.020         &    0.265\sym{*}  \\
                &  (0.007)         &  (0.006)         &  (0.007)         &  (0.006)         &  (0.068)         &  (0.158)         \\
Mulatto - White &   -0.553\sym{***}&   -0.368\sym{***}&   -0.453\sym{***}&   -0.343\sym{***}&                  &                  \\
                &  (0.006)         &  (0.006)         &  (0.006)         &  (0.006)         &                  &                  \\
(Year)X(Age+Birth Order) FE &      Yes         &      Yes         &      Yes         &      Yes         &      Yes         &      Yes         \\
(Year)X(State) FE &       No         &      Yes         &       No         &       No         &       No         &       No         \\
(Year)X(Enum. Dist.) FE &       No         &       No         &      Yes         &      Yes         &      Yes         &       No         \\
(Year)X(Fam. Char.) &       No         &       No         &       No         &      Yes         &      Yes         &       No         \\
BMD Interactions &       No         &       No         &       No         &       No         &      Yes         &       No         \\
(Year)X(Fam.) FE &       No         &       No         &       No         &       No         &       No         &      Yes         \\
\midrule
Observations    &5,216,487         &5,216,487         &5,215,503         &5,215,503         &5,215,503         &1,122,427         \\
R-Squared       &    0.052         &    0.110         &    0.220         &    0.238         &    0.238         &    0.644         \\
Count Black     &  107,514         &  107,514         &  107,460         &  107,460         &  107,460         &    5,618         \\
Count Mulatto   &   34,446         &   34,446         &   34,426         &   34,426         &   34,426         &    2,982         \\
Count White     &5,074,527         &5,074,527         &5,073,617         &5,073,617         &5,073,617         &1,113,827         \\
Mean[Outcome $ | $ Black]&    6.285         &    6.285         &    6.285         &    6.285         &    6.285         &    6.233         \\
Mean[Outcome $ | $ Mulatto]&    6.416         &    6.416         &    6.416         &    6.416         &    6.416         &    6.463         \\
Mean[Outcome $ | $ White]&    6.956         &    6.956         &    6.956         &    6.956         &    6.956         &    6.917         \\
Count Mixed Siblings&      871         &      871         &      871         &      871         &      871         &       69         \\
\bottomrule
\multicolumn{7}{l}{\footnotesize Standard errors in parentheses}\\
\multicolumn{7}{l}{\footnotesize \sym{*} \(p<0.10\), \sym{**} \(p<0.05\), \sym{***} \(p<0.01\)}\\
\end{tabular}
}

%% file: input/Tables/regressions_scorewgCWadj_plus20_byuauto_combined_black_mulatto2_wgtd_wfamfe.tex
{
\def\sym#1{\ifmmode^{#1}\else\(^{#1}\)\fi}
\begin{tabular}{l*{6}{c}}
\toprule
                &\multicolumn{1}{c}{(1)}         &\multicolumn{1}{c}{(2)}         &\multicolumn{1}{c}{(3)}         &\multicolumn{1}{c}{(4)}         &\multicolumn{1}{c}{(5)}         &\multicolumn{1}{c}{(6)}         \\
\midrule
Black - Mulatto &   -0.091\sym{***}&   -0.056\sym{***}&   -0.047\sym{***}&   -0.026\sym{***}&   -0.013         &    0.004         \\
                &  (0.002)         &  (0.001)         &  (0.001)         &  (0.001)         &  (0.008)         &  (0.011)         \\
Mulatto - White &   -0.457\sym{***}&   -0.248\sym{***}&   -0.276\sym{***}&   -0.209\sym{***}&                  &                  \\
                &  (0.002)         &  (0.001)         &  (0.001)         &  (0.001)         &                  &                  \\
(Year)X(Age+Birth Order) FE &      Yes         &      Yes         &      Yes         &      Yes         &      Yes         &      Yes         \\
(Year)X(State) FE &       No         &      Yes         &       No         &       No         &       No         &       No         \\
(Year)X(Enum. Dist.) FE &       No         &       No         &      Yes         &      Yes         &      Yes         &       No         \\
(Year)X(Fam. Char.) &       No         &       No         &       No         &      Yes         &      Yes         &       No         \\
BMS Interactions &       No         &       No         &       No         &       No         &      Yes         &       No         \\
(Year)X(Fam.) FE &       No         &       No         &       No         &       No         &       No         &      Yes         \\
\midrule
Observations    &23,708,500         &23,708,500         &23,708,079         &23,708,079         &23,708,079         &10,122,764         \\
R-Squared       &    0.232         &    0.368         &    0.424         &    0.444         &    0.444         &    0.752         \\
Count Black     &1,168,063         &1,168,063         &1,168,048         &1,168,048         &1,168,048         &  318,563         \\
Count Mulatto   &  287,719         &  287,719         &  287,715         &  287,715         &  287,715         &   87,307         \\
Count White     &22,252,718         &22,252,718         &22,252,316         &22,252,316         &22,252,316         &9,716,894         \\
Mean[Outcome $ | $ Black]&    6.192         &    6.192         &    6.192         &    6.192         &    6.192         &    6.141         \\
Mean[Outcome $ | $ Mulatto]&    6.283         &    6.283         &    6.283         &    6.283         &    6.283         &    6.246         \\
Mean[Outcome $ | $ White]&    6.753         &    6.753         &    6.753         &    6.753         &    6.753         &    6.713         \\
Count Mixed Siblings&   12,820         &   12,820         &   12,820         &   12,820         &   12,820         &    5,241         \\
\bottomrule
\end{tabular}
}

%% file: input/Tables/regressions_women_SscorewgCWadj_plus20_byuauto_combined_black_mulatto2_wgtd_wfamfe.tex
{
\def\sym#1{\ifmmode^{#1}\else\(^{#1}\)\fi}
\begin{tabular}{l*{6}{c}}
\toprule
                &\multicolumn{1}{c}{(1)}         &\multicolumn{1}{c}{(2)}         &\multicolumn{1}{c}{(3)}         &\multicolumn{1}{c}{(4)}         &\multicolumn{1}{c}{(5)}         &\multicolumn{1}{c}{(6)}         \\
\midrule
Black - Mulatto &   -0.094\sym{***}&   -0.053\sym{***}&   -0.048\sym{***}&   -0.026\sym{***}&   -0.062\sym{***}&   -0.018         \\
                &  (0.003)         &  (0.002)         &  (0.002)         &  (0.002)         &  (0.020)         &  (0.043)         \\
Mulatto - White &   -0.397\sym{***}&   -0.212\sym{***}&   -0.249\sym{***}&   -0.183\sym{***}&                  &                  \\
                &  (0.003)         &  (0.002)         &  (0.002)         &  (0.002)         &                  &                  \\
(Year)X(Age+Birth Order) FE &      Yes         &      Yes         &      Yes         &      Yes         &      Yes         &      Yes         \\
(Year)X(State) FE &       No         &      Yes         &       No         &       No         &       No         &       No         \\
(Year)X(Enum. Dist.) FE &       No         &       No         &      Yes         &      Yes         &      Yes         &       No         \\
(Year)X(Fam. Char.) &       No         &       No         &       No         &      Yes         &      Yes         &       No         \\
BMD Interactions &       No         &       No         &       No         &       No         &      Yes         &       No         \\
(Year)X(Fam.) FE &       No         &       No         &       No         &       No         &       No         &      Yes         \\
\midrule
Observations    &10,048,623         &10,048,623         &10,047,678         &10,047,676         &10,047,676         &2,949,466         \\
R-Squared       &    0.115         &    0.283         &    0.353         &    0.373         &    0.373         &    0.695         \\
Count Black     &  204,402         &  204,402         &  204,358         &  204,358         &  204,358         &   15,448         \\
Count Mulatto   &   62,702         &   62,702         &   62,688         &   62,687         &   62,687         &    7,460         \\
Count White     &9,781,519         &9,781,519         &9,780,632         &9,780,631         &9,780,631         &2,926,558         \\
Mean[Outcome $ | $ Black]&    6.347         &    6.347         &    6.347         &    6.347         &    6.347         &    6.302         \\
Mean[Outcome $ | $ Mulatto]&    6.446         &    6.446         &    6.446         &    6.446         &    6.446         &    6.460         \\
Mean[Outcome $ | $ White]&    6.838         &    6.838         &    6.838         &    6.838         &    6.838         &    6.786         \\
Count Mixed Siblings&    2,086         &    2,086         &    2,086         &    2,086         &    2,086         &      280         \\
\bottomrule
\multicolumn{7}{l}{\footnotesize Standard errors in parentheses}\\
\multicolumn{7}{l}{\footnotesize \sym{*} \(p<0.10\), \sym{**} \(p<0.05\), \sym{***} \(p<0.01\)}\\
\end{tabular}
}

%% file: appendix_figures.tex
\begin{figure}[H]
\begin{centering}
\caption{\label{fig:Geographic-frequency-of-sons}Number of differently-classified
brothers, by census cohort and county}
\par\end{centering}
\begin{centering}
\subfloat[1870 cohort]{\centering{}\includegraphics[width=0.49\textwidth]{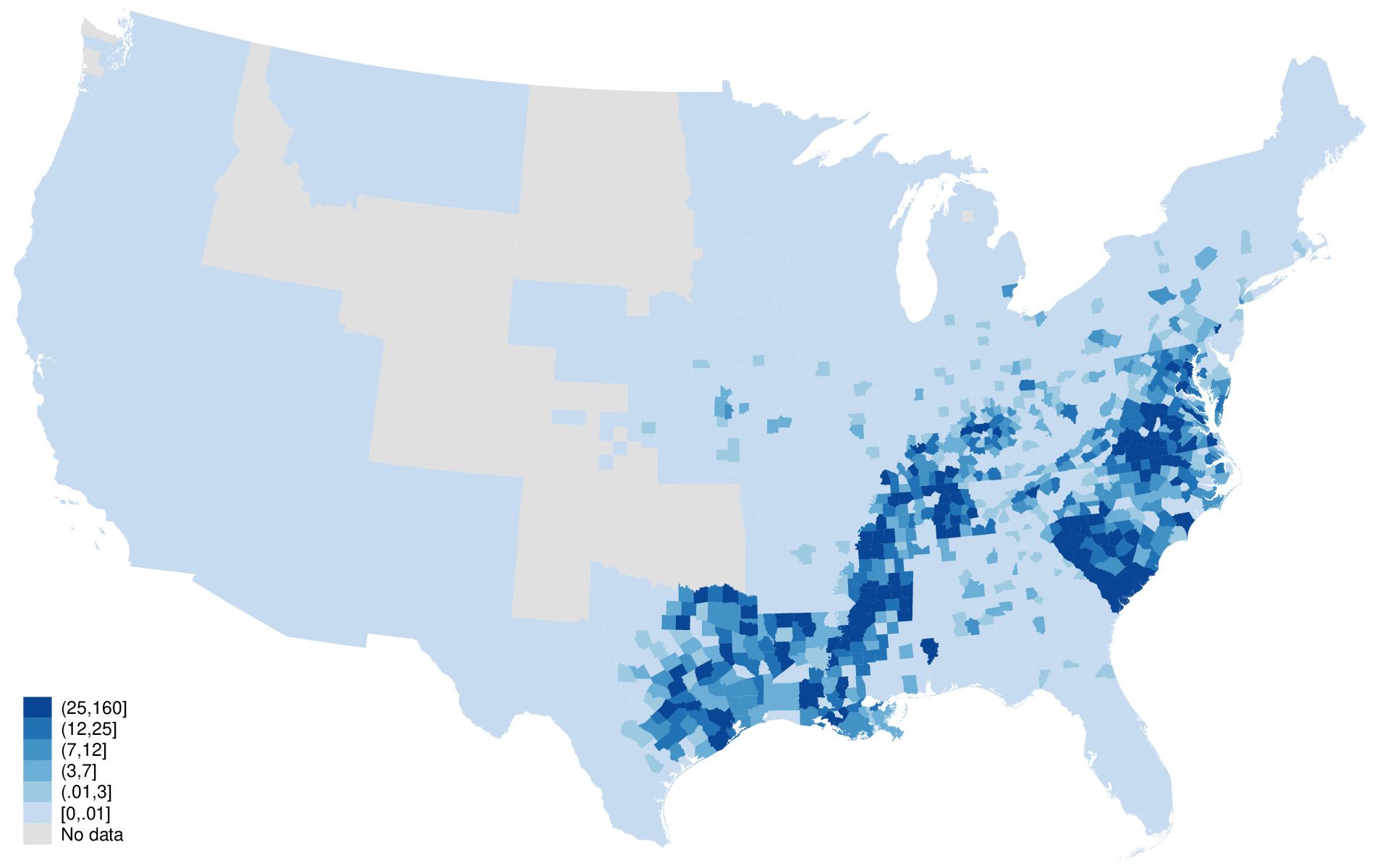}}\subfloat[1880 cohort]{\centering{}\includegraphics[width=0.49\textwidth]{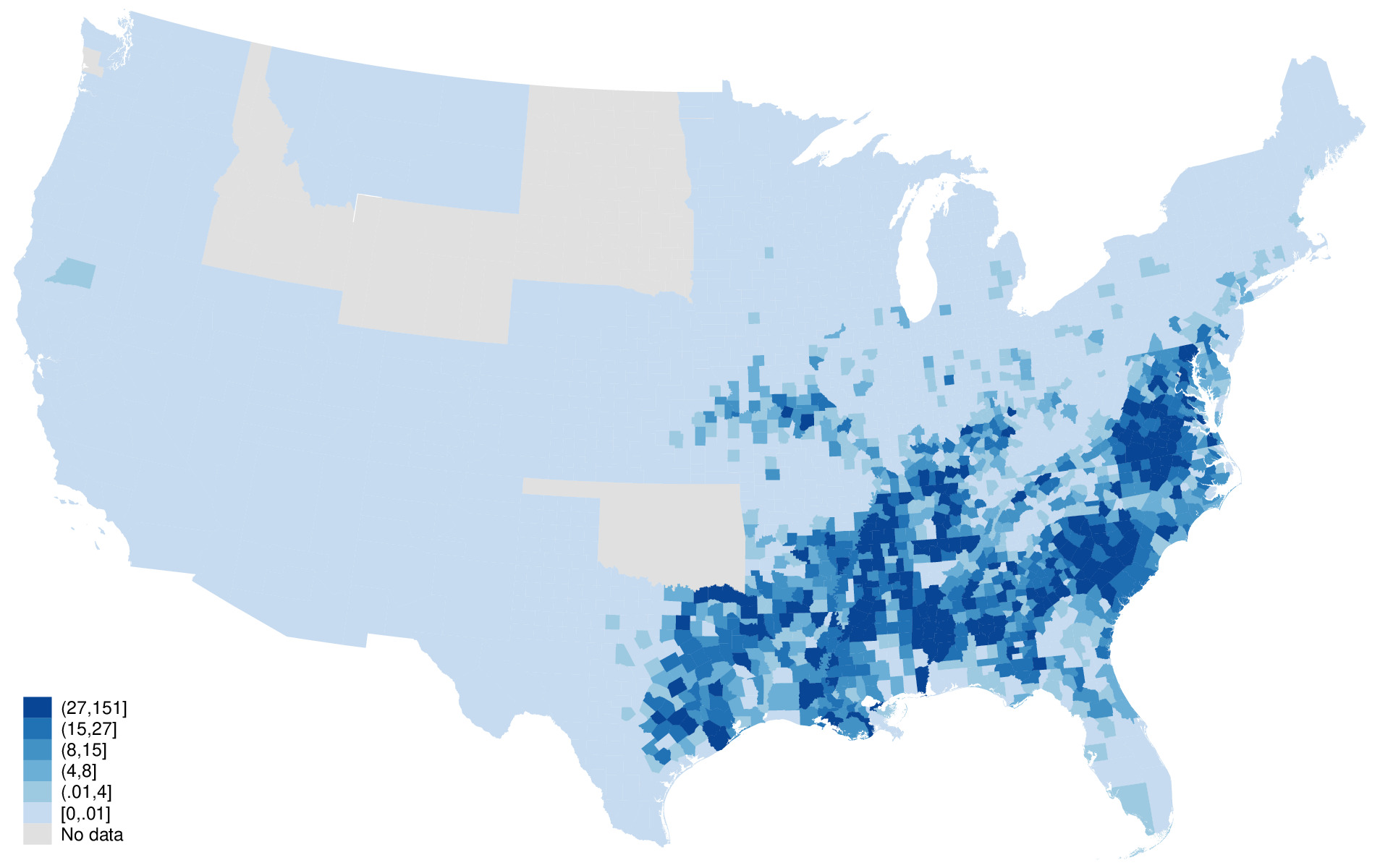}}
\par\end{centering}
\begin{centering}
\subfloat[1910 cohort]{\begin{centering}
\par\end{centering}
\centering{}\includegraphics[width=0.49\textwidth]{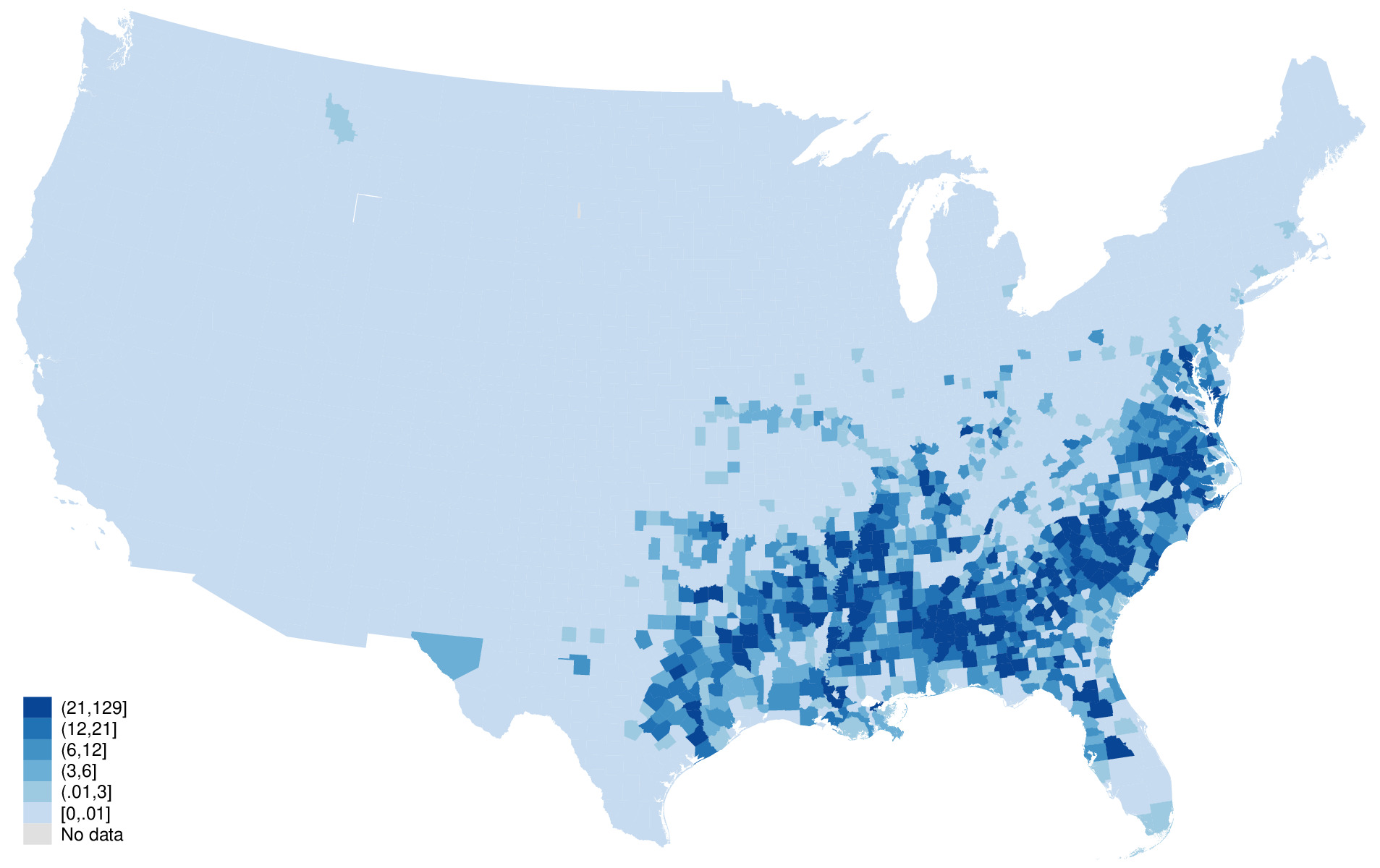}}\subfloat[1920 cohort]{\centering{}\includegraphics[width=0.49\textwidth]{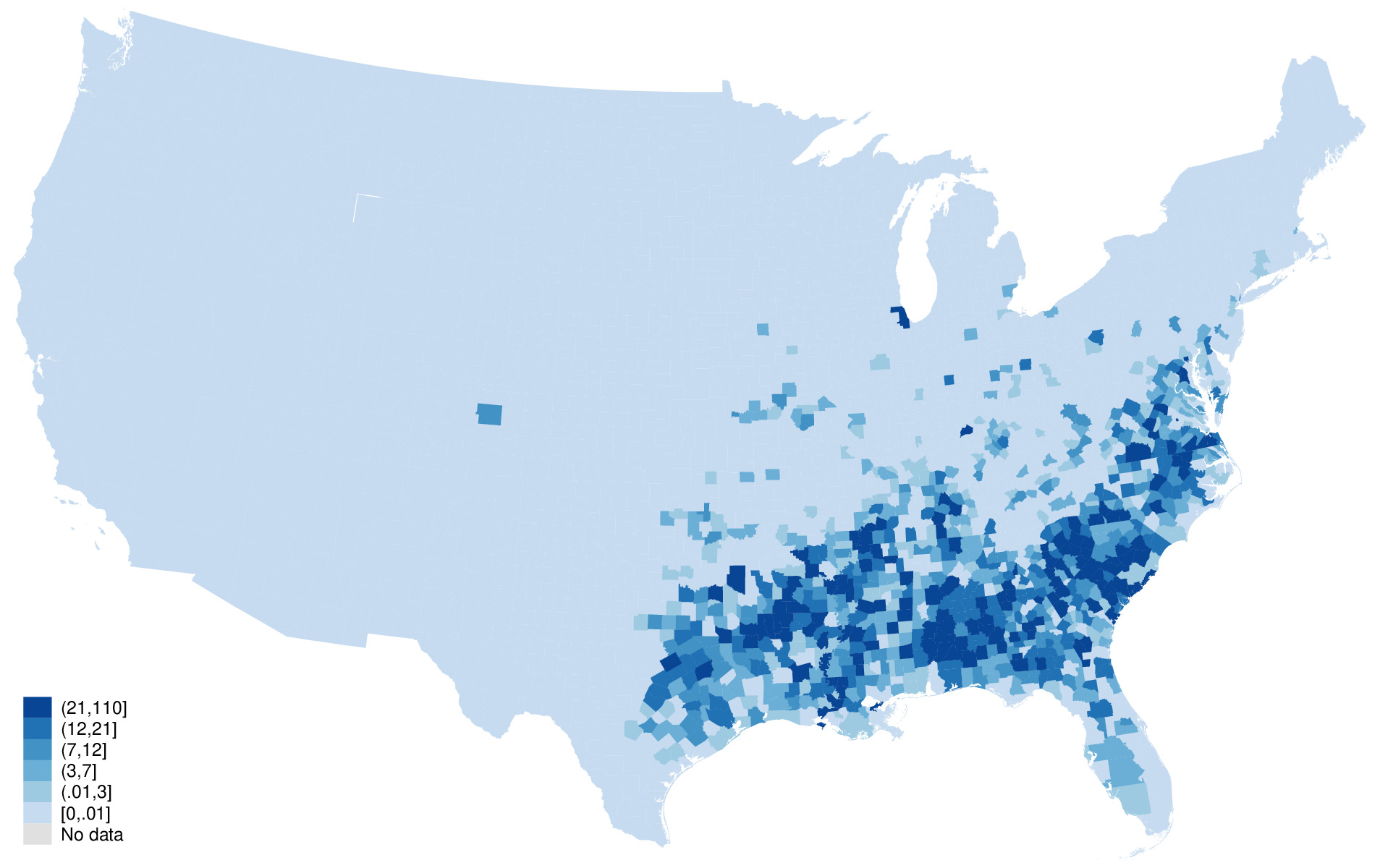}}
\par\end{centering}
{\small Note: Figure maps the number of differently-classified sons
by county and census cohort. Our Black-and-Mulatto-sons (BMS) sample
consists of households where some son aged 3--18 was classified Mulatto
and another Black. We restrict to cases where these sons shared two
parents who were both present within the household. Sample construction
is described in Section~\ref{subsec:child-cohorts}.}{\small\par}
\end{figure}
\pagebreak{}
\begin{figure}[H]
\begin{centering}
\caption{\label{fig:algorithm-robustness}Robustness of \emph{Black--Mulatto}
gaps in sons' linked adult outcomes across linking algorithms}
\par\end{centering}
\begin{centering}
\includegraphics[width=1\textwidth]{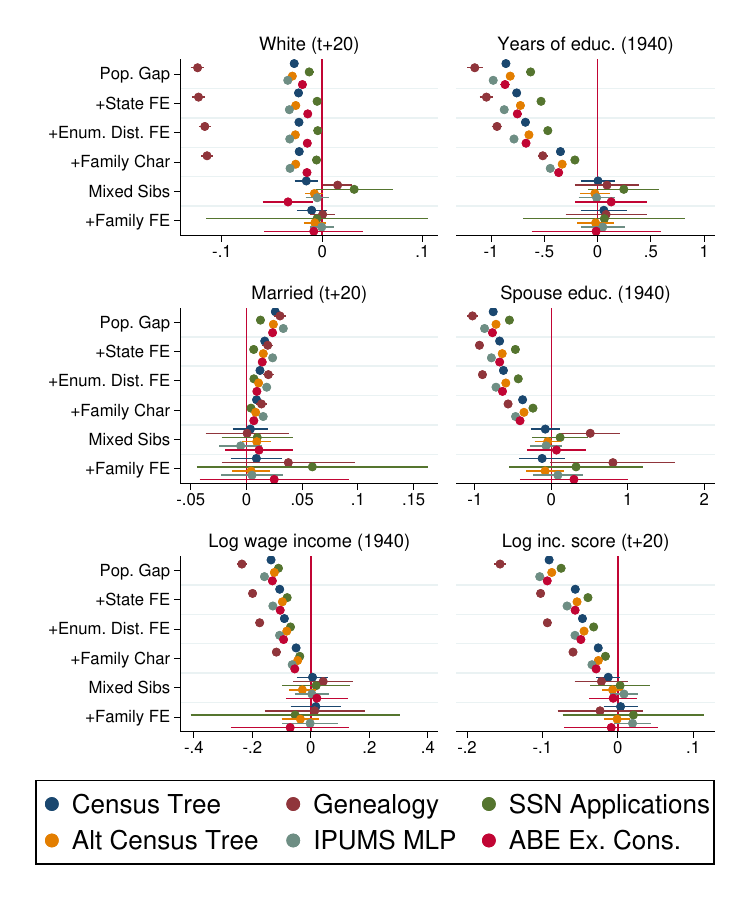}
\par\end{centering}
{\small Note: Figure shows robustness of our linked adult outcome results
to our choice of matching algorithm by showing regression coefficients
and 95\% confidence intervals documenting sons' }{\small\emph{Black--Mulatto}}{\small{}
gaps for each of our main linked outcomes, for each set of controls,
and for six different linking algorithms. We repeat results using
our preferred Census Tree links (as shown in Figures~\ref{fig:white-yearseduc}--\ref{fig:income}
and as used throughout the paper unless noted otherwise), which have
a linking rate to }{\small\emph{t+20}}{\small{} of $\MatchRatebyuauto$\%.
We now also show analogous results when individuals are linked by
genealogical links from FamilySearch users ($\MatchRategeneal$\%
linking rate), Social Security Number application-based links ($\MatchRatessnapp$\%),
public Census Tree V1.0 links ($\MatchRatectany$\%), IPUMS MLP V2.0
links ($\MatchRatemlp$\%), or the ABE Exact Conservative method ($\MatchRateABEexcon$\%).
These alternative linking methods (other than ABE Exact Conservative) often rely more heavily
on racial concordance across censuses, which can reduce perceived skin
tone differences between linked Black- and Mulatto-coded individuals
and complicate interpretation in our setting. See
Section~\ref{subsec:Matching-across-censuses} for further discussion
of these linking methods. We omit coefficients that would be based
on a comparison of fewer than $\algorithmrobustnessmincount$ individuals.}{\small\par}
\end{figure}
\pagebreak{}

\begin{figure}[H]
\begin{centering}
\caption{\label{fig:algorithm-robustness-women}Robustness of \emph{Black--Mulatto}
gaps in daughters' linked adult outcomes across linking algorithms}
\par\end{centering}
\begin{centering}
\includegraphics[width=1\textwidth]{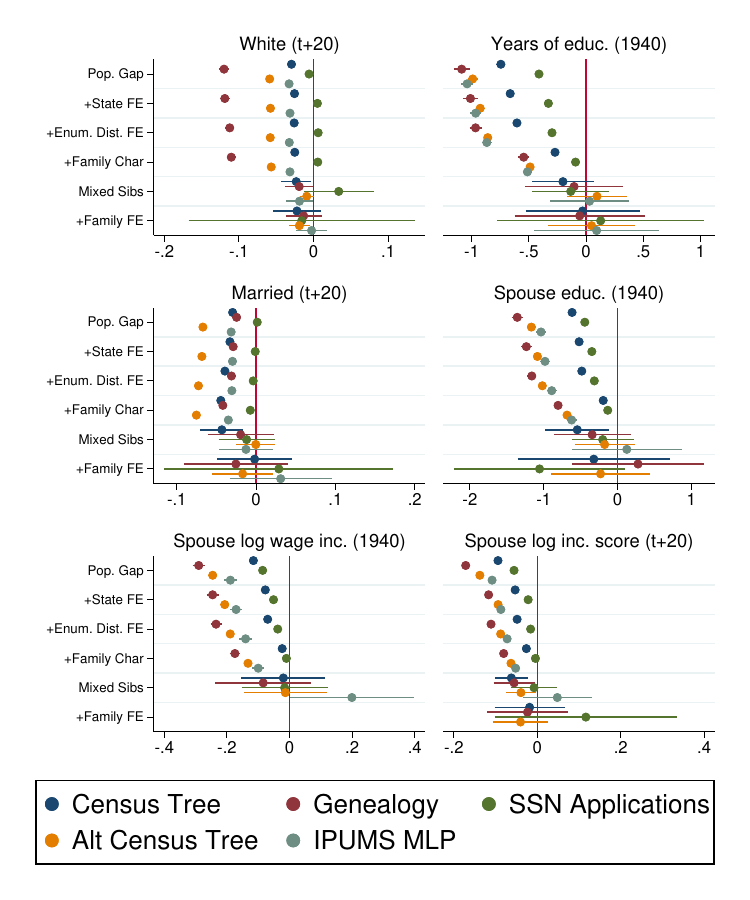}
\par\end{centering}
{\small Note: Figure shows robustness of our linked adult outcome results
to our choice of matching algorithm by showing regression coefficients
and 95\% confidence intervals documenting daughters' }{\small\emph{Black--Mulatto}}{\small{}
gaps for each of our main linked outcomes, for each set of controls,
and for five different matching algorithms. We repeat results using
our preferred Census Tree links (as shown in Figures \ref{fig:white-yearseduc}--\ref{fig:income}
and as used throughout the paper unless noted otherwise), which have
a linking rate to }{\small\emph{t+20}}{\small{} of $\MatchRatewomenbyuauto$\%.
We now also show analogous results when individuals are linked by
genealogical links from FamilySearch users ($\MatchRatewomengeneal$\%
linking rate), Social Security Number application-based links ($\MatchRatewomenssnapp$\%),
public Census Tree V1.0 links ($\MatchRatewomenctany$\%), or IPUMS
MLP V2.0 links ($\MatchRatewomenmlp$\%).
These alternative linking methods often rely more heavily
on racial concordance across censuses, which can reduce perceived skin
tone differences between linked Black- and Mulatto-coded individuals
and complicate interpretation in our setting. See
Section~\ref{subsec:Matching-across-censuses}
for further discussion of these linking methods. We omit coefficients
that would be based on a comparison of fewer than $\algorithmrobustnessmincount$
individuals.}{\small\par}
\end{figure}
\pagebreak{}

\begin{figure}[H]
\begin{centering}
\caption{\label{fig:incscore-robustness}Robustness of \emph{Black--Mulatto}
gaps across construction of income scores}
\par\end{centering}
\begin{centering}
\includegraphics[width=1\textwidth]{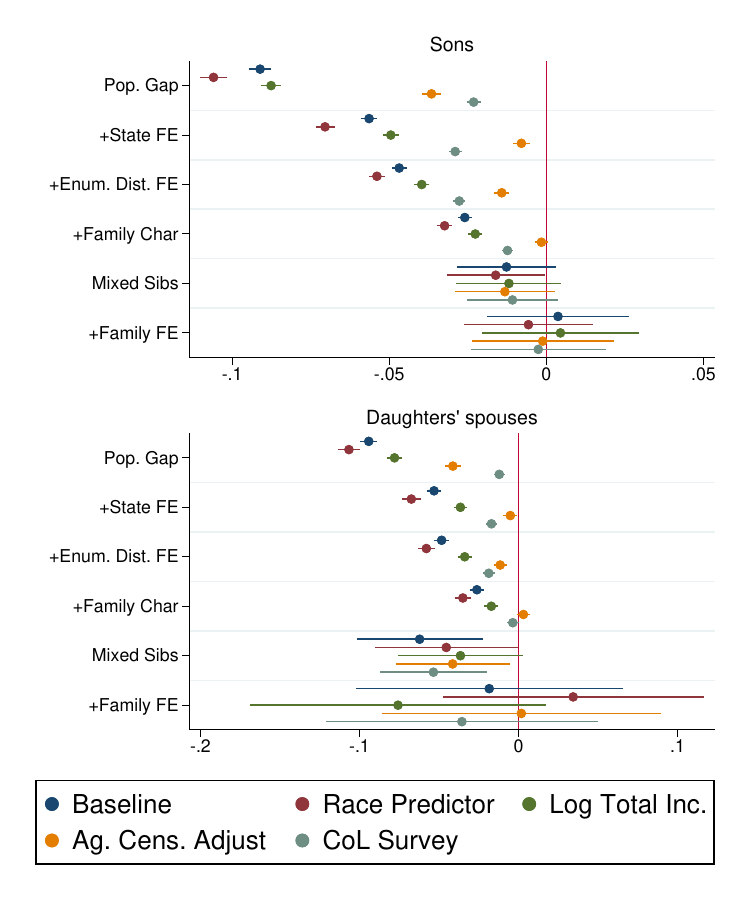}
\par\end{centering}
{\small Note: Figure shows robustness of our linked adult income score
outcomes to the construction of income scores. We do so by showing
regression coefficients and 95\% confidence intervals documenting
}{\small\emph{Black--Mulatto}}{\small{} gaps in the income scores of
sons and of daughters' spouses, for each set of controls and for different
construction methods. We repeat results using our baseline income
scores (as shown in Figure~\ref{fig:income} Panel~B), while now
also showing analogous results when income scores are modified to
either include race as an income predictor, predict log total income
(as defined in Section~\ref{subsec:adult_income}, rather than predicting
log wage income), use the 1900 Census of Agriculture to adjust incomes
for farm managers and farmer laborers (rather than the 1960-decennial-census-based
adjustment suggested by \citealp{collins2022african}), or use the
1901 Cost of Living survey (rather than decennial census data) as
our income data source. See Section~\ref{subsec:income-prediction}
for further discussion of these construction methods.}{\small\par}
\end{figure}
\pagebreak{}
\begin{figure}[H]
\begin{centering}
\caption{\label{fig:child-education-bycohort-mulattolesswhite}\emph{Mulatto--White}
gap in childhood education, by child cohort and gender}
\par\end{centering}
\begin{centering}
\subfloat[\label{subfig:inschool-bycohort-mulattolesswhite}School attendance]{
\centering{}\includegraphics[width=0.7\textwidth]{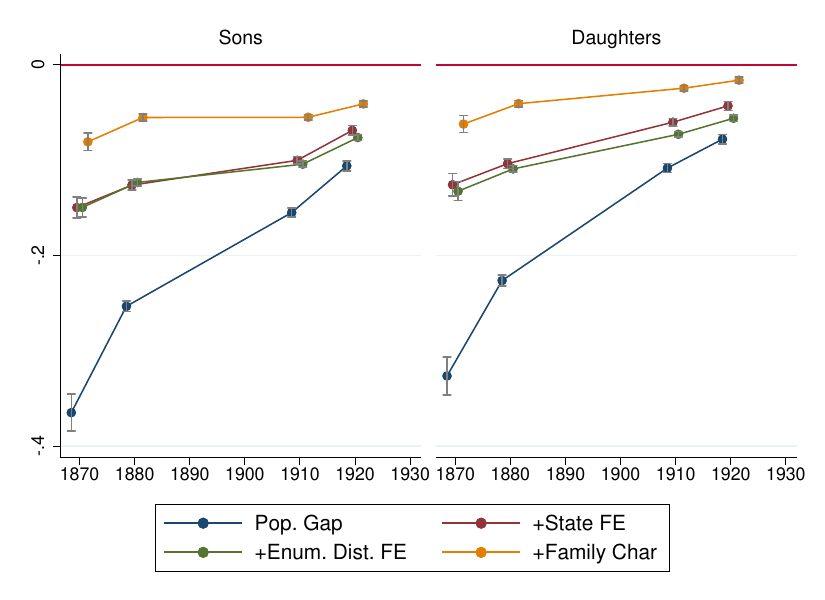}}
\par\end{centering}
\begin{centering}
\subfloat[\label{subfig:literate-bycohort-mulattolesswhite}Literacy]{
\centering{}\includegraphics[width=0.7\textwidth]{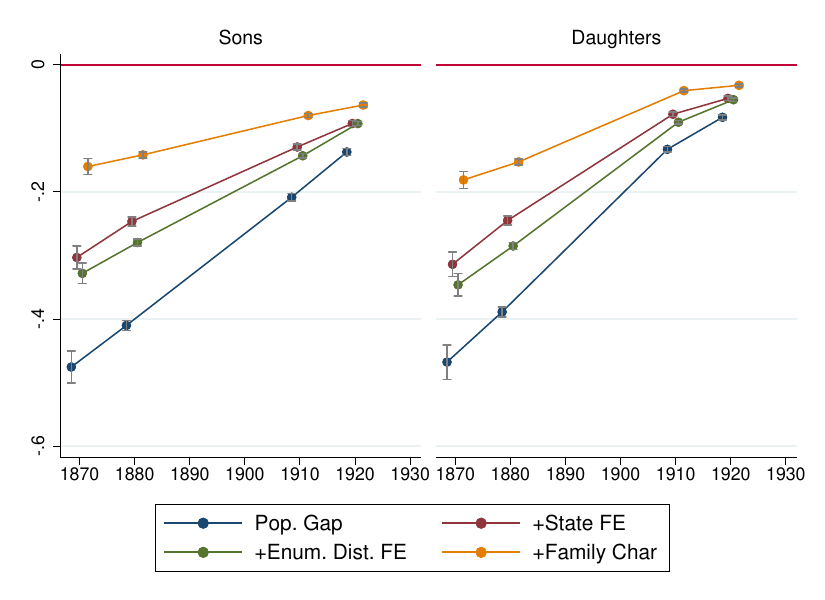}}
\par\end{centering}
{\small Note: Figure shows regression coefficients and 95\% confidence
intervals documenting }{\small\emph{Mulatto--White}}{\small{} gaps
in childhood school attendance (Panel~A) and literacy (Panel~B)
by gender and child cohort. See Section~\ref{subsec:child-education}
for detail.}{\small\par}
\end{figure}
\pagebreak{}
\begin{figure}[H]
\begin{centering}
\caption{\label{fig:child-education-bycohort-bmlesswhite}\emph{Black\slash Mulatto--White}
gap in childhood education, by child cohort and gender}
\par\end{centering}
\begin{centering}
\subfloat[\label{subfig:inschool-bycohort-bmlesswhite}School attendance]{
\centering{}\includegraphics[width=0.7\textwidth]{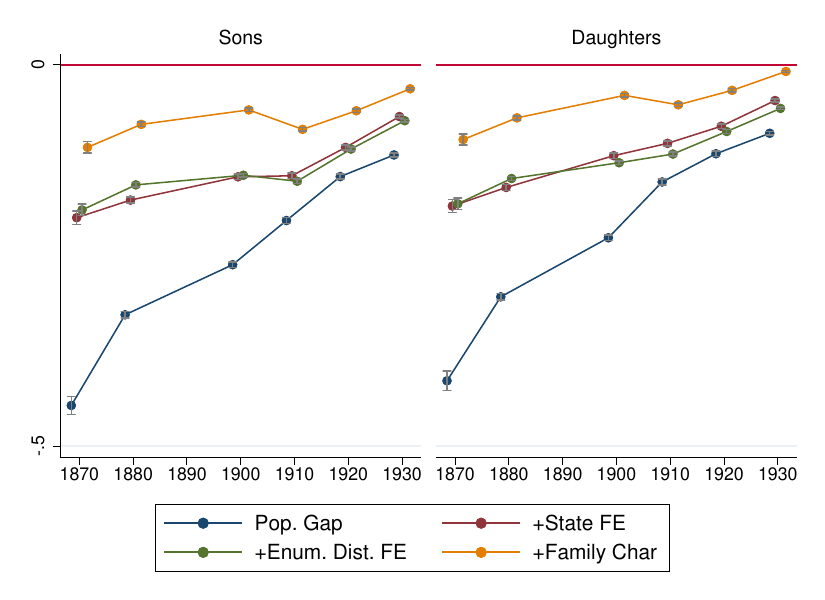}}
\par\end{centering}
\begin{centering}
\subfloat[\label{subfig:literate-bycohort-bmlesswhite}Literacy]{
\centering{}\includegraphics[width=0.7\textwidth]{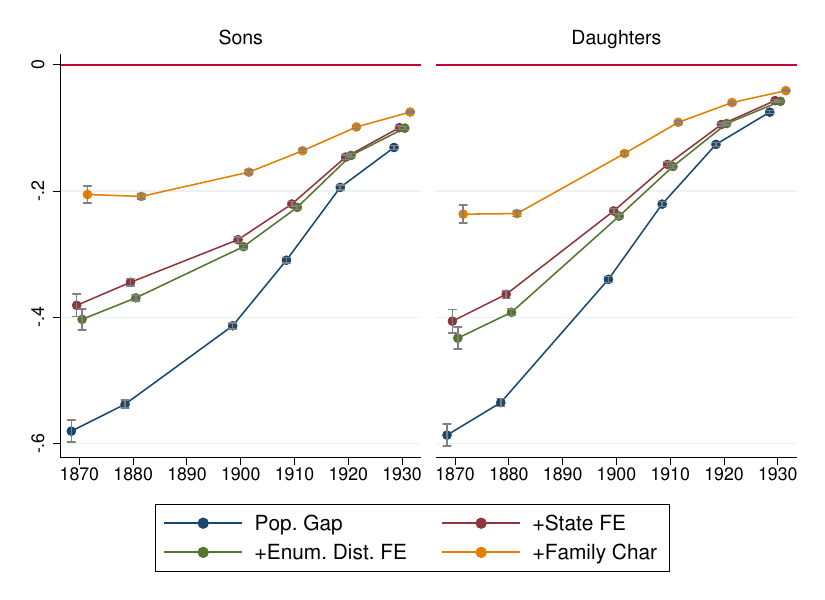}}
\par\end{centering}
{\small Note: Figure shows regression coefficients and 95\% confidence
intervals documenting }{\small\emph{Black\slash Mulatto--White}}{\small{}
gaps in childhood school attendance (Panel~A) and literacy (Panel~B)
by gender and child cohort. The regression specification modifies
Equation~\ref{eq:naive-discrimination-estimation} by pooling Mulatto-
and Black-coded children together into a single Black/Mulatto category,
which allows us to also include 1900 and 1930 child cohorts (where
the Black vs.\ Mulatto distinction was not made by census enumerators).}{\small\par}
\end{figure}

\pagebreak{}
\begin{figure}[H]
\begin{centering}
\caption{\label{fig:white-other}Additional gaps in classification as White,
by perceived child race and gender}
\par\end{centering}
\begin{centering}
\subfloat[\label{subfig:white-plus20-mulattowhite}Classification as White in
\emph{t+20 }(\emph{Mulatto--White} gap)]{
\centering{}\includegraphics[width=0.8\textwidth]{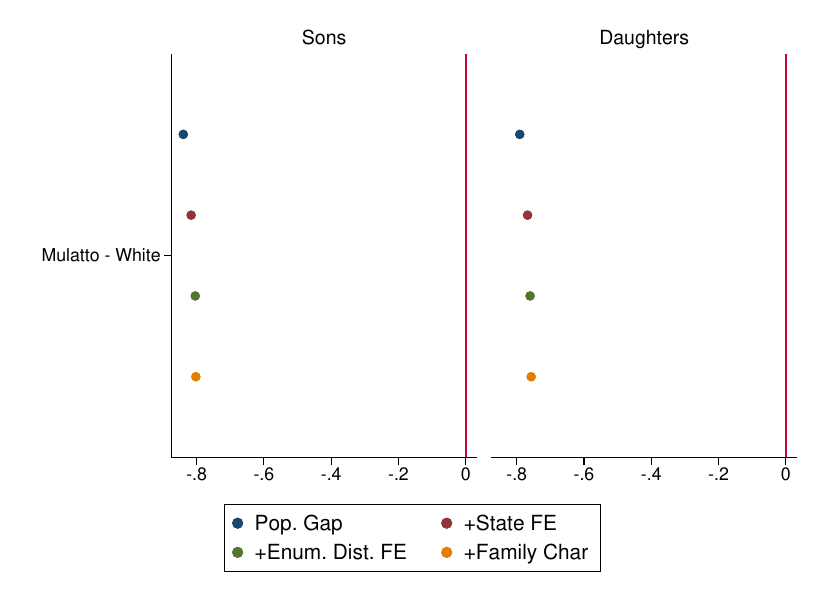}}
\par\end{centering}
\begin{centering}
\subfloat[\label{subfig:Swhite-plus20}Spouses' classification as White in \emph{t+20}]{
\begin{centering}
\includegraphics[width=0.49\textwidth]{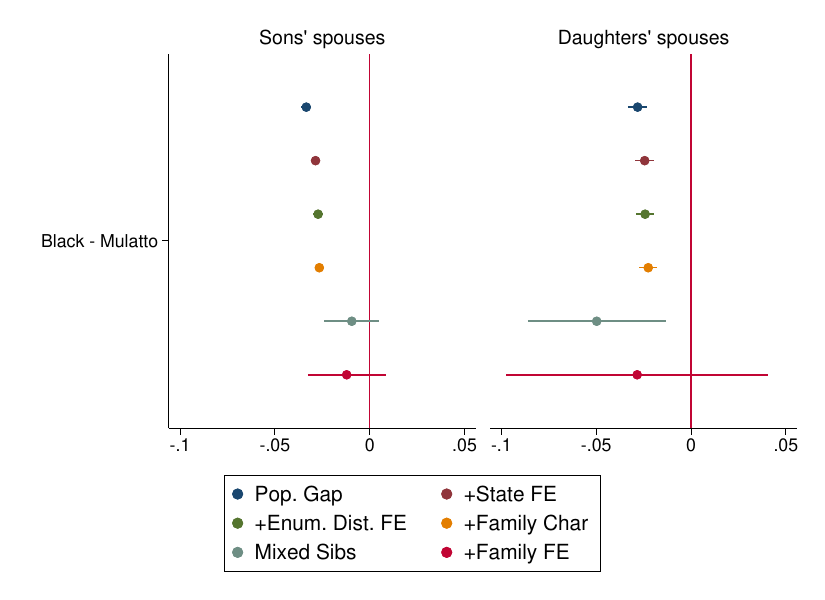}
\par\end{centering}
\centering{}\includegraphics[width=0.49\textwidth]{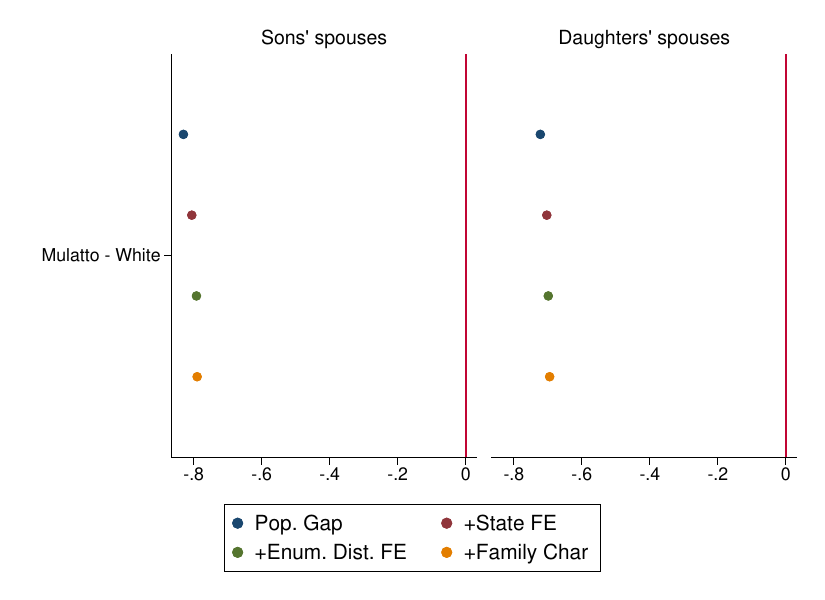}}
\par\end{centering}
{\small Note: Figure shows regression coefficients and 95\% confidence
intervals documenting gaps by perceived childhood race in an indicator
for being classified as a White adult twenty years later. Panel~A
shows the }{\small\emph{Mulatto--White}}{\small{} gap in the likelihood
that a child is him- or herself classified as White as an adult (see
Table~\ref{tab:white-plus20} for the analogous }{\small\emph{Black--Mulatto}}{\small{}
gap). Panel~B shows differences in the rate at which a child's spouse
is classified as White as an adult (showing both the }{\small\emph{Black--Mulatto}}{\small{}
and }{\small\emph{Mulatto--White}}{\small{} gaps in side-by-side plots
on different x-scales). See Sections \ref{subsec:white-adult} and
\ref{subsec:adult_educ_married} for detail and Tables \ref{tab:white-plus20}
and \ref{tab:Swhite-plus20} for corresponding regression tables.}{\small\par}
\end{figure}
\pagebreak{}
\begin{figure}[H]
\begin{centering}
\caption{\label{fig:migration}Gaps in migration by \emph{t+20}, by perceived
child race and gender}
\par\end{centering}
\begin{centering}
\subfloat[\label{subfig:mvstate-plus20}Different state in \emph{t+20} vs.\ \emph{t}]{
\centering{}\includegraphics[width=0.8\textwidth]{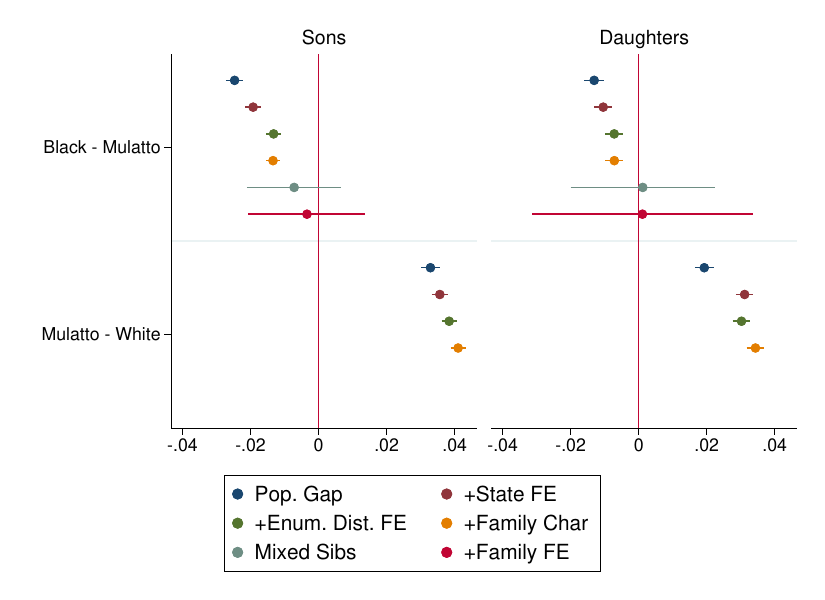}}
\par\end{centering}
\begin{centering}
\subfloat[\label{subfig:south-plus20}Lives in southern state in \emph{t+20}]{
\centering{}\includegraphics[width=0.8\textwidth]{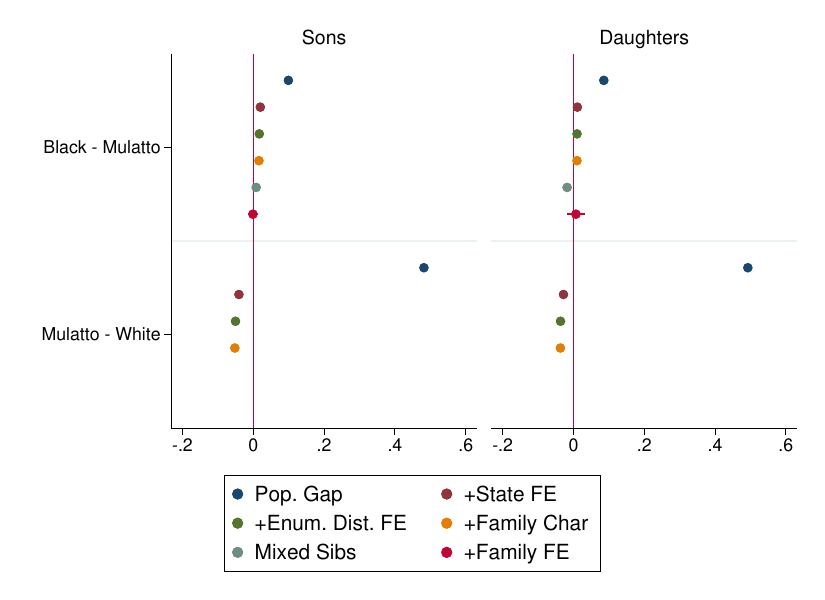}}
\par\end{centering}
{\small Note: Figure shows regression coefficients and 95\% confidence
intervals documenting gaps by perceived childhood race in migration
by }{\small\emph{t+20}}{\small . Panel~A shows differences in an indicator
for living in a state as an adult in year }{\small\emph{t+20}}{\small{}
that is different from the child's state in }{\small\emph{t}}{\small .
Panel~B shows differences in an indicator for living in a southern
state in }{\small\emph{t+20}}{\small . See Section~\ref{subsec:migration}
for detail and Tables \ref{tab:mvstate-plus20} and \ref{tab:south-plus20}
for corresponding regression tables.}{\small\par}
\end{figure}

\pagebreak{}
\begin{figure}[H]
\begin{centering}
\caption{\label{fig:age-robustness-sons}Comparison of \emph{Black--Mulatto}
gaps among sons, across child age groups}
\par\end{centering}
\begin{centering}
\includegraphics[width=1\textwidth]{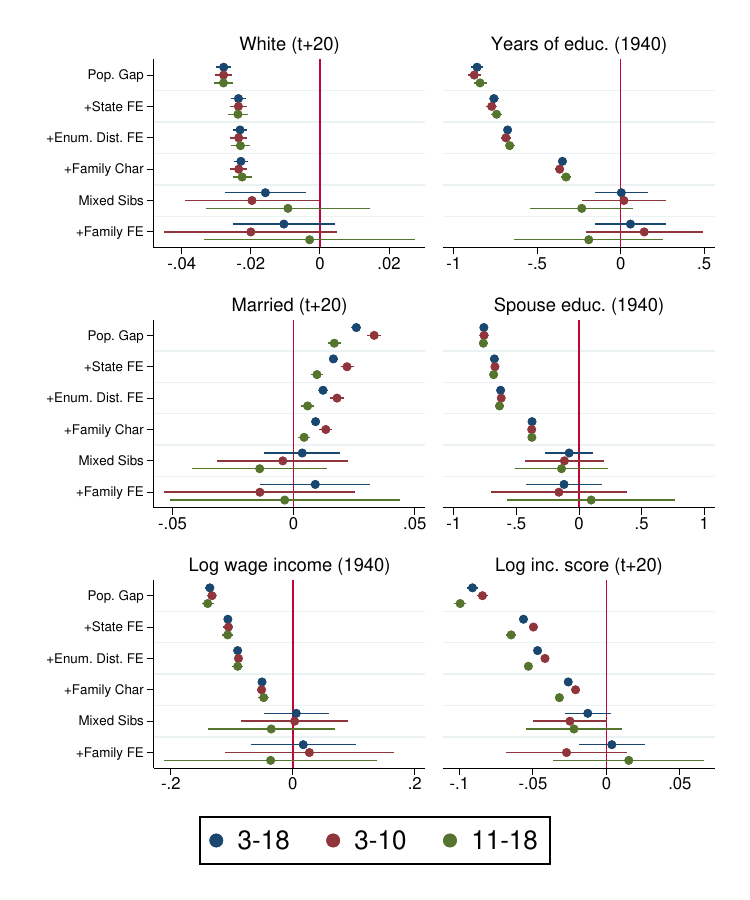}
\par\end{centering}
{\small Note: Figure shows regression coefficients and 95\% confidence
intervals documenting sons' }{\small\emph{Black--Mulatto}}{\small{}
gaps for each of our main adult outcomes, for each set of controls,
and for three samples. We repeat results using our baseline sample
(ages 3}--{\small 18 at classification, as shown in Figures~\ref{fig:child-education}--\ref{fig:income}
and as used throughout the paper unless otherwise noted), while adding
analogous results restricting to sons age 3}--{\small 10 or to 11}--{\small 18
at the time of classification. To be included in our differently-classified
(BMS) sample, two brothers must both be within the age restriction.}{\small\par}
\end{figure}
\pagebreak{}
\begin{figure}[H]
\begin{centering}
\caption{\label{fig:age-robustness-daughters}Comparison of \emph{Black--Mulatto}
gaps among daughters, across child age groups}
\par\end{centering}
\begin{centering}
\includegraphics[width=1\textwidth]{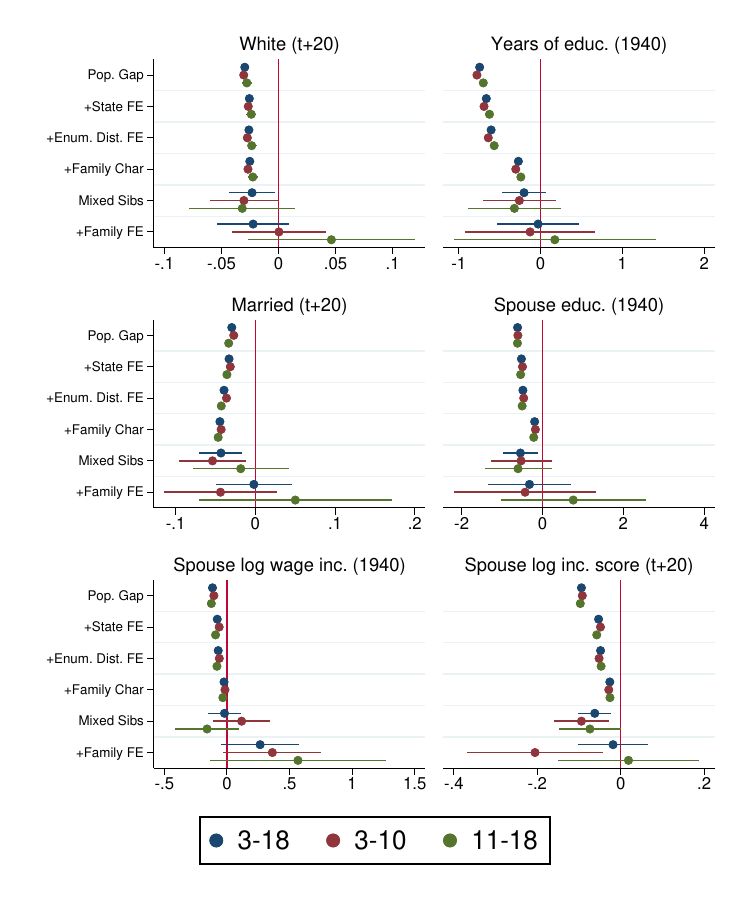}
\par\end{centering}
{\small Note: Figure shows regression coefficients and 95\% confidence
intervals documenting daughters' }{\small\emph{Black--Mulatto}}{\small{}
gaps for each of our main adult outcomes, for each set of controls,
and for three samples (as in Figure~\ref{fig:age-robustness-sons}).
We repeat results using our baseline sample (ages 3}--{\small 18 at
classification, as shown in Figures~\ref{fig:child-education}--\ref{fig:income}
and as used throughout the paper unless otherwise noted), while adding
analogous results restricting to daughters age 3}--{\small 10 or to
11}--{\small 18 at the time of classification. To be included in our
differently-classified (BMD) sample, two sisters must both be within
the age restriction.}{\small\par}
\end{figure}

\pagebreak{}
\begin{figure}[H]
\begin{centering}
\caption{\label{fig:share-mulatto-sons-byage}Proportion of sons coded as Mulatto,
by age}
\par\end{centering}
\begin{centering}
\subfloat[\label{subfig:all-share-mulatto-sons-byage}Proportion of sons coded
as Mulatto by age, among all sons coded as Black or Mulatto]{
\centering{}\includegraphics[width=0.75\textwidth]{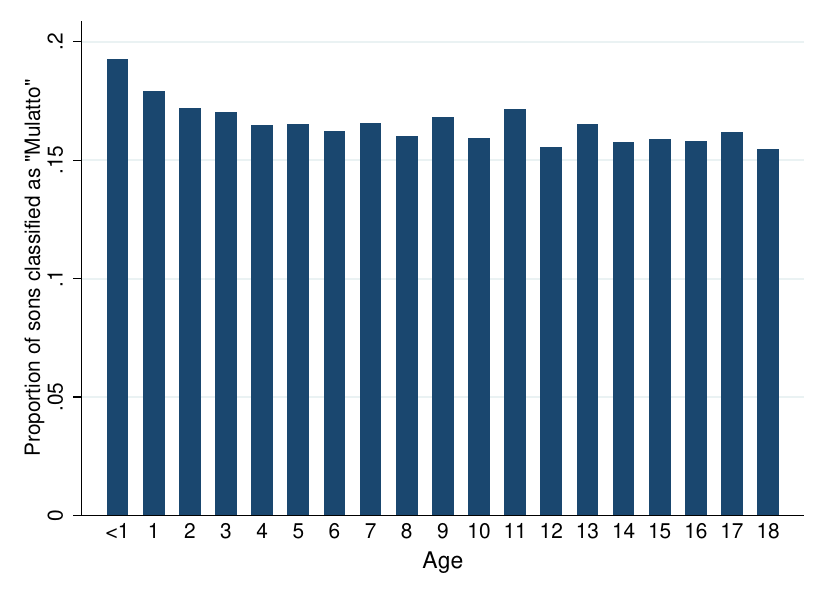}}
\par\end{centering}
\begin{centering}
\subfloat[\label{subfig:bms-share-mulatto-sons-byage}Proportion of sons coded
as Mulatto by age, among Black-and-Mulatto-sons (BMS) sample]{
\centering{}\includegraphics[width=0.75\textwidth]{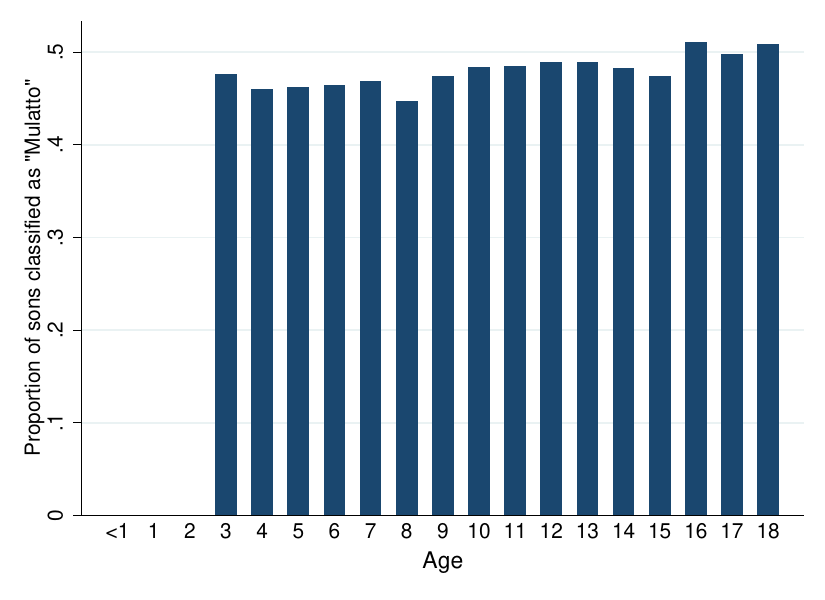}}
\par\end{centering}
{\small Note: Figure shows the share of African American sons classified
as Mulatto (vs. Black) by age. Panel~A shows these proportions among
all Black- and\slash or Mulatto-coded sons. Panel~B shows these proportions
among our sample of differently-classified sons (BMS) sample, where
at least one son in a family was classified as Black while his brother
was classified as Mulatto. See Section~\ref{subsec:child-cohorts}
for the complete definition of this sample.}{\small\par}
\end{figure}

%% file: appendix_tables.tex
\begin{sidewaystable}[H]
\begin{centering}
\caption{\label{tab:match-characteristics-daughters}Characteristics of linked
vs.\ all searched daughters}
\par\end{centering}
\begin{centering}
\input{input/Tables/match_characteristics_plus20_bw_combined_byuautossnapp_women.tex}
\par\end{centering}
{\small Note: Table shows differences in mean characteristics for our
initial and linked samples of daughters. Column (1) reports counts
and mean statistics among our initial sample prior to linking, including
children from 1870, 1880, 1910, and 1920. Columns (2) and (4) denote
mean characteristics under our Census Tree links when linking to }{\small\emph{t+20}}{\small ,
without and with inverse probability match weights, respectively (see
Section~\ref{subsec:Matching-across-censuses} for detail on linking
and weights). Columns (3) and (5) report differences between our initial
and linked means, without and with weights, respectively. Columns
(6)--(9) repeats the above for the Social Security Number application-based
linking algorithm. Standard errors are clustered by (enumeration district)$\times$(year).
See Table~\ref{tab:match-characteristics} for an analogous table
comparing the characteristics of linked vs.\ all searched sons.}{\small\par}
\end{sidewaystable}
\pagebreak{}
\begin{table}[H]
\begin{centering}
\caption{\label{tab:bms-match-rate-diff}Differences in match rate among differently-classified
siblings, by perceived child race and gender}
\par\end{centering}
\begin{centering}
\subfloat[\label{tab:bms-match-rate-diff-sons}Sons]{
\centering{}{\footnotesize\setlength{\tabcolsep}{2.5pt}\input{input/Tables/match_diff_mixed_sibs_plus20.tex}}}
\par\end{centering}
\begin{centering}
\subfloat[\label{tab:bms-match-rate-diff-daughters}Daughters]{
\centering{}{\footnotesize\setlength{\tabcolsep}{2.5pt}\input{input/Tables/match_diff_mixed_sibs_women_plus20.tex}}}
\par\end{centering}
{\small Note: Table shows differences in linking rates among Black-
vs.\ Mulatto-coded siblings in our differently-classified (BMS\slash BMD)
samples to }{\small\emph{t+20}}{\small , by linking algorithm. Observations
are weighted proportional to the inverse share of their perceived
race within a family (see Sections~\ref{sec:specifications} and
\ref{subsec:Matching-across-censuses} for detail on these weights
and linking procedures, respectively). Standard errors are clustered
by (enumeration district)$\times$(year). Panels A and B analyze differences
among sons and daughters, respectively.}{\small\par}
\end{table}

\pagebreak{}
\begin{table}[H]
\begin{centering}
\caption{\label{tab:white-ssn}Gaps in classification as White in SSN application,
by perceived child race and gender}
\subfloat[\label{tab:white-ssn-sons}Sons]{
\centering{}\resizebox{0.9\textwidth}{!}{\input{input/Tables/regressions_numwhitef_1940_byuauto_combined_black_mulatto2_wgtd_wfamfe.tex}}}
\par\end{centering}
\begin{centering}
\vspace{-0.3cm}\subfloat[\label{tab:white-ssn-daughters}Daughters]{
\centering{}\resizebox{0.9\textwidth}{!}{\input{input/Tables/regressions_women_numwhitef_1940_byuauto_combined_black_mulatto2_wgtd_wfamfe.tex}}}
\par\end{centering}
{\small Note: Table shows regressions measuring gaps in an indicator
for (self-reported) classification as White when an individual first
applied for a Social Security Number, by perceived child race among
children age 3--18 in 1910 and 1920. See Table~\ref{tab:white-plus20}
for detail regarding the empirical specification. Social Security
Number application data is provided and linked to 1940 U.S.\ decennial
census individuals by \citet{goldstein2021censoc}. Observations are
weighted by their inverse link probabilities, multiplying together
the inverse probability of linking an individual from their child
cohort to their 1940 adult self (as described in Section~\ref{subsec:Matching-across-censuses})
with the inverse probability of linking their 1940 adult self to a
Social Security Number application (as calculated by \citealp{goldstein2021censoc}).
Observations in which a child was identified as a race other than
White or Black as an adult are dropped. All controls come from the
child's census year, rather than from their linked adult self.}{\small\par}
\end{table}
\pagebreak{}
\begin{table}[H]
\begin{centering}
\caption{\label{tab:Swhite-plus20}Gaps in spouses' classification as White
in \emph{t+20}, by perceived child race and gender}
\subfloat[\label{tab:Swhite-plus20-sons}Sons' spouses]{
\centering{}\resizebox{0.9\textwidth}{!}{\input{input/Tables/regressions_Swhite_plus20_byuauto_combined_black_mulatto2_wgtd_wfamfe.tex}}}
\par\end{centering}
\begin{centering}
\vspace{-0.3cm}\subfloat[\label{tab:Swhite-plus20-daughters}Daughters' spouses]{
\centering{}\resizebox{0.9\textwidth}{!}{\input{input/Tables/regressions_women_Swhite_plus20_byuauto_combined_black_mulatto2_wgtd_wfamfe.tex}}}
\par\end{centering}
{\small Note: Table shows regressions measuring gaps in a child's spouse
being classified as White in }{\small\emph{t+20}}{\small , by perceived
child race among children age 3--18 in 1870, 1880, 1910, and 1920.
Columns (1)--(4) follow Equation~\ref{eq:naive-discrimination-estimation},
starting with age and birth order fixed effects and adding controls
for state fixed effects, enumeration district fixed effects, and family
characteristics in columns (2)--(4), respectively. See Footnote~\ref{fn:family-characteristics}
for detail regarding the set of family characteristics included as
controls. Column (5) implements Equation~\ref{eq:within-BMS-estimation}
by further adding BMS\slash BMD race interaction terms and weights
such that our }{\small\emph{Black--Mulatto}}{\small{} coefficient measures
differences among differently-classified siblings with approximately
equal weighted distributions of family characteristics. Panels A and
B analyze differences among sons' and daughters' spouses, respectively.
The bottom part of each panel presents observation counts and mean
outcome variables by perceived child race. All non-race controls are
interacted with cohort fixed effects. Observations are weighted by
their inverse link probabilities as described in Section~\ref{subsec:Matching-across-censuses}.
Standard errors are clustered by (enumeration district)$\times$(year).
Observations in which a child's spouse was identified as a race other
than White or Black as an adult are dropped. All controls come from
the child's census year, rather than from their linked adult self.
Restricts to married individuals.}{\small\par}
\end{table}

\pagebreak{}
\begin{table}[H]
\begin{centering}
\caption{\label{tab:loginctot-1940}Gaps in log total income in 1940, by perceived
child race}
\par\end{centering}
\begin{centering}
\subfloat[\label{tab:loginctot-1940-sons}Sons]{
\centering{}\resizebox{0.9\textwidth}{!}{\input{input/Tables/regressions_log_inctot_1940_byuauto_combined_black_mulatto2_wgtd_wfamfe.tex}}}
\par\end{centering}
\begin{centering}
\subfloat[\label{tab:loginctot-1940-daughters-spouse}Daughters' spouses]{
\centering{}\resizebox{0.9\textwidth}{!}{\input{input/Tables/regressions_women_Slog_inctot_1940_byuauto_combined_black_mulatto2_wgtd_wfamfe.tex}}}
\par\end{centering}
{\small Note: Table modifies Table~\ref{tab:loginc-1940} by combining
1940 wage income (as in the prior table) with estimated non-wage income
(not included in the prior table). Non-wage income is estimated as
described in Section~\ref{subsec:adult_income}.}{\small\par}
\end{table}

\pagebreak{}
\begin{table}[H]
\begin{centering}
\caption{\label{tab:mvstate-plus20}Gaps in living in different state in \emph{t+20}
vs.\ \emph{t}, by perceived child race and gender}
\par\end{centering}
\begin{centering}
\subfloat[\label{tab:mvstate-plus20-sons}Sons]{
\centering{}\resizebox{0.9\textwidth}{!}{\input{input/Tables/regressions_mvstate_plus20_byuauto_combined_black_mulatto2_wgtd_wfamfe.tex}}}
\par\end{centering}
\begin{centering}
\subfloat[\label{tab:mvstate-plus20-daughters}Daughters]{
\centering{}\resizebox{0.9\textwidth}{!}{\input{input/Tables/regressions_women_mvstate_plus20_byuauto_combined_black_mulatto2_wgtd_wfamfe.tex}}}
\par\end{centering}
{\small Note: Table shows regressions measuring gaps in an indicator
for living in a different state twenty years later, by perceived child
race among children age 3--18 in 1870, 1880, 1910, and 1920. Columns
(1)--(4) follow Equation~\ref{eq:naive-discrimination-estimation},
starting with age and birth order fixed effects and adding controls
for state fixed effects, enumeration district fixed effects, and family
characteristics in columns (2)--(4), respectively. See Footnote \ref{fn:family-characteristics}
for detail regarding the set of family characteristics included as
controls. Column (5) implements Equation~\ref{eq:within-BMS-estimation}
by further adding BMS\slash BMD race interaction terms and weights
such that our }{\small\emph{Black--Mulatto}}{\small{} coefficient measures
differences among differently-classified siblings with approximately
equal weighted distributions of family characteristics. Panels A and
B analyze differences among sons and daughters, respectively. The
bottom part of each panel presents observation counts and mean outcome
variables by perceived child race. All non-race controls are interacted
with cohort fixed effects. Observations are weighted by their inverse
link probabilities as described in Section~\ref{subsec:Matching-across-censuses}.
Standard errors are clustered by (enumeration district)$\times$(year).
All controls come from the child's census year, rather than from their
linked adult self.}{\small\par}
\end{table}
\pagebreak{}
\begin{table}[H]
\begin{centering}
\caption{\label{tab:south-plus20}Gaps in living in a southern state in \emph{t+20},
by perceived child race and gender}
\par\end{centering}
\begin{centering}
\subfloat[\label{tab:south-plus20-sons}Sons]{
\centering{}\resizebox{0.9\textwidth}{!}{\input{input/Tables/regressions_south_plus20_byuauto_combined_black_mulatto2_wgtd_wfamfe.tex}}}
\par\end{centering}
\begin{centering}
\subfloat[\label{tab:south-plus20-sons-daughters}Daughters]{
\centering{}\resizebox{0.9\textwidth}{!}{\input{input/Tables/regressions_women_south_plus20_byuauto_combined_black_mulatto2_wgtd_wfamfe.tex}}}
\par\end{centering}
{\small Note: Table shows regressions measuring gaps in an indicator
for living in a southern state twenty years later, by perceived child
race among children age 3--18 in 1870, 1880, 1910, and 1920. Columns
(1)--(4) follow Equation~\ref{eq:naive-discrimination-estimation},
starting with age and birth order fixed effects and adding controls
for state fixed effects, enumeration district fixed effects, and family
characteristics in columns (2)--(4), respectively. See Footnote \ref{fn:family-characteristics}
for detail regarding the set of family characteristics included as
controls. Column (5) implements Equation~\ref{eq:within-BMS-estimation}
by further adding BMS\slash BMD race interaction terms and weights
such that our }{\small\emph{Black--Mulatto}}{\small{} coefficient measures
differences among differently-classified siblings with approximately
equal weighted distributions of family characteristics. Panels A and
B analyze differences among sons and daughters, respectively. The
bottom part of each panel presents observation counts and mean outcome
variables by perceived child race. All non-race controls are interacted
with cohort fixed effects. Observations are weighted by their inverse
link probabilities as described in Section~\ref{subsec:Matching-across-censuses}.
Standard errors are clustered by (enumeration district)$\times$(year).
All controls come from the child's census year, rather than from their
linked adult self. Southern states are defined as: Texas, Louisiana,
Mississippi, Alabama, Georgia, South Carolina, Florida, North Carolina,
Tennessee, Virginia, Maryland, Arkansas, and the District of Columbia.}{\small\par}
\end{table}
\pagebreak{}
\begin{table}[H]
\begin{centering}
\caption{\label{tab:distinctive-names}Differences in distinctively African
American names among differently-classified siblings, by perceived
child race and gender}
\par\end{centering}
\begin{centering}
\resizebox{0.7\textwidth}{!}{\input{input/Tables/distinctivename_diff_mixed_sibs.tex}}
\par\end{centering}
{\small Note: Table shows differences in $\mathds{P}[Black\cup Mulatto|FirstName,Gender]$
among Black- vs.\ Mulatto-coded children in the general population
and in our differently-classified siblings (BMS\slash BMD) samples.
The dependent variable is calculated as the share of children who
are coded as Black or Mulatto among all same-gender children age 3--18
with the same first name in a child's census cohort (restricting to
those coded as Black, White, or Mulatto). Columns (1) and (2) show
differences among children in the general population. Columns (3)
and (4) show differences among BMS\slash BMD families, with observations
weighted proportional to the inverse share of their perceived race
within a family (see Section~\ref{sec:specifications} for detail
on these weights). Differences for sons are shown in columns (1) and
(3) while differences for daughters are shown in columns (2) and (4).
Standard errors are clustered by (enumeration district)$\times$(year).}{\small\par}
\end{table}

%% file: input/Tables/match_characteristics_plus20_bw_combined_byuautossnapp_women.tex
\begin{tabular}{l|c|cccc|cccc} \toprule  & All obs & \multicolumn{4}{|c}{Census Tree} & \multicolumn{4}{|c}{SSN Applications} \\ \midrule Observations (count) & \multicolumn{1}{|c}{49,034,333} & \multicolumn{4}{|c}{17,040,436} & \multicolumn{4}{|c}{4,112,990}  \\ Black+Mulatto in county & \multicolumn{1}{|c}{6,334,137} & \multicolumn{4}{|c}{683,480} & \multicolumn{4}{|c}{328,452}  \\ Black+Mulatto in family & \multicolumn{1}{|c}{53,762} & \multicolumn{4}{|c}{1,440} & \multicolumn{4}{|c}{289} \\ \midrule   &  & \multicolumn{2}{|c}{Unweighted} & \multicolumn{2}{c}{Weighted} & \multicolumn{2}{|c}{Unweighted} & \multicolumn{2}{c}{Weighted} \\  \cmidrule(lr){3-4} \cmidrule(lr){5-6} \cmidrule(lr){7-8} \cmidrule(lr){9-10} Child characteristics & Mean & Mean & Difference & Mean & Difference & Mean & Difference & Mean & Difference \\ \midrule  
& (1) & (2) & (3) & (4) & (5) & (6) & (7) & (8) & (9) \\ \midrule
Black & .106 & .0316 & -.0747*** & .0914 & -.015*** & .0645 & -.0419*** & .0993 & -.00707*** \\
& & & (.000215) & & (.000696) & & (.000415) & & (.000847) \\
Mulatto & .0231 & .00898 & -.0141*** & .0244 & .00129*** & .0173 & -.00584*** & .0251 & .00199*** \\
& & & (.0000782) & & (.000227) & & (.000154) & & (.000244) \\
Age & 10.2 & 10 & -.145*** & 10.2 & -.0204*** & 10.5 & .297*** & 10.2 & -.00624* \\
& & & (.00194) & & (.00213) & & (.0028) & & (.0032) \\
Able to write & .906 & .952 & .0464*** & .921 & .0154*** & .958 & .0521*** & .915 & .00882*** \\
& & & (.000424) & & (.000751) & & (.000351) & & (.000935) \\
Daughters in household & 2.12 & 2.19 & .0787*** & 2.13 & .0131*** & 2.13 & .0105*** & 2.12 & .00521*** \\
& & & (.00108) & & (.00123) & & (.00103) & & (.00143) \\
Head is literate & .854 & .919 & .0651*** & .864 & .0102*** & .906 & .0525*** & .858 & .00413*** \\
& & & (.000365) & & (.000707) & & (.00039) & & (.000837) \\ Head's occupation: & & & & & & & & & \\
- Farm laborer & .0487 & .0319 & -.0169*** & .0434 & -.0053*** & .0375 & -.0113*** & .0469 & -.00186*** \\
& & & (.000188) & & (.000343) & & (.000232) & & (.000496) \\
- Farmer & .368 & .409 & .0411*** & .369 & .00104 & .37 & .00151 & .37 & .00137 \\
& & & (.0012) & & (.00156) & & (.00111) & & (.00192) \\
- Laborer (not farm) & .102 & .0826 & -.0198*** & .103 & .000498 & .0967 & -.00575*** & .102 & -.0000637 \\
& & & (.000267) & & (.000396) & & (.000327) & & (.000491) \\
County urban & .376 & .367 & -.00965*** & .362 & -.0145*** & .408 & .0313*** & .366 & -.0105*** \\
& & & (.0014) & & (.0022) & & (.00125) & & (.00291) \\
Census Year & 1902 & 1904 & 2.3*** & 1902 & .581*** & 1910 & 8.78*** & 1902 & .234 \\
& & & (.108) & & (.17) & & (.0786) & & (.195) \\
\bottomrule \end{tabular}

%% file: input/Tables/match_diff_mixed_sibs_plus20.tex
{
\def\sym#1{\ifmmode^{#1}\else\(^{#1}\)\fi}
\begin{tabular}{l*{6}{c}}
\toprule
                &\multicolumn{1}{c}{(1)}&\multicolumn{1}{c}{(2)}&\multicolumn{1}{c}{(3)}&\multicolumn{1}{c}{(4)}&\multicolumn{1}{c}{(5)}&\multicolumn{1}{c}{(6)}\\
                &\multicolumn{1}{c}{Census Tree}&\multicolumn{1}{c}{Genealogy}&\multicolumn{1}{c}{SSN-Apps}&\multicolumn{1}{c}{Alt Census Tree}&\multicolumn{1}{c}{IPUMS MLP}&\multicolumn{1}{c}{ABE Ex. Cons.}\\
\midrule
Black - Mulatto &   -0.002         &    0.000         &   -0.002         &    0.008\sym{*}  &    0.009\sym{***}&    0.001         \\
                &  (0.004)         &  (0.002)         &  (0.002)         &  (0.004)         &  (0.003)         &  (0.002)         \\
Constant        &    0.277\sym{***}&    0.045\sym{***}&    0.059\sym{***}&    0.375\sym{***}&    0.185\sym{***}&    0.074\sym{***}\\
                &  (0.003)         &  (0.002)         &  (0.002)         &  (0.004)         &  (0.003)         &  (0.002)         \\
\midrule
Observations    &   50,333         &   50,333         &   50,333         &   50,333         &   50,333         &   50,333         \\
R-Squared       &    0.000         &    0.000         &    0.000         &    0.000         &    0.000         &    0.000         \\
Count Black     &   26,427         &   26,427         &   26,427         &   26,427         &   26,427         &   26,427         \\
Count Mulatto   &   23,906         &   23,906         &   23,906         &   23,906         &   23,906         &   23,906         \\
Mean[Outcome $ | $ Black]&    0.275         &    0.046         &    0.057         &    0.383         &    0.194         &    0.075         \\
Mean[Outcome $ | $ Mulatto]&    0.277         &    0.045         &    0.059         &    0.375         &    0.185         &    0.074         \\
\bottomrule
\end{tabular}
}

%% file: input/Tables/match_diff_mixed_sibs_women_plus20.tex
{
\def\sym#1{\ifmmode^{#1}\else\(^{#1}\)\fi}
\begin{tabular}{l*{5}{c}}
\toprule
                &\multicolumn{1}{c}{(1)}&\multicolumn{1}{c}{(2)}&\multicolumn{1}{c}{(3)}&\multicolumn{1}{c}{(4)}&\multicolumn{1}{c}{(5)}\\
                &\multicolumn{1}{c}{Census Tree}&\multicolumn{1}{c}{Genealogy}&\multicolumn{1}{c}{SSN-Apps}&\multicolumn{1}{c}{Alt Census Tree}&\multicolumn{1}{c}{IPUMS MLP}\\
\midrule
Black - Mulatto &    0.001         &   -0.004\sym{***}&   -0.001         &    0.002         &    0.006\sym{***}\\
                &  (0.003)         &  (0.001)         &  (0.002)         &  (0.003)         &  (0.002)         \\
Constant        &    0.106\sym{***}&    0.033\sym{***}&    0.048\sym{***}&    0.123\sym{***}&    0.058\sym{***}\\
                &  (0.002)         &  (0.001)         &  (0.002)         &  (0.002)         &  (0.002)         \\
\midrule
Observations    &   53,762         &   53,762         &   53,762         &   53,762         &   53,762         \\
R-Squared       &    0.000         &    0.000         &    0.000         &    0.000         &    0.000         \\
Count Black     &   27,958         &   27,958         &   27,958         &   27,958         &   27,958         \\
Count Mulatto   &   25,804         &   25,804         &   25,804         &   25,804         &   25,804         \\
Mean[Outcome $ | $ Black]&    0.107         &    0.029         &    0.047         &    0.124         &    0.064         \\
Mean[Outcome $ | $ Mulatto]&    0.106         &    0.033         &    0.048         &    0.123         &    0.058         \\
\bottomrule
\multicolumn{6}{l}{\footnotesize Standard errors in parentheses}\\
\multicolumn{6}{l}{\footnotesize \sym{*} \(p<0.10\), \sym{**} \(p<0.05\), \sym{***} \(p<0.01\)}\\
\end{tabular}
}

%% file: input/Tables/regressions_numwhitef_1940_byuauto_combined_black_mulatto2_wgtd_wfamfe.tex
{
\def\sym#1{\ifmmode^{#1}\else\(^{#1}\)\fi}
\begin{tabular}{l*{6}{c}}
\toprule
                &\multicolumn{1}{c}{(1)}         &\multicolumn{1}{c}{(2)}         &\multicolumn{1}{c}{(3)}         &\multicolumn{1}{c}{(4)}         &\multicolumn{1}{c}{(5)}         &\multicolumn{1}{c}{(6)}         \\
\midrule
Black - Mulatto &   -0.016\sym{**} &   -0.013\sym{**} &   -0.006         &   -0.007         &   -0.054         &    0.001         \\
                &  (0.006)         &  (0.006)         &  (0.007)         &  (0.007)         &  (0.071)         &  (0.007)         \\
Mulatto - White &   -0.860\sym{***}&   -0.822\sym{***}&   -0.829\sym{***}&   -0.830\sym{***}&                  &                  \\
                &  (0.006)         &  (0.006)         &  (0.007)         &  (0.007)         &                  &                  \\
(Year)X(Age+Birth Order) FE &      Yes         &      Yes         &      Yes         &      Yes         &      Yes         &      Yes         \\
(Year)X(State) FE &       No         &      Yes         &       No         &       No         &       No         &       No         \\
(Year)X(Enum. Dist.) FE &       No         &       No         &      Yes         &      Yes         &      Yes         &       No         \\
(Year)X(Fam. Char.) &       No         &       No         &       No         &      Yes         &      Yes         &       No         \\
BMS Interactions &       No         &       No         &       No         &       No         &      Yes         &       No         \\
(Year)X(Fam.) FE &       No         &       No         &       No         &       No         &       No         &      Yes         \\
\midrule
Observations    &  565,541         &  565,541         &  540,013         &  540,011         &  540,011         &   30,073         \\
R-Squared       &    0.779         &    0.784         &    0.855         &    0.855         &    0.855         &    0.950         \\
Count Black     &   20,244         &   20,244         &   18,651         &   18,651         &   18,651         &      607         \\
Count Mulatto   &    5,229         &    5,229         &    4,789         &    4,789         &    4,789         &      179         \\
Count White     &  540,068         &  540,068         &  516,573         &  516,571         &  516,571         &   29,287         \\
Mean[Outcome $ | $ Black]&    0.111         &    0.111         &    0.108         &    0.108         &    0.108         &    0.056         \\
Mean[Outcome $ | $ Mulatto]&    0.127         &    0.127         &    0.120         &    0.120         &    0.121         &    0.080         \\
Mean[Outcome $ | $ White]&    0.987         &    0.987         &    0.988         &    0.988         &    0.988         &    0.994         \\
Count Mixed Siblings&      183         &      183         &      172         &      172         &      172         &        4         \\
\bottomrule
\end{tabular}
}

%% file: input/Tables/regressions_women_numwhitef_1940_byuauto_combined_black_mulatto2_wgtd_wfamfe.tex
{
\def\sym#1{\ifmmode^{#1}\else\(^{#1}\)\fi}
\begin{tabular}{l*{6}{c}}
\toprule
                &\multicolumn{1}{c}{(1)}         &\multicolumn{1}{c}{(2)}         &\multicolumn{1}{c}{(3)}         &\multicolumn{1}{c}{(4)}         &\multicolumn{1}{c}{(5)}         &\multicolumn{1}{c}{(6)}         \\
\midrule
Black - Mulatto &   -0.011         &   -0.007         &   -0.004         &   -0.005         &    0.100         &   -1.000\sym{***}\\
                &  (0.009)         &  (0.009)         &  (0.011)         &  (0.011)         &  (0.100)         &  (0.010)         \\
Mulatto - White &   -0.655\sym{***}&   -0.613\sym{***}&   -0.609\sym{***}&   -0.612\sym{***}&                  &                  \\
                &  (0.008)         &  (0.008)         &  (0.010)         &  (0.010)         &                  &                  \\
(Year)X(Age+Birth Order) FE &      Yes         &      Yes         &      Yes         &      Yes         &      Yes         &      Yes         \\
(Year)X(State) FE &       No         &      Yes         &       No         &       No         &       No         &       No         \\
(Year)X(Enum. Dist.) FE &       No         &       No         &      Yes         &      Yes         &      Yes         &       No         \\
(Year)X(Fam. Char.) &       No         &       No         &       No         &      Yes         &      Yes         &       No         \\
BMD Interactions &       No         &       No         &       No         &       No         &      Yes         &       No         \\
(Year)X(Fam.) FE &       No         &       No         &       No         &       No         &       No         &      Yes         \\
\midrule
Observations    &  658,518         &  658,518         &  635,810         &  635,808         &  635,808         &   27,478         \\
R-Squared       &    0.534         &    0.544         &    0.703         &    0.704         &    0.704         &    0.847         \\
Count Black     &   14,753         &   14,753         &   13,787         &   13,787         &   13,787         &      240         \\
Count Mulatto   &    4,103         &    4,103         &    3,837         &    3,837         &    3,837         &       93         \\
Count White     &  639,662         &  639,662         &  618,186         &  618,184         &  618,184         &   27,145         \\
Mean[Outcome $ | $ Black]&    0.317         &    0.317         &    0.314         &    0.314         &    0.314         &    0.222         \\
Mean[Outcome $ | $ Mulatto]&    0.329         &    0.329         &    0.323         &    0.323         &    0.323         &    0.297         \\
Mean[Outcome $ | $ White]&    0.984         &    0.984         &    0.984         &    0.984         &    0.984         &    0.988         \\
Count Mixed Siblings&      112         &      112         &      102         &      102         &      102         &        2         \\
\bottomrule
\multicolumn{7}{l}{\footnotesize Standard errors in parentheses}\\
\multicolumn{7}{l}{\footnotesize \sym{*} \(p<0.10\), \sym{**} \(p<0.05\), \sym{***} \(p<0.01\)}\\
\end{tabular}
}

%% file: input/Tables/regressions_Swhite_plus20_byuauto_combined_black_mulatto2_wgtd_wfamfe.tex
{
\def\sym#1{\ifmmode^{#1}\else\(^{#1}\)\fi}
\begin{tabular}{l*{6}{c}}
\toprule
                &\multicolumn{1}{c}{(1)}         &\multicolumn{1}{c}{(2)}         &\multicolumn{1}{c}{(3)}         &\multicolumn{1}{c}{(4)}         &\multicolumn{1}{c}{(5)}         &\multicolumn{1}{c}{(6)}         \\
\midrule
Black - Mulatto &   -0.033\sym{***}&   -0.029\sym{***}&   -0.027\sym{***}&   -0.027\sym{***}&   -0.009         &   -0.012         \\
                &  (0.001)         &  (0.001)         &  (0.001)         &  (0.001)         &  (0.007)         &  (0.010)         \\
Mulatto - White &   -0.830\sym{***}&   -0.806\sym{***}&   -0.792\sym{***}&   -0.790\sym{***}&                  &                  \\
                &  (0.001)         &  (0.001)         &  (0.001)         &  (0.002)         &                  &                  \\
(Year)X(Age+Birth Order) FE &      Yes         &      Yes         &      Yes         &      Yes         &      Yes         &      Yes         \\
(Year)X(State) FE &       No         &      Yes         &       No         &       No         &       No         &       No         \\
(Year)X(Enum. Dist.) FE &       No         &       No         &      Yes         &      Yes         &      Yes         &       No         \\
(Year)X(Fam. Char.) &       No         &       No         &       No         &      Yes         &      Yes         &       No         \\
BMS Interactions &       No         &       No         &       No         &       No         &      Yes         &       No         \\
(Year)X(Fam.) FE &       No         &       No         &       No         &       No         &       No         &      Yes         \\
\midrule
Observations    &17,005,106         &17,005,106         &17,004,528         &17,004,528         &17,004,528         &6,040,344         \\
R-Squared       &    0.791         &    0.794         &    0.799         &    0.799         &    0.799         &    0.928         \\
Count Black     &  829,589         &  829,589         &  829,567         &  829,567         &  829,567         &  173,824         \\
Count Mulatto   &  195,950         &  195,950         &  195,943         &  195,943         &  195,943         &   44,870         \\
Count White     &15,979,567         &15,979,567         &15,979,018         &15,979,018         &15,979,018         &5,821,650         \\
Mean[Outcome $ | $ Black]&    0.128         &    0.128         &    0.128         &    0.128         &    0.128         &    0.083         \\
Mean[Outcome $ | $ Mulatto]&    0.162         &    0.162         &    0.162         &    0.162         &    0.162         &    0.133         \\
Mean[Outcome $ | $ White]&    0.992         &    0.992         &    0.992         &    0.992         &    0.992         &    0.995         \\
Count Mixed Siblings&    9,103         &    9,103         &    9,103         &    9,103         &    9,103         &    2,775         \\
\bottomrule
\end{tabular}
}

%% file: input/Tables/regressions_women_Swhite_plus20_byuauto_combined_black_mulatto2_wgtd_wfamfe.tex
{
\def\sym#1{\ifmmode^{#1}\else\(^{#1}\)\fi}
\begin{tabular}{l*{6}{c}}
\toprule
                &\multicolumn{1}{c}{(1)}         &\multicolumn{1}{c}{(2)}         &\multicolumn{1}{c}{(3)}         &\multicolumn{1}{c}{(4)}         &\multicolumn{1}{c}{(5)}         &\multicolumn{1}{c}{(6)}         \\
\midrule
Black - Mulatto &   -0.028\sym{***}&   -0.024\sym{***}&   -0.024\sym{***}&   -0.023\sym{***}&   -0.050\sym{***}&   -0.028         \\
                &  (0.003)         &  (0.003)         &  (0.002)         &  (0.002)         &  (0.019)         &  (0.035)         \\
Mulatto - White &   -0.722\sym{***}&   -0.703\sym{***}&   -0.699\sym{***}&   -0.695\sym{***}&                  &                  \\
                &  (0.003)         &  (0.002)         &  (0.002)         &  (0.002)         &                  &                  \\
(Year)X(Age+Birth Order) FE &      Yes         &      Yes         &      Yes         &      Yes         &      Yes         &      Yes         \\
(Year)X(State) FE &       No         &      Yes         &       No         &       No         &       No         &       No         \\
(Year)X(Enum. Dist.) FE &       No         &       No         &      Yes         &      Yes         &      Yes         &       No         \\
(Year)X(Fam. Char.) &       No         &       No         &       No         &      Yes         &      Yes         &       No         \\
BMD Interactions &       No         &       No         &       No         &       No         &      Yes         &       No         \\
(Year)X(Fam.) FE &       No         &       No         &       No         &       No         &       No         &      Yes         \\
\midrule
Observations    &10,854,275         &10,854,275         &10,853,400         &10,853,398         &10,853,398         &3,327,409         \\
R-Squared       &    0.672         &    0.676         &    0.697         &    0.698         &    0.698         &    0.885         \\
Count Black     &  217,810         &  217,810         &  217,773         &  217,773         &  217,773         &   17,083         \\
Count Mulatto   &   67,048         &   67,048         &   67,034         &   67,033         &   67,033         &    8,244         \\
Count White     &10,569,417         &10,569,417         &10,568,593         &10,568,592         &10,568,592         &3,302,082         \\
Mean[Outcome $ | $ Black]&    0.244         &    0.244         &    0.244         &    0.244         &    0.244         &    0.174         \\
Mean[Outcome $ | $ Mulatto]&    0.273         &    0.273         &    0.273         &    0.273         &    0.273         &    0.291         \\
Mean[Outcome $ | $ White]&    0.995         &    0.995         &    0.995         &    0.995         &    0.995         &    0.997         \\
Count Mixed Siblings&    2,239         &    2,239         &    2,239         &    2,239         &    2,239         &      321         \\
\bottomrule
\multicolumn{7}{l}{\footnotesize Standard errors in parentheses}\\
\multicolumn{7}{l}{\footnotesize \sym{*} \(p<0.10\), \sym{**} \(p<0.05\), \sym{***} \(p<0.01\)}\\
\end{tabular}
}

%% file: input/Tables/regressions_log_inctot_1940_byuauto_combined_black_mulatto2_wgtd_wfamfe.tex
{
\def\sym#1{\ifmmode^{#1}\else\(^{#1}\)\fi}
\begin{tabular}{l*{6}{c}}
\toprule
                &\multicolumn{1}{c}{(1)}         &\multicolumn{1}{c}{(2)}         &\multicolumn{1}{c}{(3)}         &\multicolumn{1}{c}{(4)}         &\multicolumn{1}{c}{(5)}         &\multicolumn{1}{c}{(6)}         \\
\midrule
Black - Mulatto &   -0.102\sym{***}&   -0.075\sym{***}&   -0.065\sym{***}&   -0.033\sym{***}&    0.016         &    0.052         \\
                &  (0.003)         &  (0.003)         &  (0.003)         &  (0.002)         &  (0.021)         &  (0.032)         \\
Mulatto - White &   -0.598\sym{***}&   -0.397\sym{***}&   -0.440\sym{***}&   -0.357\sym{***}&                  &                  \\
                &  (0.003)         &  (0.003)         &  (0.003)         &  (0.003)         &                  &                  \\
(Year)X(Age+Birth Order) FE &      Yes         &      Yes         &      Yes         &      Yes         &      Yes         &      Yes         \\
(Year)X(State) FE &       No         &      Yes         &       No         &       No         &       No         &       No         \\
(Year)X(Enum. Dist.) FE &       No         &       No         &      Yes         &      Yes         &      Yes         &       No         \\
(Year)X(Fam. Char.) &       No         &       No         &       No         &      Yes         &      Yes         &       No         \\
BMS Interactions &       No         &       No         &       No         &       No         &      Yes         &       No         \\
(Year)X(Fam.) FE &       No         &       No         &       No         &       No         &       No         &      Yes         \\
\midrule
Observations    &15,009,125         &15,009,125         &15,008,698         &15,008,698         &15,008,698         &5,915,738         \\
R-Squared       &    0.094         &    0.139         &    0.198         &    0.208         &    0.208         &    0.629         \\
Count Black     &  672,091         &  672,091         &  672,075         &  672,075         &  672,075         &  169,435         \\
Count Mulatto   &  180,317         &  180,317         &  180,313         &  180,313         &  180,313         &   50,273         \\
Count White     &14,156,717         &14,156,717         &14,156,310         &14,156,310         &14,156,310         &5,696,030         \\
Mean[Outcome $ | $ Black]&    6.129         &    6.129         &    6.129         &    6.129         &    6.129         &    6.090         \\
Mean[Outcome $ | $ Mulatto]&    6.244         &    6.244         &    6.244         &    6.244         &    6.244         &    6.219         \\
Mean[Outcome $ | $ White]&    6.825         &    6.825         &    6.825         &    6.825         &    6.825         &    6.795         \\
Count Mixed Siblings&    5,718         &    5,718         &    5,718         &    5,718         &    5,718         &    2,163         \\
\bottomrule
\end{tabular}
}

%% file: input/Tables/regressions_women_Slog_inctot_1940_byuauto_combined_black_mulatto2_wgtd_wfamfe.tex
{
\def\sym#1{\ifmmode^{#1}\else\(^{#1}\)\fi}
\begin{tabular}{l*{6}{c}}
\toprule
                &\multicolumn{1}{c}{(1)}         &\multicolumn{1}{c}{(2)}         &\multicolumn{1}{c}{(3)}         &\multicolumn{1}{c}{(4)}         &\multicolumn{1}{c}{(5)}         &\multicolumn{1}{c}{(6)}         \\
\midrule
Black - Mulatto &   -0.084\sym{***}&   -0.047\sym{***}&   -0.042\sym{***}&   -0.009\sym{**} &   -0.028         &    0.187\sym{*}  \\
                &  (0.005)         &  (0.005)         &  (0.005)         &  (0.005)         &  (0.049)         &  (0.107)         \\
Mulatto - White &   -0.476\sym{***}&   -0.273\sym{***}&   -0.330\sym{***}&   -0.250\sym{***}&                  &                  \\
                &  (0.005)         &  (0.004)         &  (0.004)         &  (0.004)         &                  &                  \\
(Year)X(Age+Birth Order) FE &      Yes         &      Yes         &      Yes         &      Yes         &      Yes         &      Yes         \\
(Year)X(State) FE &       No         &      Yes         &       No         &       No         &       No         &       No         \\
(Year)X(Enum. Dist.) FE &       No         &       No         &      Yes         &      Yes         &      Yes         &       No         \\
(Year)X(Fam. Char.) &       No         &       No         &       No         &      Yes         &      Yes         &       No         \\
BMD Interactions &       No         &       No         &       No         &       No         &      Yes         &       No         \\
(Year)X(Fam.) FE &       No         &       No         &       No         &       No         &       No         &      Yes         \\
\midrule
Observations    &6,617,097         &6,617,097         &6,616,230         &6,616,230         &6,616,230         &1,724,642         \\
R-Squared       &    0.044         &    0.104         &    0.183         &    0.195         &    0.195         &    0.600         \\
Count Black     &  142,804         &  142,804         &  142,768         &  142,768         &  142,768         &    9,645         \\
Count Mulatto   &   44,959         &   44,959         &   44,945         &   44,945         &   44,945         &    4,731         \\
Count White     &6,429,334         &6,429,334         &6,428,517         &6,428,517         &6,428,517         &1,710,266         \\
Mean[Outcome $ | $ Black]&    6.334         &    6.334         &    6.334         &    6.334         &    6.334         &    6.282         \\
Mean[Outcome $ | $ Mulatto]&    6.426         &    6.426         &    6.426         &    6.426         &    6.426         &    6.453         \\
Mean[Outcome $ | $ White]&    6.895         &    6.895         &    6.895         &    6.895         &    6.895         &    6.850         \\
Count Mixed Siblings&    1,155         &    1,155         &    1,155         &    1,155         &    1,155         &      135         \\
\bottomrule
\multicolumn{7}{l}{\footnotesize Standard errors in parentheses}\\
\multicolumn{7}{l}{\footnotesize \sym{*} \(p<0.10\), \sym{**} \(p<0.05\), \sym{***} \(p<0.01\)}\\
\end{tabular}
}

%% file: input/Tables/regressions_mvstate_plus20_byuauto_combined_black_mulatto2_wgtd_wfamfe.tex
{
\def\sym#1{\ifmmode^{#1}\else\(^{#1}\)\fi}
\begin{tabular}{l*{6}{c}}
\toprule
                &\multicolumn{1}{c}{(1)}         &\multicolumn{1}{c}{(2)}         &\multicolumn{1}{c}{(3)}         &\multicolumn{1}{c}{(4)}         &\multicolumn{1}{c}{(5)}         &\multicolumn{1}{c}{(6)}         \\
\midrule
Black - Mulatto &   -0.025\sym{***}&   -0.019\sym{***}&   -0.013\sym{***}&   -0.013\sym{***}&   -0.007         &   -0.003         \\
                &  (0.001)         &  (0.001)         &  (0.001)         &  (0.001)         &  (0.007)         &  (0.009)         \\
Mulatto - White &    0.033\sym{***}&    0.036\sym{***}&    0.038\sym{***}&    0.041\sym{***}&                  &                  \\
                &  (0.001)         &  (0.001)         &  (0.001)         &  (0.001)         &                  &                  \\
(Year)X(Age+Birth Order) FE &      Yes         &      Yes         &      Yes         &      Yes         &      Yes         &      Yes         \\
(Year)X(State) FE &       No         &      Yes         &       No         &       No         &       No         &       No         \\
(Year)X(Enum. Dist.) FE &       No         &       No         &      Yes         &      Yes         &      Yes         &       No         \\
(Year)X(Fam. Char.) &       No         &       No         &       No         &      Yes         &      Yes         &       No         \\
BMS Interactions &       No         &       No         &       No         &       No         &      Yes         &       No         \\
(Year)X(Fam.) FE &       No         &       No         &       No         &       No         &       No         &      Yes         \\
\midrule
Observations    &25,842,325         &25,842,325         &25,841,935         &25,841,935         &25,841,935         &11,497,769         \\
R-Squared       &    0.007         &    0.043         &    0.087         &    0.094         &    0.094         &    0.696         \\
Count Black     &1,263,923         &1,263,923         &1,263,911         &1,263,911         &1,263,911         &  360,022         \\
Count Mulatto   &  314,335         &  314,335         &  314,330         &  314,330         &  314,330         &  100,041         \\
Count White     &24,264,067         &24,264,067         &24,263,694         &24,263,694         &24,263,694         &11,037,706         \\
Mean[Outcome $ | $ Black]&    0.226         &    0.226         &    0.226         &    0.226         &    0.226         &    0.180         \\
Mean[Outcome $ | $ Mulatto]&    0.252         &    0.252         &    0.252         &    0.252         &    0.251         &    0.210         \\
Mean[Outcome $ | $ White]&    0.219         &    0.219         &    0.219         &    0.219         &    0.219         &    0.199         \\
Count Mixed Siblings&   13,974         &   13,974         &   13,974         &   13,974         &   13,974         &    6,104         \\
\bottomrule
\end{tabular}
}

%% file: input/Tables/regressions_women_mvstate_plus20_byuauto_combined_black_mulatto2_wgtd_wfamfe.tex
{
\def\sym#1{\ifmmode^{#1}\else\(^{#1}\)\fi}
\begin{tabular}{l*{6}{c}}
\toprule
                &\multicolumn{1}{c}{(1)}         &\multicolumn{1}{c}{(2)}         &\multicolumn{1}{c}{(3)}         &\multicolumn{1}{c}{(4)}         &\multicolumn{1}{c}{(5)}         &\multicolumn{1}{c}{(6)}         \\
\midrule
Black - Mulatto &   -0.013\sym{***}&   -0.010\sym{***}&   -0.007\sym{***}&   -0.007\sym{***}&    0.001         &    0.001         \\
                &  (0.001)         &  (0.001)         &  (0.001)         &  (0.001)         &  (0.011)         &  (0.017)         \\
Mulatto - White &    0.019\sym{***}&    0.031\sym{***}&    0.030\sym{***}&    0.034\sym{***}&                  &                  \\
                &  (0.001)         &  (0.001)         &  (0.001)         &  (0.001)         &                  &                  \\
(Year)X(Age+Birth Order) FE &      Yes         &      Yes         &      Yes         &      Yes         &      Yes         &      Yes         \\
(Year)X(State) FE &       No         &      Yes         &       No         &       No         &       No         &       No         \\
(Year)X(Enum. Dist.) FE &       No         &       No         &      Yes         &      Yes         &      Yes         &       No         \\
(Year)X(Fam. Char.) &       No         &       No         &       No         &      Yes         &      Yes         &       No         \\
BMD Interactions &       No         &       No         &       No         &       No         &      Yes         &       No         \\
(Year)X(Fam.) FE &       No         &       No         &       No         &       No         &       No         &      Yes         \\
\midrule
Observations    &17,040,433         &17,040,433         &17,039,884         &17,039,883         &17,039,883         &6,277,543         \\
R-Squared       &    0.003         &    0.044         &    0.092         &    0.097         &    0.097         &    0.701         \\
Count Black     &  538,194         &  538,194         &  538,173         &  538,173         &  538,173         &   90,979         \\
Count Mulatto   &  153,073         &  153,073         &  153,066         &  153,066         &  153,066         &   30,905         \\
Count White     &16,349,166         &16,349,166         &16,348,645         &16,348,644         &16,348,644         &6,155,659         \\
Mean[Outcome $ | $ Black]&    0.203         &    0.203         &    0.203         &    0.203         &    0.203         &    0.148         \\
Mean[Outcome $ | $ Mulatto]&    0.219         &    0.219         &    0.219         &    0.219         &    0.219         &    0.170         \\
Mean[Outcome $ | $ White]&    0.201         &    0.201         &    0.201         &    0.201         &    0.201         &    0.187         \\
Count Mixed Siblings&    5,781         &    5,781         &    5,781         &    5,781         &    5,781         &    1,440         \\
\bottomrule
\multicolumn{7}{l}{\footnotesize Standard errors in parentheses}\\
\multicolumn{7}{l}{\footnotesize \sym{*} \(p<0.10\), \sym{**} \(p<0.05\), \sym{***} \(p<0.01\)}\\
\end{tabular}
}

%% file: input/Tables/regressions_south_plus20_byuauto_combined_black_mulatto2_wgtd_wfamfe.tex
{
\def\sym#1{\ifmmode^{#1}\else\(^{#1}\)\fi}
\begin{tabular}{l*{6}{c}}
\toprule
                &\multicolumn{1}{c}{(1)}         &\multicolumn{1}{c}{(2)}         &\multicolumn{1}{c}{(3)}         &\multicolumn{1}{c}{(4)}         &\multicolumn{1}{c}{(5)}         &\multicolumn{1}{c}{(6)}         \\
\midrule
Black - Mulatto &    0.099\sym{***}&    0.020\sym{***}&    0.017\sym{***}&    0.016\sym{***}&    0.008         &   -0.001         \\
                &  (0.002)         &  (0.001)         &  (0.001)         &  (0.001)         &  (0.005)         &  (0.007)         \\
Mulatto - White &    0.482\sym{***}&   -0.041\sym{***}&   -0.050\sym{***}&   -0.052\sym{***}&                  &                  \\
                &  (0.003)         &  (0.001)         &  (0.001)         &  (0.001)         &                  &                  \\
(Year)X(Age+Birth Order) FE &      Yes         &      Yes         &      Yes         &      Yes         &      Yes         &      Yes         \\
(Year)X(State) FE &       No         &      Yes         &       No         &       No         &       No         &       No         \\
(Year)X(Enum. Dist.) FE &       No         &       No         &      Yes         &      Yes         &      Yes         &       No         \\
(Year)X(Fam. Char.) &       No         &       No         &       No         &      Yes         &      Yes         &       No         \\
BMS Interactions &       No         &       No         &       No         &       No         &      Yes         &       No         \\
(Year)X(Fam.) FE &       No         &       No         &       No         &       No         &       No         &      Yes         \\
\midrule
Observations    &25,842,325         &25,842,325         &25,841,935         &25,841,935         &25,841,935         &11,497,769         \\
R-Squared       &    0.172         &    0.799         &    0.807         &    0.808         &    0.808         &    0.938         \\
Count Black     &1,263,923         &1,263,923         &1,263,911         &1,263,911         &1,263,911         &  360,022         \\
Count Mulatto   &  314,335         &  314,335         &  314,330         &  314,330         &  314,330         &  100,041         \\
Count White     &24,264,067         &24,264,067         &24,263,694         &24,263,694         &24,263,694         &11,037,706         \\
Mean[Outcome $ | $ Black]&    0.788         &    0.788         &    0.788         &    0.788         &    0.788         &    0.799         \\
Mean[Outcome $ | $ Mulatto]&    0.688         &    0.688         &    0.688         &    0.688         &    0.688         &    0.689         \\
Mean[Outcome $ | $ White]&    0.207         &    0.207         &    0.207         &    0.207         &    0.207         &    0.210         \\
Count Mixed Siblings&   13,974         &   13,974         &   13,974         &   13,974         &   13,974         &    6,104         \\
\bottomrule
\end{tabular}
}

%% file: input/Tables/regressions_women_south_plus20_byuauto_combined_black_mulatto2_wgtd_wfamfe.tex
{
\def\sym#1{\ifmmode^{#1}\else\(^{#1}\)\fi}
\begin{tabular}{l*{6}{c}}
\toprule
                &\multicolumn{1}{c}{(1)}         &\multicolumn{1}{c}{(2)}         &\multicolumn{1}{c}{(3)}         &\multicolumn{1}{c}{(4)}         &\multicolumn{1}{c}{(5)}         &\multicolumn{1}{c}{(6)}         \\
\midrule
Black - Mulatto &    0.086\sym{***}&    0.011\sym{***}&    0.010\sym{***}&    0.010\sym{***}&   -0.018\sym{**} &    0.007         \\
                &  (0.003)         &  (0.001)         &  (0.001)         &  (0.001)         &  (0.007)         &  (0.013)         \\
Mulatto - White &    0.493\sym{***}&   -0.028\sym{***}&   -0.036\sym{***}&   -0.037\sym{***}&                  &                  \\
                &  (0.003)         &  (0.001)         &  (0.001)         &  (0.001)         &                  &                  \\
(Year)X(Age+Birth Order) FE &      Yes         &      Yes         &      Yes         &      Yes         &      Yes         &      Yes         \\
(Year)X(State) FE &       No         &      Yes         &       No         &       No         &       No         &       No         \\
(Year)X(Enum. Dist.) FE &       No         &       No         &      Yes         &      Yes         &      Yes         &       No         \\
(Year)X(Fam. Char.) &       No         &       No         &       No         &      Yes         &      Yes         &       No         \\
BMD Interactions &       No         &       No         &       No         &       No         &      Yes         &       No         \\
(Year)X(Fam.) FE &       No         &       No         &       No         &       No         &       No         &      Yes         \\
\midrule
Observations    &17,040,433         &17,040,433         &17,039,884         &17,039,883         &17,039,883         &6,277,543         \\
R-Squared       &    0.160         &    0.823         &    0.831         &    0.831         &    0.831         &    0.947         \\
Count Black     &  538,194         &  538,194         &  538,173         &  538,173         &  538,173         &   90,979         \\
Count Mulatto   &  153,073         &  153,073         &  153,066         &  153,066         &  153,066         &   30,905         \\
Count White     &16,349,166         &16,349,166         &16,348,645         &16,348,644         &16,348,644         &6,155,659         \\
Mean[Outcome $ | $ Black]&    0.794         &    0.794         &    0.794         &    0.794         &    0.794         &    0.787         \\
Mean[Outcome $ | $ Mulatto]&    0.707         &    0.707         &    0.707         &    0.707         &    0.708         &    0.683         \\
Mean[Outcome $ | $ White]&    0.213         &    0.213         &    0.213         &    0.213         &    0.213         &    0.217         \\
Count Mixed Siblings&    5,781         &    5,781         &    5,781         &    5,781         &    5,781         &    1,440         \\
\bottomrule
\multicolumn{7}{l}{\footnotesize Standard errors in parentheses}\\
\multicolumn{7}{l}{\footnotesize \sym{*} \(p<0.10\), \sym{**} \(p<0.05\), \sym{***} \(p<0.01\)}\\
\end{tabular}
}

%% file: input/Tables/distinctivename_diff_mixed_sibs.tex
{
\def\sym#1{\ifmmode^{#1}\else\(^{#1}\)\fi}
\begin{tabular}{l*{4}{c}}
\toprule
                &\multicolumn{2}{c}{\shortstack{General Population}}&\multicolumn{2}{c}{\shortstack{Differently-Classified\\ Siblings}}\\\cmidrule(lr){2-3}\cmidrule(lr){4-5}
                &\multicolumn{1}{c}{(1)}&\multicolumn{1}{c}{(2)}&\multicolumn{1}{c}{(3)}&\multicolumn{1}{c}{(4)}\\
                &\multicolumn{1}{c}{Sons}&\multicolumn{1}{c}{Daughters}&\multicolumn{1}{c}{Brothers}&\multicolumn{1}{c}{Sisters}\\
\midrule
Black - Mulatto &    0.018\sym{***}&    0.014\sym{***}&    0.002         &    0.001         \\
                &  (0.000)         &  (0.000)         &  (0.002)         &  (0.002)         \\
Constant        &    0.211\sym{***}&    0.210\sym{***}&    0.229\sym{***}&    0.222\sym{***}\\
                &  (0.000)         &  (0.000)         &  (0.001)         &  (0.001)         \\
\midrule
Observations    &6,253,944         &6,346,235         &   50,333         &   53,762         \\
R-Squared       &    0.001         &    0.001         &    0.000         &    0.000         \\
Count Black     &5,202,057         &5,213,710         &   26,427         &   27,958         \\
Count Mulatto   &1,051,887         &1,132,525         &   23,906         &   25,804         \\
Mean[Outcome $ | $ Black]&    0.228         &    0.224         &    0.231         &    0.223         \\
Mean[Outcome $ | $ Mulatto]&    0.211         &    0.210         &    0.229         &    0.222         \\
\bottomrule
\end{tabular}
}

%% file: appendix_race_instructions.tex
\begin{itemize}
\item \textbf{1870: }\emph{``Color }White (W), black (B), mulatto (M),
Chinese (C), Indian (I)... It must not be assumed that, where nothing
is written in this column, ``White'' is to be understood. The column
is always to be filled. Be particularly careful in reporting the class
\emph{Mulatto}. The word is here generic, and includes quadroons,
octoroons, and all persons having any perceptible trace of African
blood. Important scientific results depend upon the correct determination
of this class in schedules 1 and 2.''
\item \textbf{1880: }\emph{``Color }It must not be assumed that, where
nothing is written in this column, ``white'' is to be understood.
The column is always to be filled. Be particularly careful in reporting
the class mulatto. The word is here generic, and includes quadroons,
octoroons, and all persons having any perceptible trace of African
blood. Important scientific results depend upon the correct determination
of this class in schedules 1 and 5.''
\item \textbf{1900: }\emph{``Color or race}\textbf{ }Write ``W'' for
white; ``B'' for black (negro or of negro descent); ``Ch'' for
Chinese; ``JP'' for Japanese, and ``In'' for Indian, as the case
may be.''
\item \textbf{1910:} ``\emph{Color or race} Write ``W'' for white; ``B''
for black; ``Mu'' for mulatto; ``Ch'' for Chinese; ``Jp'' for
Japanese; ``In'' for Indian. For all persons not falling within
one of these classes, write ``Ot'' (for other), and write on the
left-hand margin of the schedule the race of the person so indicated...
For census purposes, the term ``black'' (B) includes all persons
who are evidently full-blooded negroes, while the term ``mulatto''
(Mu) includes all other persons having some proportion or perceptible
trace of negro blood.''
\item \textbf{1920:} ``\emph{Color or race} Write ``W'' for white, ``B''
for black; ``Mu'' for mulatto; ``In'' for Indian; ``Ch'' for
Chinese; ``Jp'' for Japanese; ``Fil'' for Filipino; ``Hin''
for Hindu; ``Kor'' for Korean. for all persons not falling within
one of these classes, write ``Ot'' (for other), and write on the
left-hand margin of the schedule the race of the person so indicated...
For census purposes the term ``black'' (B) includes all Negroes
of full blood, while the term ``mulatto'' (Mu) includes all Negroes
having some proportion of white blood.''
\item \textbf{1930: }``\emph{Color or race }Write ``W'' for white, ``B''
for black; ``Mex'' for Mexican; ``In'' for Indian; ``Ch'' for
Chinese; ``Jp'' for Japanese; ``Fil'' for Filipino; ``Hin''
for Hindu; ``Kor'' for Korean. For a person of any other race, write
the race in full.
\begin{itemize}
\item \emph{Negroes. }A person of mixed white and Negro blood should be
returned as a Negro, no matter how small the percentage of Negro blood.
Both black and mulatto persons are to be returned as Negroes, without
distinction. A person of mixed Indian and Negro blood should be returned
a Negro, unless the Indian blood predominates and the status as an
Indian is generally accepted in the community.
\item \emph{Indians. }A person of mixed white and Indian blood should be
returned as Indian, except where the percentage of Indian blood is
very small, or where he is regarded as a white person by those in
the community where he lives. (See par. 151 for mixed Indian and Negro.)
For a person reported as Indian in column 12, report is to be made
in column 19 as to whether ``full blood'' or ``mixed blood,''
and in column 20 the name of the tribe is to be reported. For Indians,
columns 19 and 20 are thus to be used to indicate the degree of Indian
blood and the tribe, instead of the birthplace of father and mother.
\item \emph{Mexicans. }Practically all Mexican laborers are of a racial
mixture difficult to classify, though usually well recognized in the
localities where they are found. In order to obtain separate figures
for this racial group, it has been decided that all person born in
Mexico, or having parents born in Mexico, who are not definitely white,
Negro, Indian, Chinese, or Japanese, should be returned as Mexican
(``Mex'').
\item \emph{Other mixed races. }Any mixture of white and nonwhite should
be reported according to the nonwhite parent. Mixtures of colored
races should be reported according to the race of the father, except
Negro-Indian (see par. 151).''
\end{itemize}
\item \textbf{1940: }\emph{Color or Race} Write ``W'' for white; ``Neg''
for Negro; ``In'' for Indian; ``Chi'' for Chinese; ``Jp'' for
Japanese; ``Fil'' for Filipino; ``Hi'' for Hindu; and ``Kor''
for Korean. For a person of any other race, write the race in full.
\begin{itemize}
\item \emph{Mexicans.} Mexicans are to be regarded as white unless definitely
of Indian or other nonwhite race.
\item \emph{Negroes. }A person of mixed white and Negro blood should be
returned as Negro, no matter how small a percentage of Negro blood.
Both black and mulatto persons are to be returned as Negroes, without
distinction. A person of mixed Indian and Negro blood should be returned
as a Negro, unless the Indian blood very definitely predominates and
he is universally accepted in the community as an Indian.
\item \emph{Indians. }A person of mixed white and Indian blood should be
returned as an Indian, if enrolled on an Indian agency or reservation
roll, or if not so enrolled, if the proportion of Indian blood is
one-fourth or more, or if the person is regarded as an Indian in the
community where he lives.
\item \emph{Mixed Races.} Any mixture of white and nonwhite should be reported
according to the nonwhite parent. Mixtures of nonwhite races should
be reported according to the race of the father, except that Negro-Indian
should be reported as Negro.''
\end{itemize}
\end{itemize}

%% file: input/Tables/usct_complexion.tex
{
\def\sym#1{\ifmmode^{#1}\else\(^{#1}\)\fi}
\begin{tabular}{lcccc}
\toprule
 & (1) & (2) & (3) & (4) \\
\midrule
Mulatto - Black & 0.109\sym{***} & 0.116\sym{***} & 0.112\sym{***} & 0.116\sym{***} \\
 & (0.030) & (0.029) & (0.033) & (0.032) \\
Company fixed effects & No & Yes & No & Yes \\
\midrule
Observations &     2,330 &     2,330 &     1,672 &     1,672 \\
R-Squared & 0.015 & 0.267 & 0.015 & 0.286 \\
Count Black &     2,020 &     2,020 &     1,459 &     1,459 \\
Count Mulatto &       310 &       310 &       213 &       213 \\
Mean[Outcome $ | $ Black] & 0.091 & 0.091 & 0.090 & 0.090 \\
Mean[Outcome $ | $ Mulatto] & 0.200 & 0.200 & 0.202 & 0.202 \\
Companies &       117 &       117 &       117 &       117 \\
Census-link quality & 1--2 & 1--2 & 1 & 1 \\
\bottomrule
\multicolumn{5}{l}{\footnotesize Standard errors in parentheses}\\
\multicolumn{5}{l}{\footnotesize \sym{*} \(p<0.10\), \sym{**} \(p<0.05\), \sym{***} \(p<0.01\)}\\
\end{tabular}
}
\par\smallskip
\begin{minipage}{\linewidth}\small
Note: Table shows regressions measuring gaps in an indicator for brown/light complexion in Union Army records, by 1870 Census classification among veterans in the Expanded U.S. Colored Troops data. Columns (1) and (2) retain Census-link quality 1--2, and columns (3) and (4) retain quality 1. Columns (2) and (4) add company fixed effects. The Early Indicators project's four-point quality scale runs from 1 (strongest) to 4 (weakest). The bottom part of the table presents observation counts and mean outcomes by Census classification for each column's estimation sample. Observations are unweighted. Standard errors are clustered by military company. The outcome equals one for brown/light (Army codes BR/LT) and zero for black/dark (BK/DK); other and missing complexion labels are excluded. Complexion groups use only wartime records. Documented Census-link criteria do not list complexion or Census color, although complete blinding of manual linkage cannot be established.
\end{minipage}